\documentclass[twocolumn]{aastex631}
\accepted{March 31, 2023}

\submitjournal{ApJ}

\usepackage{amsmath,environ}
\usepackage{listings}
\usepackage{float}
\usepackage{color}
\usepackage{mathtools}
\usepackage{rotating}

\defcitealias{lelli2016}{SPARC}
\defcitealias{springob2005}{S05}
\defcitealias{tifft1988}{T88}
\defcitealias{rots1980}{R80}
\defcitealias{koribalski2004}{HIPASS}
\defcitealias{haynes18}{ALFALFA}
\defcitealias{courtois2009}{EDD}
\defcitealias{papastergis16}{P16}

\newcommand{\hi}{H\,{\sc i}\ }
\newcommand{\wfvty}{$W50$\,}
\newcommand{\wfvtym}{$W50_\mathrm{model}$\,}
\newcommand{\vf}{$v_\mathrm{flat}$\,}
\newcommand{\cii}{[C\,{\sc ii}]\,}

\begin{document}

\title{Parametrized Asymmetric Neutral hydrogen Disk Integrated Spectrum Characterization (PANDISC) I: Introduction to A Physically Motivated \hi Model}

\correspondingauthor{Bo Peng}
\email{bp392@cornell.edu}

\author[0000-0002-1605-0032]{Bo Peng}
\affiliation{Department of Astronomy, Cornell University, Ithaca, NY 14853, USA}

\author[0000-0001-5334-5166]{Martha P. Haynes}
\affiliation{Cornell Center for Astrophysics and Planetary Science, Space Sciences Building, Cornell University, Ithaca, NY 14853, USA}

\author[0000-0002-1895-0528]{Catie J. Ball}
\affiliation{Department of Astronomy, Cornell University, Ithaca, NY 14853, USA}

\author[0000-0002-5434-4904]{Michael G. Jones}
\affiliation{Steward Observatory, University of Arizona, 933 North Cherry Avenue, Tucson, AZ 85721-0065, USA}

\begin{abstract}

Modelling the integrated \hi spectra of galaxies has been a difficult task due to their diverse shapes, but more dynamical information is waiting to be explored in \hi line profiles. 
Based on simple assumptions, we construct a physically motivated model for the integrated \hi spectra: Parametrized Asymmetric Neutral hydrogen Disk Integrated Spectrum Characterization (PANDISC). 
The model shows great flexibility in reproducing the diverse \hi profiles. 
We use Monte-Carlo Markov Chain (MCMC) for fitting the model to global \hi profiles and produce statistically robust quantitative results. 
Comparing with several samples of \hi data available in the literature , we find the model-fitted width agree with catalogued velocity widths (e.g., \wfvty) down to S/N $\lesssim$ 6. 
While dynamical information can only be extracted reliably from spectra with S/N $>$ 8.
The model is also shown to be useful for applications like the baryonic Tully-Fisher relation (BTFR) and profile-based sample control. 
By comparing the model parameter $v_r$ to \vf, we uncover how the \hi width is affected by the structure of the rotation curve, following a trend consistent with the difference in the BTFR slope. 
We also select a sample of spectra with broad wing-like features suggestive of a population of galaxies with unusual gas dynamics.
The PANDISC model bears both promise and limitations for potential use beyond \hi lines.
Further application on the whole ALFALFA sample will enable us to perform large scale ensemble studies of the \hi properties and dynamics in nearby galaxies.

\end{abstract}


\section{Introduction}
\label{sec:intro}

Single dish observations of the \hi 21~cm hyperfine structure line has enabled the study of many aspects of galaxies, such as the redshift distribution, neutral gas mass function \citep{roberts1974, jones2018}, Tully-Fisher relation \citep{tully1977} and Baryonic Tully-Fisher relation \citep[BTFR,][]{mcgaugh2000}. 
Future single dish surveys such as the ongoing CRAFTS extragalactic \hi survey with FAST \citep{zhang21} promises a wider and more complete picture of the neutral gas in the local Universe.
But even for interferometric surveys like Apertif \citep{adams22} and WALLABY \citep{koribalski20,westmeier22}, a significant fraction of the expected detections would be only marginally (less than three beams) resolved or unresolved. 
And the unresolved fraction is expected to be even higher for the future deep \hi surveys like LADUMA \citep{blyth16} and DINGO \citep{meyer09}.
Therefore, there is still strong need to develop techniques for analyzing the integrated, spatially unresolved \hi spectrum to study the distribution and kinematics of the \hi gas.

Most studies of the integrated \hi spectrum only measure the redshift, flux and width of the line \citep[e.g.][]{chengalur1993, springob2005}, but 
more pieces of information are encoded in the global \hi profile, including the asymmetry \citep{richter94}, line shape, gas dynamics, wing-like component, etc.
But because those features are more difficult to quantify and measure, their scientific potential remains to be fully explored. 
Besides, previous large \hi surveys have typically relied on human inspection in both source identification and line width measurements \citep[e.g.][]{koribalski2004,haynes18}. 
Although manual reduction performs well in handling the diverse \hi profiles, such a human-dependent approach lacks consistency and statistical rigour, and is very difficult to scale up to the number of sources that will be detected by the next generation surveys.
In contrast, a parametrized model has the benefits of (1) getting parametrized descriptions of line profiles which enable comparison and sample control; (2) extracting more dynamical information from the integrated spectrum; (3) providing statistically robust descriptions of the spectral line for ensemble study. 

Due to the complexity and diversity of global \hi line profiles, modelling the global \hi spectrum has long proved to be a difficult task.
Even for high S/N spectra, the main challenges are the ability to describe both the double horn and single peak profiles in the same framework, as well as the varying degree of asymmetry.
There has been numerous previous attempts to model the integrated \hi profile. 
Recent examples include the use of Hermite functions in \cite{saintonge2007}, using a segmented function to describe the trough and edge in \citet{springob2005,jones2018}, and the Busy function introduced by \citet{westmeier2014} which connects two damped parabolic functions on each half of the spectral line to account for varying line shape and asymmetry, making it the most versatile line model so far. 
However, these models are mostly purely mathematical descriptions, making them obscure in physical meaning.
In addition, they are based on rather arbitrary math forms and only focus on the phenomenological descriptions of line profiles, complicating the interpretation and applicability. 

In this paper we introduce PANDISC (Parametrized Asymmetric Neutral hydrogen Disk Integrated Spectrum Characterization), a physically motivated parametrized model for global \hi line profiles. 
The model is based on simple physical assumptions which combine an asymmetric co-rotating disk component with a gaussian component.
The model consists of seven parameters, with five of them controlling the shape of the profile and the other two setting the line center and total flux. 
Note that seven is also the number of parameters needed for the Busy function \citep{westmeier2014}. 
In Sec.~\ref{sec:model}, we describe the assumptions and formulation of the model.
Sec.~\ref{sec:sample_data} describes the data and galaxy samples used for different tests, followed by the Monte Carlo Markov Chain (MCMC) fitting routine in Sec.~\ref{sec:fitting} and comments on fitting quality in Sec.~\ref{sec:quality}.
In Sec.~\ref{sec:results}, we demonstrate various applications of the model, including the ability to parametrize spectra down to low S/N in Sec.~\ref{sec:demo}, application of the BTFR and profile based sample control in Sec.~\ref{sec:p16}, and the physical meaning of model fitted line width by comparison with \vf and other definition of line widths in Sec.~\ref{sec:sparc}. 
In Sec.~\ref{sec:discussion}, we discuss the caveats of the model, the broad wing candidates which it identifies, and the potential application of the model beyond \hi spectra.
Sec.~\ref{sec:summary} summarizes the capabilities, limitations and prospects of the PANDISC model which should be recognized in future applications

\section{\texorpdfstring{\hi}{H I} Line model}
\label{sec:model}

\subsection{Model assumptions}
\label{sec:assumptions}

\begin{figure*}
	\centering
	\includegraphics[width = \textwidth]{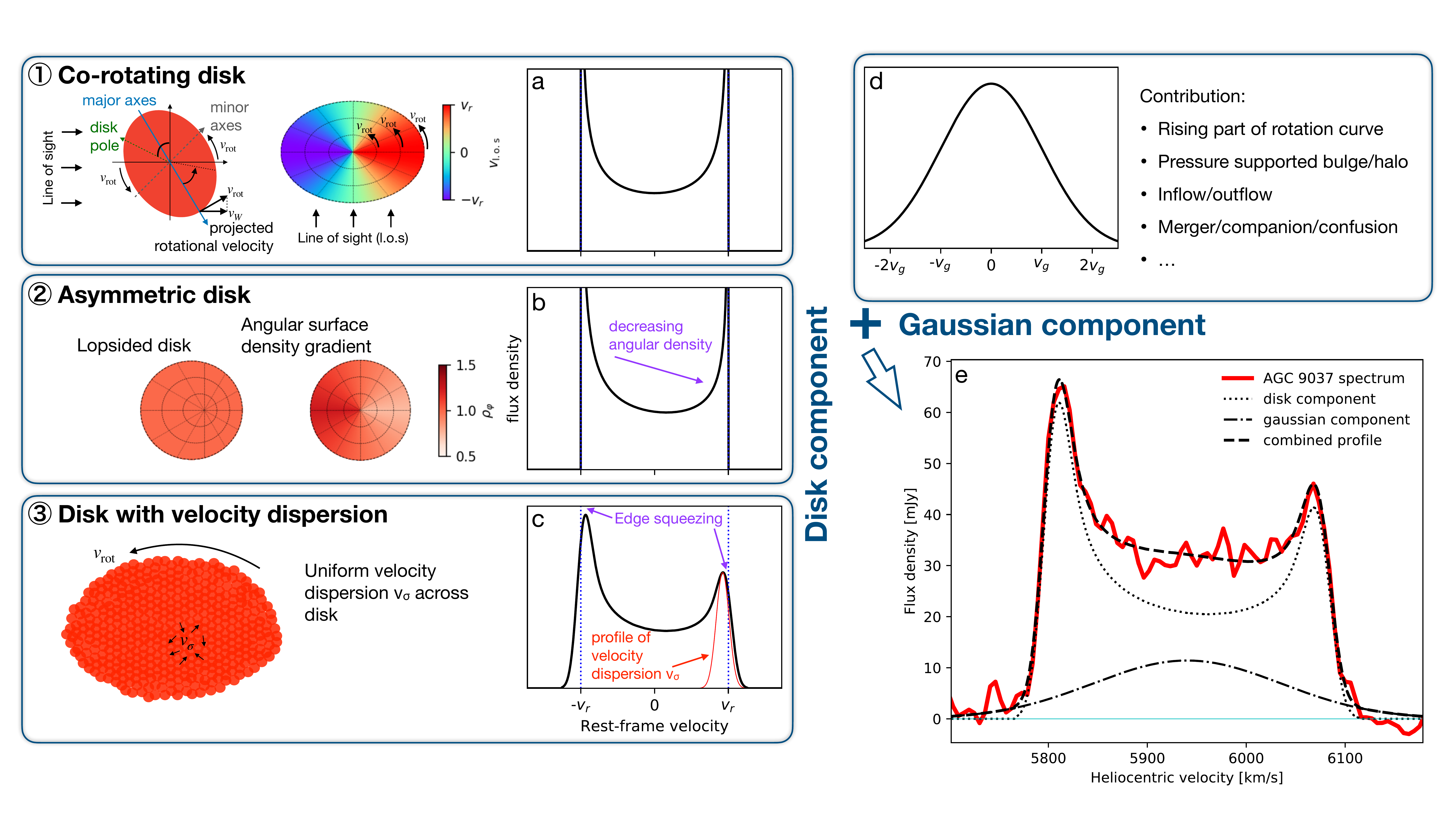}
	\caption{Construction of the model. The left panels 1 to 3 show how the disk component is built. Panel 1 plots a co-rotating disk tilted at an inclination $\theta$, with the projected velocity field shown on the right, and its global profile shown in panel a. Panel 2 demonstrates an asymmetric disk with a lopsided disk on the left, and the model assumed constant angular density gradient $k$ on the right as a mathematical approximation of the lopsided disk for our purpose. The global profile of such an asymmmetric disk is plotted in panel b. Panel 3 shows the model assumed velocity dispersion, which could be either cloud-wise random motions, or turbulence within gas clouds. The resultant profile convolved with the velocity dispersion $v_\sigma$ is plotted in panel c, with the effect of edge squeezing highlighted. Also plotted is the Gaussian peak $v_\sigma$ used for the convolution in red. The upper right panel illustrates the profile and the possible origin of the gaussian component. The lower right panel plots the disk component (black dotted), the gaussian component (black dash dotted) and the combined model (thick black dashed) over the ALFALFA spectrum of UGC 9037 (CGCG 046-060; thick red) as an example for its mildly asymmetric and double-horn shape, showing excellent agreement between the PANDISC model and the observational data.}
	\label{f:intro}
\end{figure*}

In this section we step through all the assumptions of the PANDISC model, and how the parametrized description is formulated.
A graphical explanation is shown in Fig.~\ref{f:intro} as a visual aid.

The most important assumption of this model is that the \hi disk is rotating at the same velocity at all radii. 
This assumption is motivated by the facts that the rotation curve is typically found to be flat beyond stellar disk scale length while the \hi disk is much more extended than the stellar disk, so that a significant fraction of the neutral hydrogen samples the flat part of the rotation curve \citep{catinella06}. 
This assumption greatly reduces the complexity associated with the disk modelling for the \hi gas, by ignoring the radial dependence of the velocity, and removing the need for the density information since the emission from a co-rotating disk can be considered in the same way as a rotating ring. 
The resulting spectral profile of a disk rotating at $v_\mathrm{rot}$ and inclination angle $\theta$ is given by
\begin{equation}\begin{split}
\label{equ:disk_rot}
	\frac{\mathrm{d}F}{\mathrm{d}v} \propto \left| \frac{\mathrm{d}}{\mathrm{d}v} \arccos \left( \frac{v}{v_r} \right) \right| = \frac{1}{\sqrt{v_r^2-v^2}}
\end{split}\end{equation}
where $v_r = v_\mathrm{rot} \sin \theta$ is the projected rotation velocity, and $v$ denotes the line velocity along line of sight (L.o.S.).
It is worth noting that the inclination $\theta$ is degenerate with $v_\mathrm{rot}$ throughout the whole model, thus only the projected velocity $v_r$ is used, and $\theta$ cannot be inferred from the integrated line profile alone.

The second assumption of the model aims to account for the asymmetry of the \hi profile, which is often associated with the uneven distribution of the neutral hydrogen \citep{haynes98}. 
There are many possible physical causes of the non-uniform distribution, including tails or elongated morphology due to tidal interaction \citep{toomre72}, uneven surface density associated with lopsidedness \citep{baldwin80}, unevenly distributed regions with depleted \hi such as H\,{\sc ii} regions in spiral arms, etc. 
Based on the idea of non-uniform distribution, we assume a variation of the angular distribution of the neutral gas. 
For mathematical simplicity, we assume this variation to be a constant gradient of the angular density of the \hi gas from one end of the disk projected on the sky to the other end, namely 
\begin{equation}\begin{split}
\label{equ:k}
	k = \frac{ \mathrm{d} \rho_\varphi}{\mathrm{d}\varphi} = Const.
\end{split}\end{equation}
where \(\rho_\varphi = \frac{\mathrm{d} m}{\mathrm{d} \varphi} \frac{2\pi}{M}\) is the normalized angular density at an angle $\phi$ on a disk of mass $M$, with the receding side of the rotating disk defined as the origin (see Fig.~\ref{f:intro} panel 1). 
$k$ is defined as the constant gradient, varying in the range $-2/\pi$ to $2/\pi$. 
The resulting asymmetric line profile is
\begin{equation}\begin{split}
\label{equ:disk_asym}
	\frac{\mathrm{d} F}{\mathrm{d} v} \propto \frac{1}{\sqrt{v_r^2-v^2}}  \left\{ 1 + k \left[ \arccos \left( \frac{v}{v_r} \right) - \frac{\pi}{2} \right] \right\}
\end{split}\end{equation}

Another important parameter in modelling a rotating disk is the velocity dispersion, which creates the smooth edge of the \hi line and squeezes the peak width narrower than the raw profile. 
For simplicity, we use a single variable $v_\sigma$ to describe the velocity dispersion. 
The raw line profile is hence convolved with a Gaussian kernel characterized by $v_\sigma$
\begin{equation}\begin{split}
\label{equ:disk_sigma}
	\frac{\mathrm{d}F}{\mathrm{d}v} \propto \int_0^\pi [1 + k(\varphi - \pi/2)]\exp\left[- \frac{(v_r \cos \varphi - v)^2}{2 v_\sigma^2}  \right] \mathrm{d} \varphi
\end{split}\end{equation}
Because of the \(\varphi \exp (\cos \varphi)\) term, this expression is not analytically integrable. 
It is worth noting that when applied to the observed spectrum, the fitted $v_\sigma$ will also include the instrumental smoothing due to limited spectral resolution. 

In addition to asymmetry, another major obstacle to modelling the \hi line profile is the flexibility needed to account for flat-top and sometimes single-peaked Gaussian-like shapes. 
Here we resolve the issue by simply adding a Gaussian peak in addition to the co-rotating disk. 
This ``gaussian'' component is centered at the same velocity as the ``disk'' component, and its shape is characterized by a single variable $v_g$, which is the standard deviation (STD) of this Gaussian peak, controlling the width.

The relative height contrast of the disk to the gaussian component is set by the variable $r$, defined as the fraction of the disk component flux in the total integrated line flux. 
And the absolute height of the line is set by the variable $F$, namely the integrated line flux of the model. 
Finally, we add the line center velocity $v_c$ to complete the model, and we can get the generic expression of the model flux density as a function of velocity $F_v$
\begin{equation}\begin{split}
\label{equ:model}
	F_v = &F_{v, \mathrm{disk}} + F_{v, \mathrm{gaus}} \\
	= & F \times \bigg\{  \frac{r}{\sqrt{2\pi} v_\sigma \pi} \int_0^\pi [1 + k(\varphi - \pi/2)] \\
	& \exp\left[- \frac{(v_r \cos \varphi + v_c - v)^2}{2 v_\sigma^2}  \right] \mathrm{d} \varphi + \\
	& \frac{1-r}{\sqrt{2\pi} v_g} \exp \left[-\frac{(v_c - v)^2}{2 v_g^2} \right] \bigg\}
\end{split}\end{equation}
The line model has in total 7 variables summarized below

\begin{itemize}
	\item $v_r$: the projected co-rotating velocity, characterizing the un-dispersed width of the ``disk'' component;
	\item $k$: the gradient of the angular density of the ``disk'', characterizing the asymmetry of the line;
	\item $v_\sigma$: the velocity dispersion of the ``disk'', controlling the steepness of the line edge;
	\item $v_g$: the STD of the ``gaussian'' component, controlling the width of the Gaussian peak;
	\item $r$: ``disk'' flux fraction
	\item $F$: integrated line flux
	\item $v_c$: heliocentric velocity of the line center
\end{itemize}

\subsection{Model properties}
\label{sec:properties}

\begin{figure*}
	\centering
	\includegraphics[width = \textwidth]{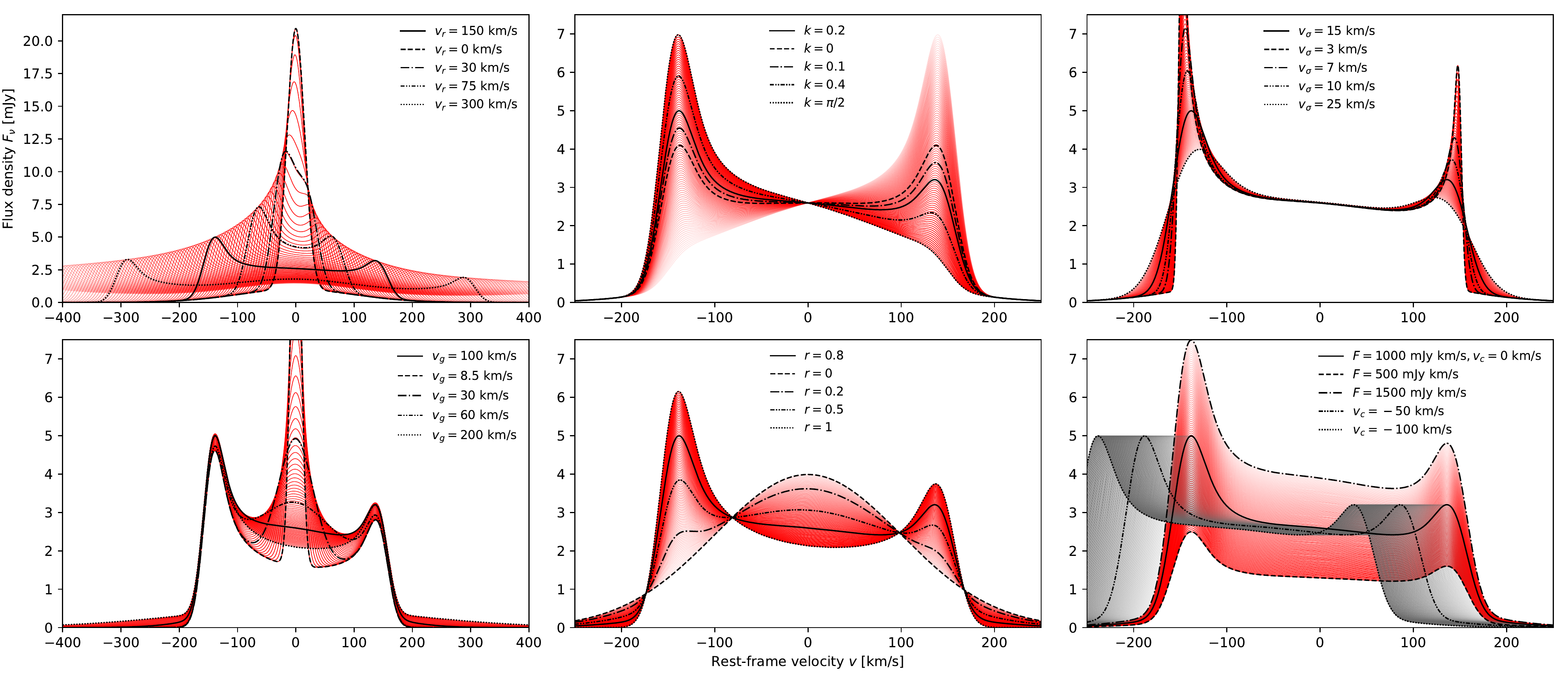}
	\caption{Effect of each variable in the model. In each panel except the lower right one, the model spectra are plotted in thin red lines with one variable being varied from the smallest value to the largest value in smoothly varying depth of color, and the models with specified values (as indicated in each pannel) are plotted as black thick lines and in different line styles to aid the reader. The lower right panel shows models with varying $F$ in thin red lines and varying $v_c$ in thin grey lines. All panels show the model with a common set of variables $v_r=150 \ \mathrm{km/s}$, $k=0.2$, $v_\sigma=15 \ \mathrm{km/s}$, $v_g=100 \ \mathrm{km/s}$, $r=0.8$, $F=1000 \ \mathrm{mJy~km/s}$ and $v_c =0 \ \mathrm{km/s}$ in thick black solid lines for comparison.}
	\label{f:parameters}
\end{figure*}

\begin{figure*}
	\centering
	\includegraphics[width= \textwidth]{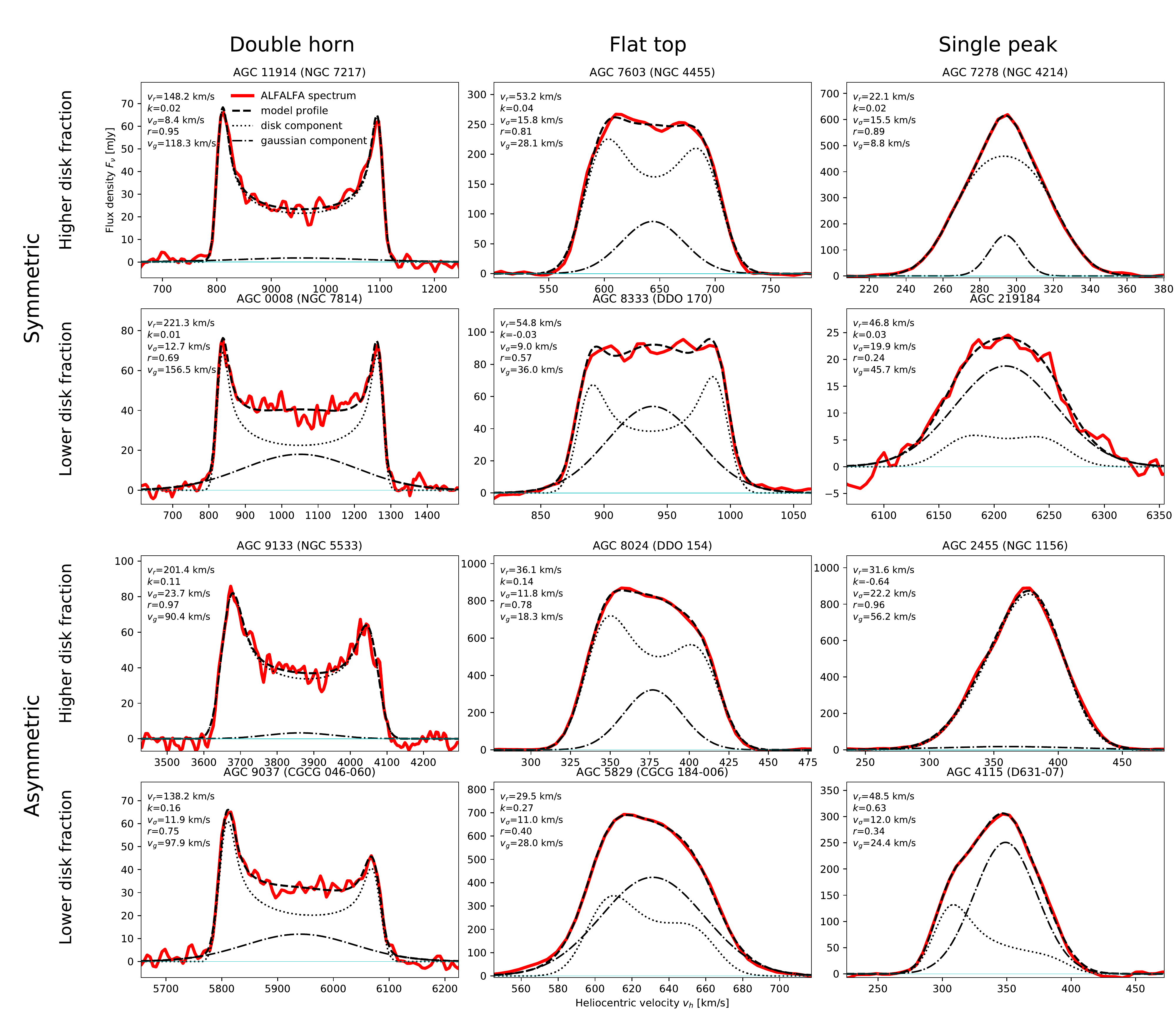}
	\caption{Atlas of the integrated \hi line profile. The raw spectra are shown in red, and the best fit model, disk, and gaussian components are plotted as black dashed, dotted, and dot-dashed lines respectively. The median fit parameters are also printed, omitting the flux and line centers which do not affect the shape. The spectra are first separated into three columns ``double horn'', ``flat top'' and ``single peak'' based on the number and the flatness of the peaks. The spectra are then classified into symmetric and asymmetric based on the model fit $k$, each occupying two rows. The spectra are further distinguished by higher and lower disk fraction based on the model fit $r$, demonstrating the effect of $r$ on the shape of the spectrum, as well as the wing-like features in some of the high disk spectra. The spectra are taken from the ALFALFA database, among the samples of galaxies used in the paper.}
	\label{f:atlas}
\end{figure*}

Because the model line profile has a non-linear dependence on most of the parameters, the effect of change in parameters is complex and is presented in Fig.~\ref{f:intro} and \ref{f:parameters}. 
As shown in the figures, the disk part of the model manifested as the double horn shape is modulated by $v_r$, $k$ and $v_\sigma$. 
But the width of the double horn is not only controlled by $v_r$, it is also affected by $v_\sigma$ by the ``edge squeezing'' effect, namely that the convolution shifts more fluxes in inner velocity channels close to the edges, hence shifting the apparent peaks away from the edge of the raw profile and narrowing the peak width (panel 2 in Fig.~\ref{f:intro}). 
In the highly asymmetric case, the width is also affected as the shape transitions from double horn to single peak. 

The purpose of including a gaussian component is to account for the flat-top and single peak profiles, which can be well described by mixing a double horn shape with a Gaussian peak. 
However, the Gaussian component also accounts for other features in the model like the flatness in the trough and broad wings extending beyond the line peaks. 

Because the past applications of integrated \hi spectra focus on the line width, we also provide here a method to estimate the commonly used \wfvty (the width at 50\% of the peak flux) using the model parameters, denoted as \wfvtym.
The formulation and derivation of \wfvtym are detailed in Appendix \ref{sec:width} along with other width estimates.

To demonstrate the power of the model in describing real \hi spectra, we plot an atlas of spectra from the ALFALFA survey \citep{haynes18} in Fig.~\ref{f:atlas}
The spectra are selected based on high S/N (to avoid, for demonstraction purpose, the impact of noise), shape, number of peaks (flux density maxima), and level of asymmetry.
They are further divided into sub-groups by the disk fraction in the model fitting results. 
The best fitted models (thick black dashed line) show great agreement with the observed spectra (red solid line).

Despite the agreement with real data, many simplifications are made in constructing the line model. 
The caveats about the model parameters and how they should be interpreted are discussed in detail in Sec.~\ref{sec:model_caveat}. 
The model is available online \footnote{The PANDISC package is available at \url{https://github.com/bpqdbpqd/pandisc}, DOI: 10.5281/zenodo.7739693} as a python package \citep{peng23} which performs the basic function of evaluating the model as well as computing the derived quantities like \wfvtym.

\section{Data and method}
\label{sec:data}

\subsection{Sample and data}
\label{sec:sample_data}

The Arecibo Legacy Fast ALFA Survey \citep[ALFALFA;][]{haynes18} 
produced a final  catalog (a.100) of $\sim$31,500 extragalactic \hi detections in the local Universe. 
The $\sim$4 arcmin Arecibo beam encloses most of the neutral hydrogen gas in a galaxy except for a few very nearby galaxies, and the integrated spectra are readily available\footnote{The ALFALFA data archive is available at \url{http://egg.astro.cornell.edu/alfalfa/data/index.php}}. 
However, the large beam size also raises the problem of source confusion, which is discussed in Appendix~\ref{sec:confusion}. 
The survey also covers a wide range of galaxy types and masses, from massive \hi disks to dwarf galaxies. 
ALFALFA is hence the largest and most comprehensive dataset available to study the integrated \hi profiles of galaxies. 
For the our purposes of demonstration and application, we selected sub-samples from the ALFALFA data and literature as described below; a future paper (Peng et al. in prep) will address the analysis of the entire ALFALFA database.

We first demonstrate the applicability of the model on a sample of high S/N data (high S/N sample).
387 galaxies are selected by the criteria such that the ALFALFA reported S/N is greater than 100, and heliocentric velocity $V_\mathrm{h}$ is not in the range $\pm 100\ \mathrm{km/s}$ to avoid confusion with Galactic \hi.
Because the data is highly reliable, we also use this sample to optimize the prior probability for the MCMC fitting (for details check Sec.~\ref{sec:fitting} and Appendix~\ref{sec:priori}), and to understand the occasional mismatches of the model and their possible causes.

To demonstrate the ability of model fitting on low S/N data, we select an un-biased random sample of ALFALFA galaxies based on line width and S/N (ALFALFA demonstration sample). 
The selection based on the line width is to mitigate the effect that the ALFALFA detections are preferentially narrow profiles.
The selection criteria are as follows: for each bin of \wfvty in the range [0, 100], [100, 160], and [160, 500] km/s, and each S/N bin of [0, 6], [6, 8], [8, 10], [10, 15], [15, 1000], 20 galaxies are randomly drawn from the catalogue, resulting in a sample of 300 galaxies in total. 

To demonstrate the application of the line model in BTFR and sample control, we applied it on the galaxy sample selected in \cite{papastergis16} \citepalias[hereafter][]{papastergis16}. 
The authors selected 97 highly inclined, gas rich galaxies detected by the ALFALFA survey to study the BTFR. 
The study also finds a dependence of the BTFR on the kurtosis of the \hi profiles, which is compared with our model based sample control.
The integrated spectra of these galaxies are readily available in ALFALFA.

To test the physical assumption of the ``projected co-rotating velocity'' for $v_r$, we apply the line model to the galaxies with rotation curves presented in \cite{lelli2016}, known as the \citetalias[hereafter][]{lelli2016} galaxies.
The SPARC sample consists of 175 disk galaxies with baryonic masses ranging from $10^8$ through $>10^{11}$ $M_\sun$. 
All the galaxies have had their rotation curves mapped with interferometric \hi observations\footnote{The rotation curve data are acquired at \url{http://astroweb.cwru.edu/SPARC/}}, with the outer flat part measured as the rotation velocity \vf. 
We did an extensive literature search for integrated, single-dish \hi observations for the \citetalias{lelli2016} sample. 
A total of 158 galaxies with \hi spectra available were cross matched with sample, including 51 in the ALFALFA catalog, 56 in \citet{springob2005} \citepalias[hereafter ][]{springob2005}, 11 in \citet{courtois2009} \citepalias[hereafter][]{courtois2009}, 10 in \citet{koribalski2004} \citepalias[hereafter][]{koribalski2004}, 27 in \citet{tifft1988} \citepalias[hereafter ][]{tifft1988} and 3 in \citet{rots1980} \citepalias[hereafter][]{rots1980}. 
The \hi spectra were collected using various instruments on several telescopes and spectrometers, with different channel size, bandwidth and noise characterization.
Therefore we use the auto correlation of the blank (line-free) channels in the spectra to infer the correlation scale of each spectrum, which is then used in the likelihood evaluation in MCMC fitting (see Sec.~\ref{sec:fitting}).

For galaxies included in \citetalias{springob2005} and \citetalias{courtois2009}, the spectra have a variety of velocity resolutions due to the diverse correlators used. 
Therefore both the rms and channel correlations are derived from fitting the blank channels of each spectrum individually.
The \citetalias{koribalski2004} and \citetalias{rots1980} spectra do not show correlation across channels, while the \citetalias{tifft1988} data are well fitted by a correlation of about 7 channels, which is the value used in the inference.

\subsection{Model fitting}
\label{sec:fitting} 

\begin{figure*}
	\centering
	\includegraphics[width = 0.7 \textwidth]{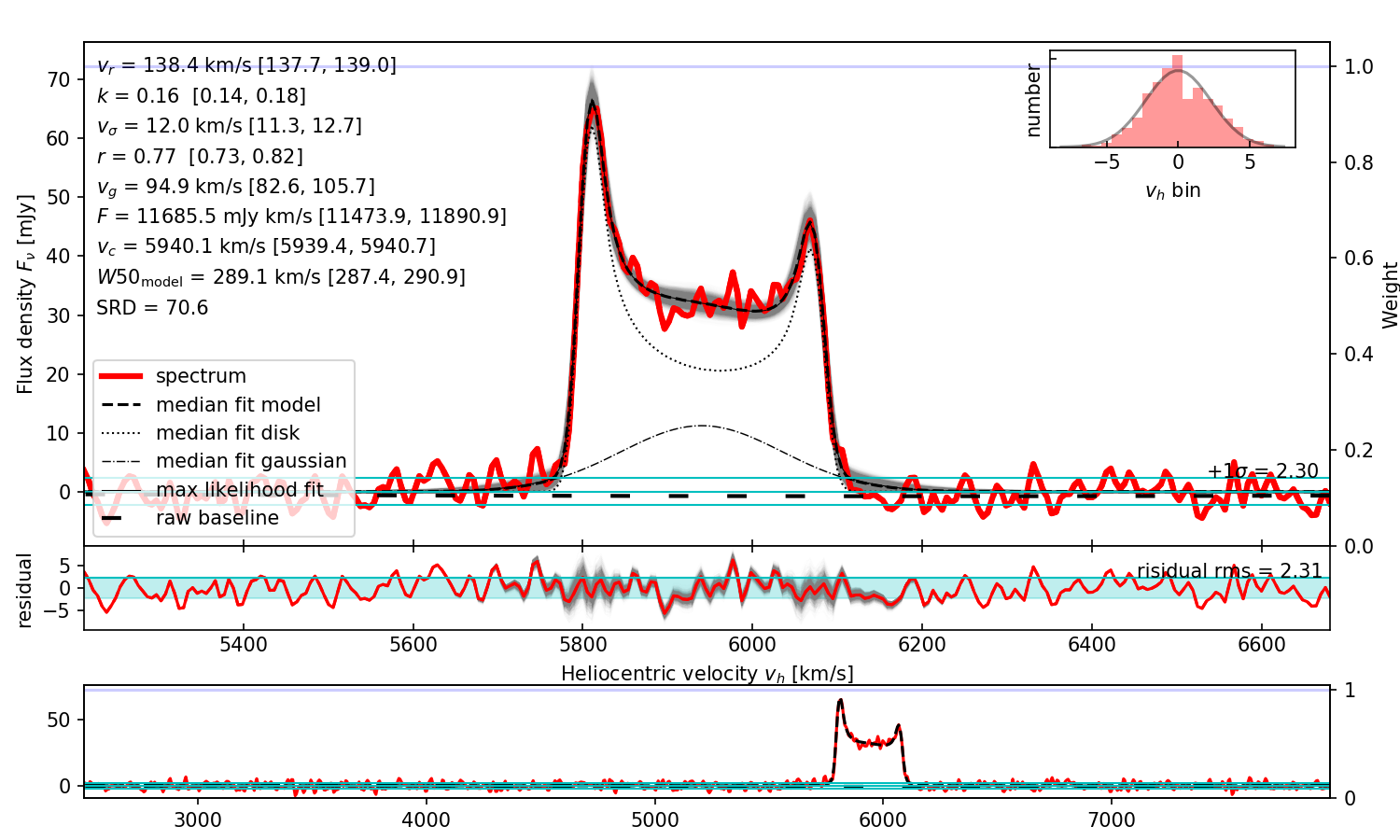}
	\includegraphics[width = 0.6 \textwidth]{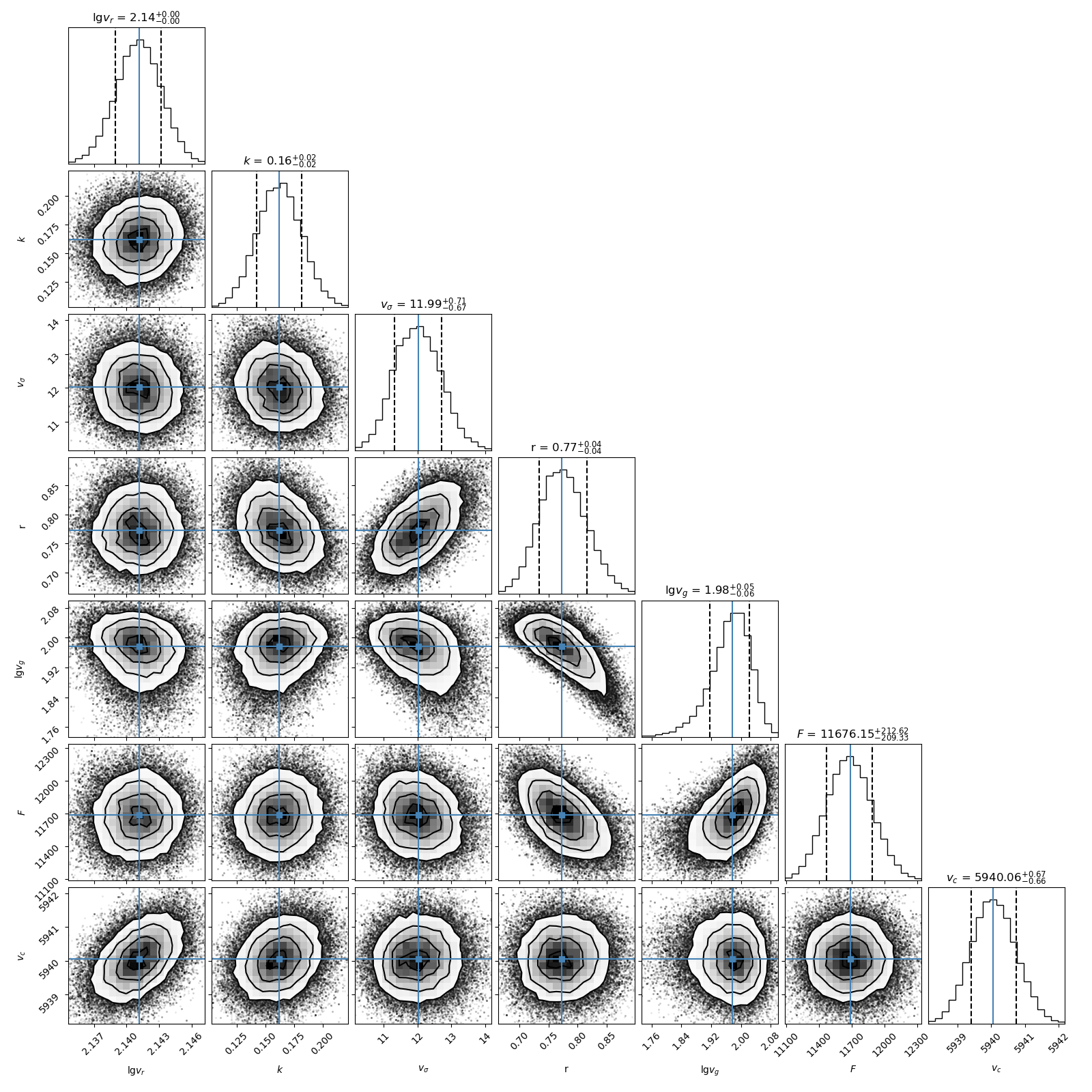}
	\caption{An example of the PANDISC model fitting result. In the upper panel of the upper figure, the spectrum of UGC 9037 (red solid if weight $>$ 0.25, red dashed if weight $\leq$ 0.25) is plotted with the model (black short dashed) and the model components (black thin dotted and dash-dotted) using the median values in the marginalized distribution. The model corresponding to the parameter set with the maximum likelihood in the recorded MCMC samples is also plotted as the thin black solid line. The subtracted raw baseline is plotted as a thick black long dashed line for reference where available. The fitted and derived values with 16th and 84th percentiles are printed on the upper left corner, and a histogram of the flux density in the blank channels not used for fitting is on the upper right to justify the rms$_\mathrm{blank}$ value. A pair of thin cyan lines are shown for $\pm \mathrm{rms_{blank}}$ with the actual value indicated. The grey shade is formed by plotting the model curves of 300 recorded MCMC samples. A thin blue line is plotted against the y-axis on the right to show the weight of the data in each channel. The middle panel shows the residual after subtracting the median fit in red, and the shades for subtracting the MCMC samples. The blue line indicates $\mathrm{rms_{residual}}$, with the blue shade showing $\pm \mathrm{rms_{blank}}$ for comparison. The lower panel shows the full velocity coverage of the spectrum, with the median fit model and the raw baseline. The lower figure shows the corner plot of the posterior distribution of the model parameters for this spectrum obtained by model fitting. The median value is indicated by the cyan line, and the 16, 84 percentile by the black dashed lines.}
	\label{f:fitting}
\end{figure*}

It is not a trivial task to fit the model to real data, partly due to the high dimensionality and the non-linear behavior of the model, and partly because the integration in the model evaluation doesn't have an analytical solution. 
Monte Carlo Markov chain (MCMC) hence becomes the most reasonable method for fitting the model. 
Besides its power in fitting a high dimensional and computationally heavy model, MCMC also provides a way to get statistically robust measurements of parameters. 
To increase the sampling efficiency, $v_r$ and $v_g$ are sampled in logarithmic space.

The data are first processed in preparation for applying the MCMC analysis.
The whole spectrum is trimmed in spectral dimension to include only the portion containing the line emission and the blank channels covering twice the line width on each side, in order to alleviate the computational burden and exclude other sources at a different redshift but in the same beam. 
Then the blank channels not selected in the previous step are used to estimate the noise rms$_\mathrm{blank}$, or auto-correlation function if there are a sufficient number of channels.

The likelihood function uses the difference between the model and the line spectrum to assess the goodness of the fit. 
In the case that the correlation of the spectral channels can not be estimated reliably, the channel-wise difference is simply compared with the blank channel rms$_\mathrm{blank}$.
In the case that the correlation can be measured, the likelihood is estimated assuming the channel data follow a Gaussian process characterized by the blank channels auto-correlation function. 
This is a more statistically sound approach, as most of the instruments have finite spectral resolution, and it is a common practice to smooth the spectrum before analysis.
Considering the correlation between channels also avoids underestimating the uncertainty of the fitting result.

In order to obtain statistically robust result on low-S/N data, we selected and tested the prior function carefully. 
The prior function used for Bayesian inference is composed of a flat prior for all parameters except for $k$ and $r$, and one special term that is used to avoid ill-shaped model fitting. 
The formulation and justification of the prior function are described in detail in Appendix~\ref{sec:priori}.

In MCMC sampling, we start with three stages of burn-in, each with 150 iterations and different moving algorithms to account for multi-modal distribution, followed by 2000 iterations of 128 walkers of sampling used for posterior inference. 
The last 250 iterations are stored for searching for the highest posterior likelihood parameter set, making figures like Fig.~\ref{f:fitting}, and potential ensemble study. 
The python package \texttt{emcee} \citep{emcee} is used for MCMC sampling, and \texttt{george} \citep{george} is used for the likelihood inference.

By default, the median value and the 16, 84 percentiles of the posterior distribution are used as the fitted value and uncertainties, respectively.
Other derived values like \wfvtym are also inferred from the posterior distribution of the model parameters.

As a by-product of the Bayesian inference, we define and use another statistical quantity with similar meaning to S/N. 
The model based Square Root Deviance (SRD) is based on the likelihood contrast of the original spectrum to that of the residual after subtracting the model, defined as
\begin{equation}\begin{split}
	\mathrm{SRD} = \sqrt{2 \ln \frac{p(\mathrm{spec} - \mathrm{model})}{p(\mathrm{spec})}}
\end{split}\end{equation}
This value quantifies the statistical significance of the existence of the spectral line compare with the noise, based on the knowledge of the noise behavior in the spectrum.
It enables us to derive a more statistically robust ``signal-to-noise ratio'' by taking into account the channel-wise correlation, e.g. in ALFALFA data.
This value has a similar statistical meaning to S/N by denoting the significance of the presence of any signal compared to pure noise, and the formula reduces to $\sqrt{\Pi \Delta_i/N\sigma}$ in the absence of correlated noise, which is the same as the definition of S/N.

\subsection{Fitting quality and sample control}
\label{sec:quality}

It is hard to assess the quality of the model fitting due to the high dimensionality and the occasional existance of a multi-modal posterior distribution. 
In this work, we define a quality factor $q$ to evaluate the model fitting, which relies on the root mean square of the residual spectrum (hereafter rms$_\mathrm{residual}$).
There are two major contributors to rms$_\mathrm{residual}$ in a good fitting result: one is the noise in the observation which should resemble the rms measured in the line-free blank channels, such that $\mathrm{rms_{residual,noise}} \propto \mathrm{rms_{blank}}$; the other is the intrinsic structures and peculiar motions of neutral gas clouds in the galaxy in addition to the rotation and velocity dispersion assumed in the model. 
The latter effect can be hypothesized as originating from a certain fraction of neutral gas, so that $\mathrm{rms_{residual,intrinsic}} \propto \bar{F_\nu} \cdot \sqrt{W} \propto \mathrm{rms_{blank}} \cdot \mathrm{S/N}$, where $\bar{F_\nu}$ is the average flux density of the line and $W$ is the line width.
The scaling with the measured flux density of RMS$_\mathrm{residual,intrinsic}$ means rms$_\mathrm{residual}$ is expected to be larger in high S/N spectra, and this is witnessed when fitting the spectra of the high S/N sample and the SPARC sample. 

Empirically, we define the quality factor $q$ as $q = \mathrm{rms_{residual}} / \mathrm{rms_{blank}}(1 + 0.003 \mathrm{S/N})$.
The empirical value of $0.003$ combines the noise contribution from both observational and intrinsic structures. 
We then set the threshold of $q$ to 1.25, namely any fitting result with $q > 1.25$ will be considered as a ``low quality fit''. 
Because we also introduced SRD as an estimate of S/N, in practice we use SRD instead of S/N to compute $q$.
Some examples of ``low quality fit'' can be found in Appendix~\ref{sec:unusual}.

The selection of samples of \hi spectra  often involves the assessment on the peakiness and symmetry of the line profiles, and these criteria can be quantified using the model fitting results.
Details of a sample control using the ``disk fit quality'', ``asymmetry'', and ``W50 discrepancy'', are discussed in Sec.~\ref{sec:p16}.

\section{Results}
\label{sec:results}

\subsection{Comparison and distribution of the ALFALFA demonstration sample}
\label{sec:demo}

\begin{figure*}
	\centering
	\includegraphics[width = \textwidth]{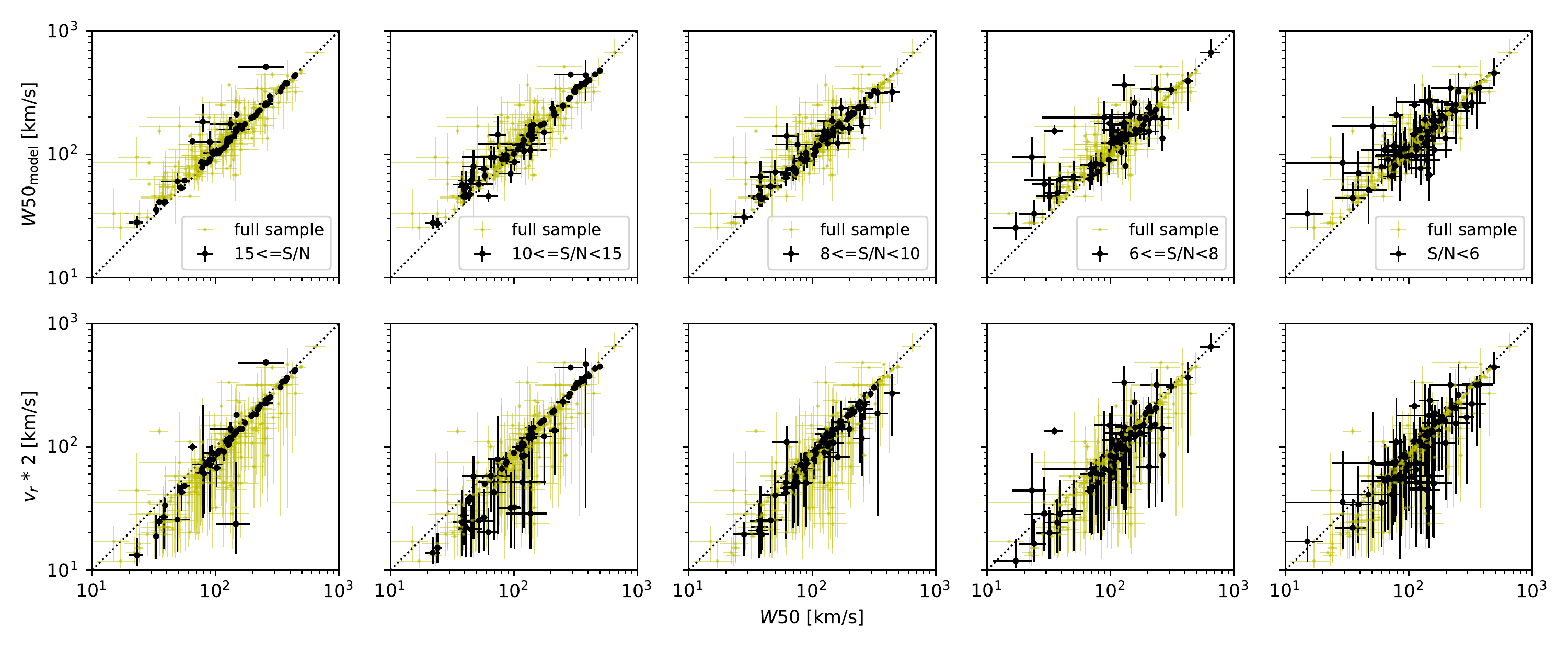}
	\caption{Width comparisons for the ALFALFA demonstration sample in different S/N bins. First row: ALFALFA \wfvty to \wfvtym comparison; second row: ALFALFA \wfvty to $2v_r$ comparison; left to right: highlight of spectra in high S/N bin to low S/N bin. In each figure, the width measurements of the whole sample are plotted in yellow, while the spectra in the highlighted S/N bin (described in legend) are plotted in black. The dotted diagonal line shows the one-to-one relation. }
	\label{f:demo_width}
\end{figure*}

We test the precision of the line width and flux measurements of the PANDISC model on the ALFALFA demonstration sample and compare them to the corresponding measurements derived by manual inspection. 
The comparison of the width is shown in Fig.~\ref{f:demo_width}.
Both $v_r$ and \wfvtym are compared in different S/N bins. 

\wfvtym shows good agreement with \wfvty for S/N $\ge$ 8 profiles. 
While the scatter increases significantly in lower S/N bins, both measurements still agree within the range marked by their error bars. 
However, there are two noticeable features in the comparison figure. 
The first is a slight overestimation of the width by \wfvtym compared to the ALFALFA \wfvty, especially in the small line width end. 
The same trend shows up weakly in the highest S/N bin, and the deviation grows towards lower S/N bins. 
The trend can be attributed to the attempt by the fitting routine to fit a broad gaussian component sitting below some of the very narrow disk profiles. 
This broad gaussian component could indicate either a common wing component which becomes more apparent in narrow single peaked profiles, or the contribution of noise or residual baseline ripple which can affect \hi spectral data. 
The common broad wing component in narrow profiles is more robustly selected in the high S/N sample and discussed in Sec.~\ref{sec:broad}.
The second noticeable feature in the comparison figure is the presence of some apparent outliers. 
These outliers always have larger error bars than other spectra with similar S/N, and their \wfvtym values are often greater than \wfvty. 
They turn out to be unusual profiles that can be sorted into three general categories: (1) asymmetric profiles for which the ALFALFA measurements only consider one peak or part of the profile (e.g. UGC 8605, AGC 123910, AGC 193902); (2) broad and low S/N profiles with clearly underestimated widths (e.g. AGC 114774); (3) poor fits caused by confusion (shoulder or wing like features) or low-quality spectra (e.g. UGC 6204, AGC 728887).
Some unusual profiles are further discussed in Appendex~\ref{sec:unusual}.

The comparison of $v_r$ shows much larger scatter and a different trend. 
Because $v_r$ is different from \wfvty by a fraction of the $v_\sigma$ as discussed in Sec.~\ref{sec:properties}, such an offset shows up clearly in all S/N bins. 
But even taking the offset into consideration, $v_r$ still tends to underestimate the width with significantly larger error bars, which is more obvious at the narrow width end and in the lower S/N bins. 
This behavior arises because the line profile resembles a single peak as the S/N and width of the profile decrease, making it harder to fit a disk component. 
Additionally, when the line profile is well matched by a gaussian component, $r$ converges to a low value and $v_r$ becomes completely unconstrained. 
The comparison shows that in the case of S/N$<$8, $v_r$ is a poor estimator of \wfvty, and this is inherent to the model assumption for $v_r$. 
It also suggests that the convergence of $v_r$ can be useful in selecting double horn profiles that are dominated by global rotation. 

\begin{figure}
	\centering
	\includegraphics[width = 0.45 \textwidth]{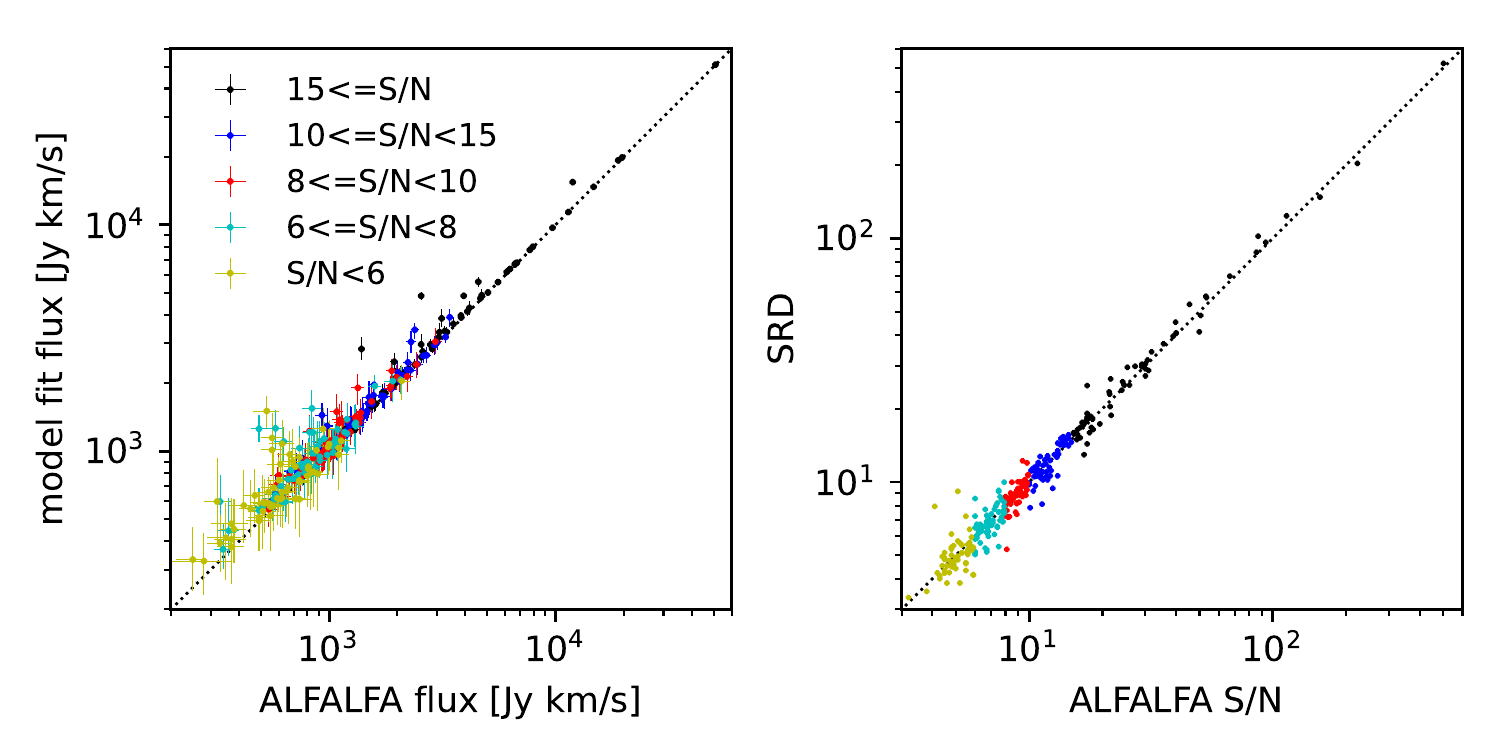}
	\caption{Comparison of the model-derived flux and SDR for the ALFALFA demonstration sample with the ALFALFA catalog measurements, in the same color scheme as Fig.~\ref{f:demo_hist}, with the one-to-one relation plotted as the dotted line.}
	\label{f:demo_flux}
\end{figure}

A comparison of the model fit flux and SRD are also shown in Fig.~\ref{f:demo_flux}.
The fluxes recovered by the model agree well with the ALFALFA measurements down to the lowest S/N bin, except for a few obvious outliers with underestimated ALFALFA fluxes. 
These outliers correspond to the same outliers in the \wfvtym to \wfvty comparison, and arise mostly because the ALFALFA measurements ignore wing- or shoulder-like features, or miss part of an asymmetric profile.
The SRD matches tightly with the ALFALFA S/N, reaching the expectation of a model-based alternative to S/N.

\begin{figure*}
	\centering
	\includegraphics[width = \textwidth]{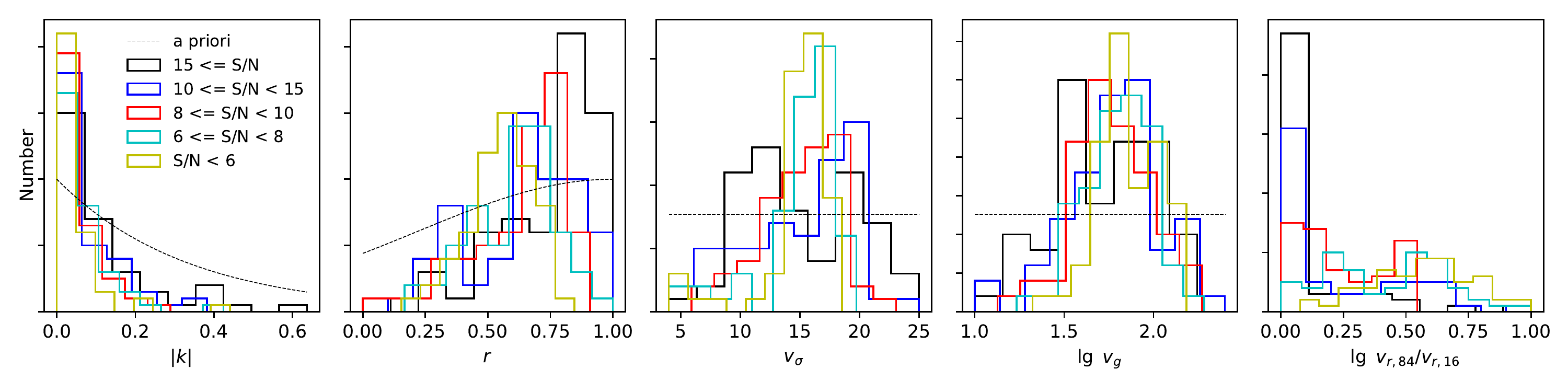}
	\caption{Distribution of the selected model fitted parameters for the ALFALFA demonstration sample to show the model the constraining power as a function of S/N. The parameters plotted are $|k|$ describing the degree of asymmetry, disk fraction $r$, velocity $v_\sigma$ controlling the steepness of the edge, $\lg v_g$ describing the width of the gaussian component, and $\lg v_{r, 84}/ v_{r, 16}$ the 84th percentile to 16th percentile posterior contrast as the quality of the disk fit. The distributions of the spectra from the highest S/N bin to the lowest S/N bin are shown in black, blue, red, cyan and yellow lines, with the prior function of the parameter (except $\lg v_{r, 84}/\lg v_{r, 16}$) plotted as the dashed line for comparison.}
	\label{f:demo_hist}
	
\end{figure*}

Fig.~\ref{f:demo_hist} shows how the spectrum S/N affects the constraining power of the model fit.
Similar to the previous discussion, the model fitting becomes less constrained as S/N gets lower, which is equivalent to saying it is more difficult to extract information from noisy spectra. 
From high to low S/N bins, the asymmetry $|k|$ transitions from a more extended distribution to being concentrated around zero, meaning that the model fitting is less likely to pick out the asymmetry of the low S/N profile.
$r$ transitions from a disk-dominated population to a lower disk fraction, clustering around $0.5$, as the profiles become more single peaked at low S/N.
$v_\sigma$ changes from a broad distribution to peaking around $15\ \mathrm{km/s}$, the median value of the prior distribution.
Additionally, $v_g$ shows a similar but weaker trend than $v_\sigma$. 
The comparison of the 84th percentile to 16th percentile of $v_r$ measurement $\lg v_{r,84}/v_{r,16}$, displays a transition from concentrating around zero, meaning a restricted $v_r$ posterior distribution, to a distribution beyond the value 0.176, meaning $v_{r,84}$ is at least $1.5 \times v_{r,16}$.
All these results show how the model fitting become less constrained as the S/N decreases. 
Thus for S/N $<$ 8 spectra, it may not be realistic to extract any additional information beyond the width and flux for individual profiles. 
At such low S/N, ensemble studies become necessary.

\subsection{Application to the BTFR}
\label{sec:p16}

To demonstrate the application of the PANDISC model, we apply the model to the gas-rich \citetalias{papastergis16} sample and use the results to fit a BTFR as did those authors.  
\begin{figure}
	\centering
	\includegraphics[width = 0.45 \textwidth]{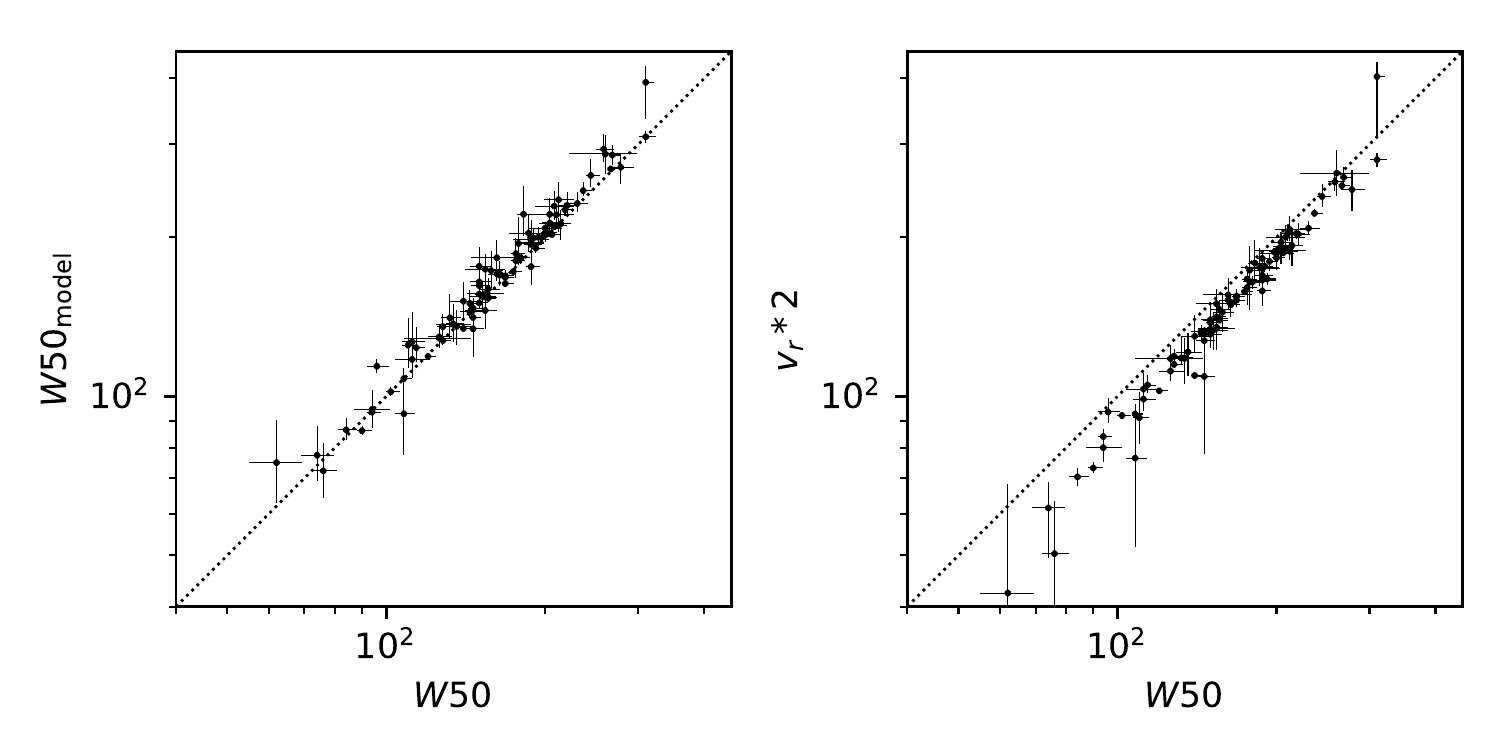}
	\caption{Width comparisons for the \citetalias{papastergis16} sample. Both \wfvtym (left) and $2v_r$ are compared with \wfvty reported in \citetalias{papastergis16}, with the dotted line showing the one-to-one relation.}
	\label{f:p16_width}
\end{figure}

As the first step, we compare the width measurements in Fig.~\ref{f:p16_width}. 
The comparisons of \wfvtym and $v_r$ with \wfvty display similar trends as described in Sec.~\ref{sec:demo}, with a few obvious outliers. 
The outliers in \wfvtym are mainly due to the asymmetric line shape (e.g. UGC 6747) and probable confusion (e.g. AGC 252877 at the largest width end). 
In contrast, the $v_r$ comparison outliers are mainly due to the unconstrained fit on single peaked profiles (e.g. AGC 122217). 

\begin{figure}
	\centering
	\includegraphics[width = 0.45 \textwidth]{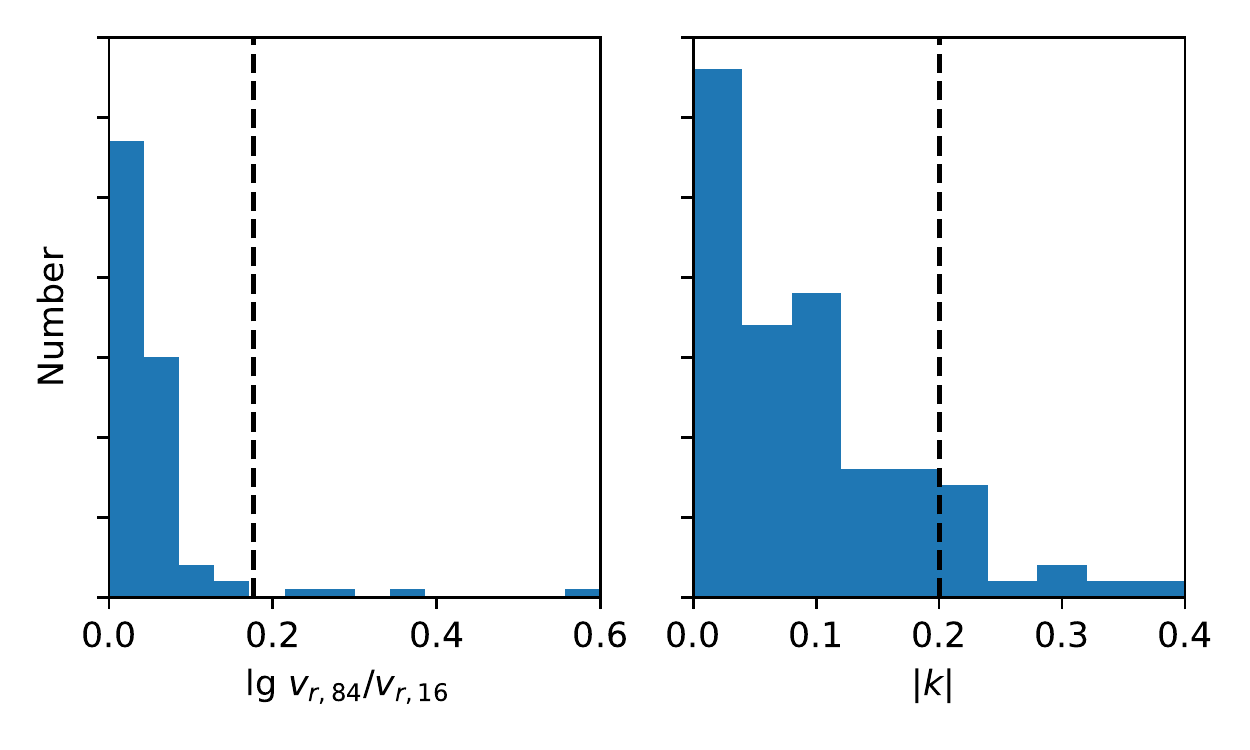}
	\caption{Distribution of the model fitted $\lg v_{r, 84}/v_{r, 16}$ (left) and $|k|$ (right) of \citetalias{papastergis16} sample as the justification of outlier selection, the  dashed lines indicate where $\lg v_{r, 84}/v_{r, 16} = \lg 1.5$ and $|k| = 0.2$, as our adopted cutoff criteria for ``unconstrained $v_r$'' and ``asymmetric'' selections.}
	\label{f:p16_hist}
\end{figure}

As a next step, we refine the sample with the model fitting parameters. 
We first exclude the profiles with low quality fits according to the $q$ factor defined in Sec.~\ref{sec:quality}.
We also exclude sources with \wfvtym to \wfvty discrepancy greater than $2\sigma$ (hereafter ``\wfvty discrepancy'' flag), which are usually spectra with unusual \hi profiles (see Appendix~\ref{sec:unusual}).

\begin{figure*}
	{\centering
	\includegraphics[width = 1.0 \textwidth]{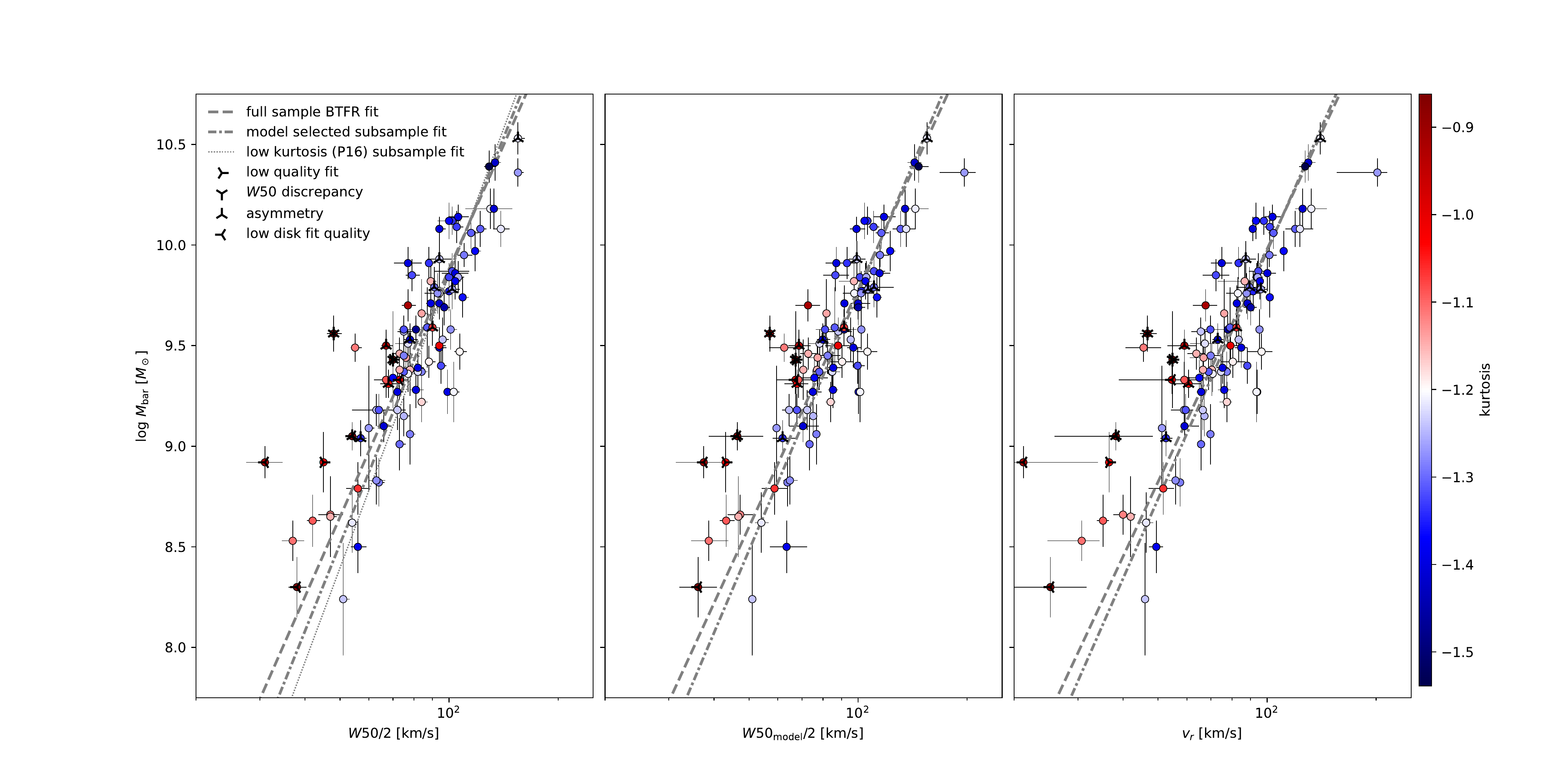}}
	\includegraphics[width = 0.475 \textwidth]{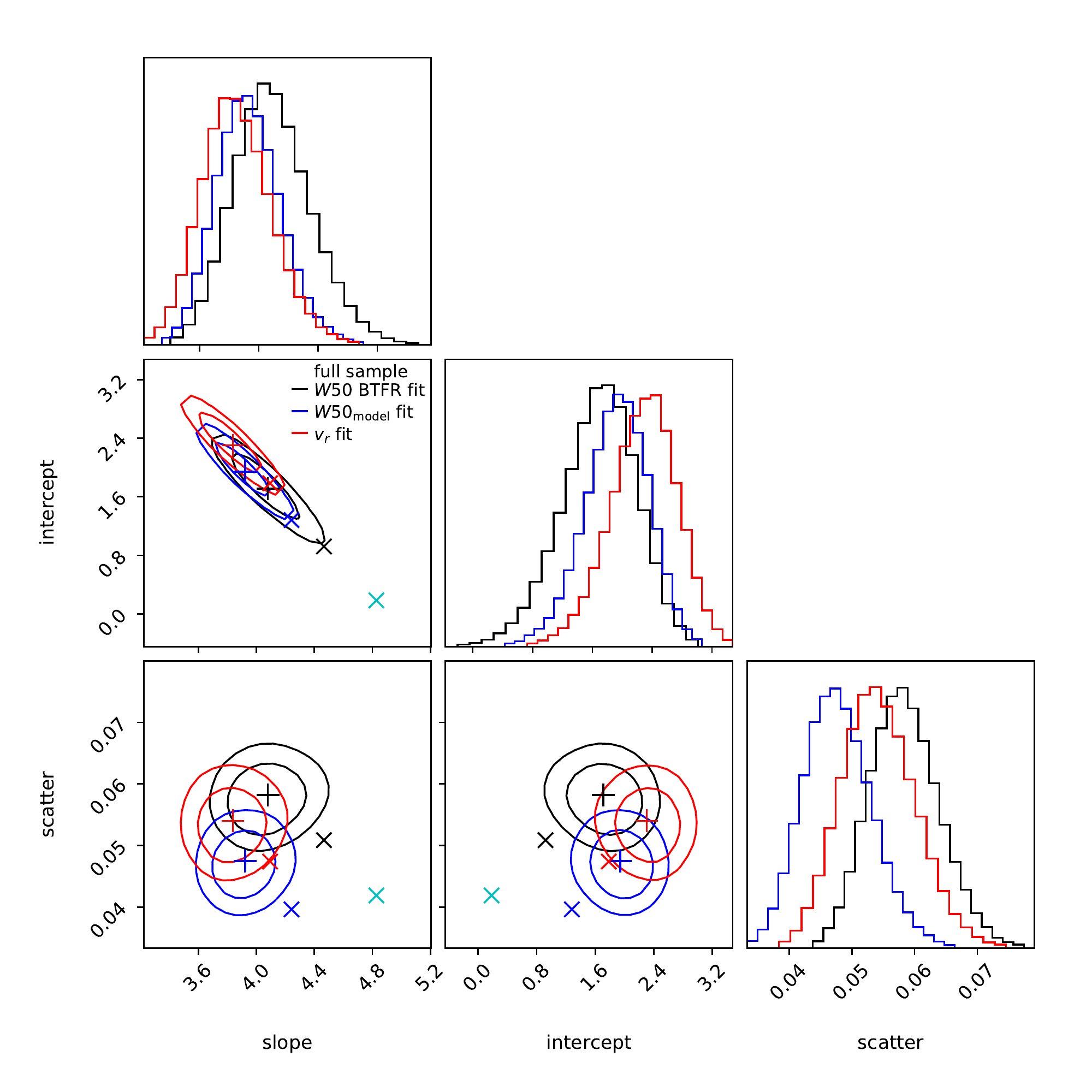}
	\includegraphics[width = 0.475 \textwidth]{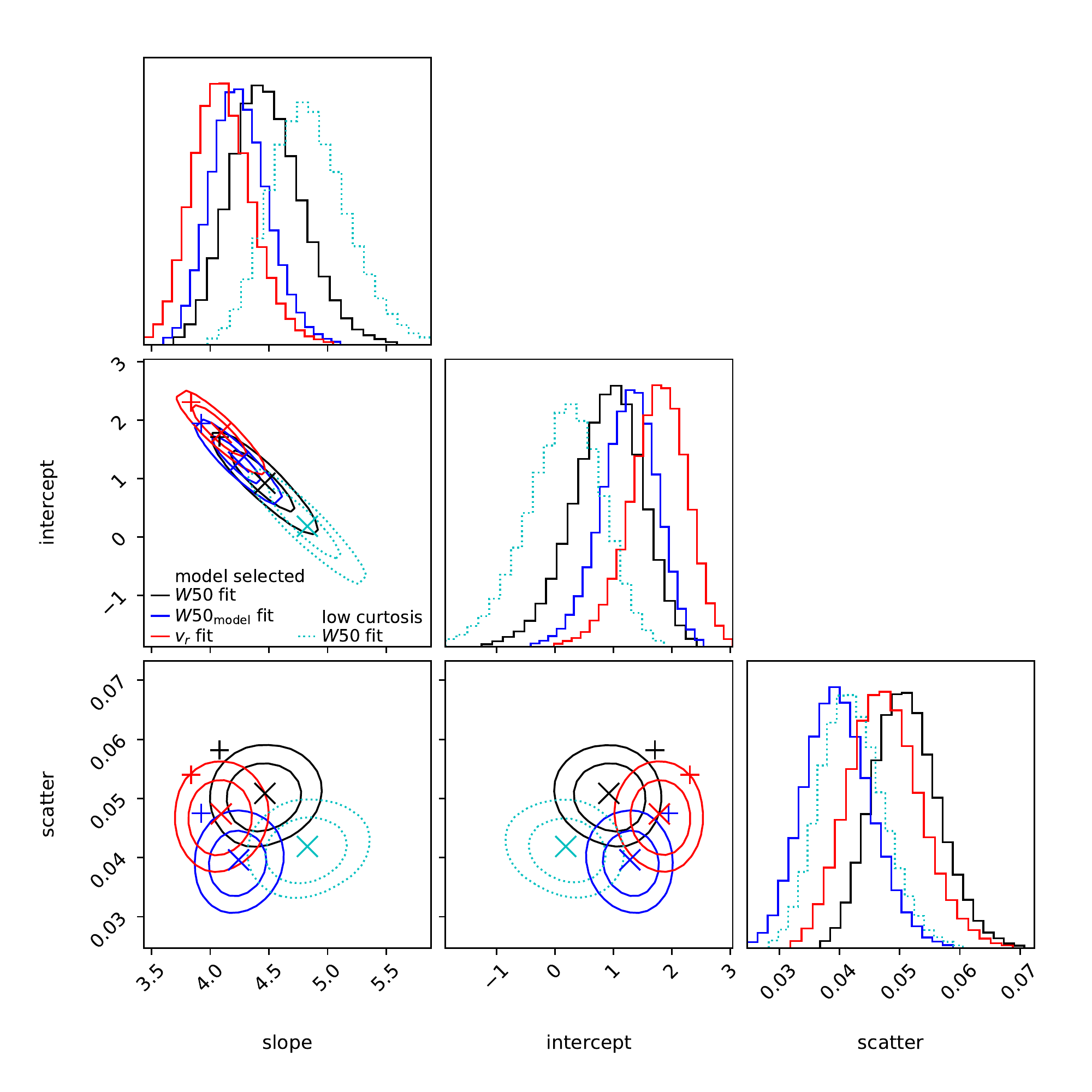}
	\caption{The BTFR relations and MCMC fit posterior distributions with different width measurements for the \citetalias{papastergis16} sample. The top three panels show the BTFR using \wfvty, \wfvtym and $v_r$. The baryonic mass and the kurtosis adopted for the color coding are taken from \citetalias{papastergis16}. The flagged data points are labelled according to the selection criteria discussed in the paper. Additionally, the BTFR fits for the full sample, the subsample after flagging, and low kurtosis subsample are plotted as dashed, dash-dotted and dotted lines, respectively. The lower panels are the corner plots of the posterior distributions of the BTFR fits, using the full sample (left) and the various subsamples (right). The distribution contours are colored by the width measurement and the subsample used, and two contour levels are shown as enclosing $39.3\%$ and $68\%$ of the posterior distributions. The median of the posteriors are labelled as ``+'' sign for the full sample fits, and ``$\times$'' sign for the subsample fits, in the color described in the legend, and in both corner plots for the purpose of comparison.}
	\label{f:p16_btfr}
\end{figure*}

The BTFR is known to depend on the tracer, galaxy mass and type, as well as the width and mass measurement methods \citep{bradford16}.
This is especially important at the low mass end, as for dwarf galaxies with still-rising rotation curves, the \hi may not sample the flat part of the rotation curve\citep{oh15}, and, for the lowest masses, the gas dynamics may become pressure-supported instead of rotation. 
In addition, the narrow \hi line profiles of these low mass galaxies are more prone to turbulence and tidal interactions. 
Thus the BTFR at low masses often displays larger scatter \citep{bradford16,brook16}, and its physical meaning may also differ from that of higher mass galaxies \citep{mcgaugh2000}.

Motivated by the goal to derive a uniform BTFR for the rotation-supported systems, we apply two naive restrictions: (1) excluding asymmetric profiles, (2) excluding profiles with unconstrained disk fits.
The first criterion stems from the concern that asymmetric profiles are likely the result of tidal interaction or source confusion \citep{haynes98,espada11}. 
The asymmetry restriction is performed by applying an empirical cut on $|k|$ at the value 0.2 (hereafter ``asymmetry'' flag).
This value is justified by the fact that in Fig.~\ref{f:p16_hist}, the $|k|$ distribution shows an excess beyond 0.2, consistent with Fig.~\ref{f:demo_hist}. 
The second criterion limits the sample to the double-horn profiles showing a clear signature of rotation; it also excludes galaxies that are face-on (which is not a concern for the \citetalias{papastergis16} sample), or those \hi profiles that are dominated by either the rising part of the rotation curve or those that are pressure-dominated.
This cut is achieved by selecting the spectra with well constrained $v_r$, such that $v_{W, 84}/v_{W, 16} <= 1.5$ (hereafter ``low disk fit quality'' flag). 
This choice is based on the 84th to 16th percentile contrast of $v_r$ in Fig.~\ref{f:p16_hist}, which shows a tight concentration below $\lg v_{W, 84}/v_{W, 16} = \lg 1.5$ , and a long tail beyond that value. 

In the mass-width diagram in Fig.~\ref{f:p16_btfr}, the data points are labelled if they are flagged by any of the criteria mentioned above. 
It can be noticed that most of the obvious outliers in the BTFR are either picked by our selection criteria, or are compensated by a large error in the width measurement (e.g. the one at the upper right corner of the figure which is a confused source F568-V01, shown as an example in Appendix~\ref{sec:confusion}). 
It is also worth noting that many of the spectra flagged by the model fitting have high kurtosis values in \citetalias{papastergis16}, especially those selected by asymmetry or $v_r$ fitting constraint.
This is because the kurtosis cut, disk fit quality and asymmetry cut all prefer spectra with clear double-horn shapes, though the kurtosis cut puts a stronger bias in selecting wider profiles than the model-based criteria.
Even if the asymmetry cut also flags several low kurtosis profiles that don't appear as outliers in the BTFR, we still exclude these profiles for consistency and physical robustness of the sample.

\begin{table}
\centering
\label{t:btfr}
\caption{}
\begin{tabular}{cccc}
	\hline
	\hline
	Width 	& slope $\alpha$ 					& intercept $\beta$ 	& scatter $\sigma_\perp$ \\ 
	\hline
	\multicolumn{4}{c}{Full \citetalias{papastergis16} sample} \\
	\hline
	\wfvty 	& $4.09^{+0.28}_{-0.24}$ 	& $1.70^{+0.47}_{-0.53}$ 	& $0.058^{+0.006}_{-0.005}$ \\
	\wfvtym	& $3.92^{+0.23}_{-0.20}$ 	& $1.94^{+0.40}_{-0.44}$  	& $0.047^{+0.006}_{-0.005}$ \\
	$v_r$	& $3.84^{+0.24}_{-0.22}$ 	& $2.30^{+0.42}_{-0.47}$ 	& $0.054^{+0.006}_{-0.006}$ \\
	\hline
	\multicolumn{4}{c}{Model selected subsample} \\
	\hline
	\wfvty	& $4.47^{+0.32}_{-0.28}$ 	& $0.92^{+0.55}_{-0.62}$ 	& $0.051^{+0.006}_{-0.005}$ \\
	\wfvtym & $4.24^{+0.26}_{-0.23}$	& $1.28^{+0.45}_{-0.50}$	& $0.040^{+0.006}_{-0.005}$ \\
	$v_r$	& $4.09^{+0.27}_{-0.24}$	& $1.79^{+0.46}_{-0.50}$ 	& $0.047^{+0.006}_{-0.006}$ \\
	\hline
	\multicolumn{4}{c}{Low-kurtosis subsample} \\
	\hline
	\wfvty 	& $4.83^{+0.35}_{-0.31}$ 	& $0.19^{+0.61}_{-0.69}$ 	& $0.042^{+0.005}_{-0.005}$\\
	\hline
\end{tabular}
\end{table}

We carry out the BTFR fit using the same formulation described in Appendix~B of \citetalias{papastergis16} with intrinsic scatter, except that the intercept is defined at $\log_{10} v_\mathrm{rot} = 0$ to get a sample-independent BTFR fit. 
We also use the \wfvty, baryonic mass and the kurtosis cut in the \citetalias{papastergis16} paper for the purpose of comparison.
The fitted BTFRs are detailed in Table.~\ref{t:btfr} and Fig.~\ref{f:p16_btfr}, along with the posterior distribution of the slope, intercept and intrinsic scatter. 
Fig.~\ref{f:p16_btfr} also compares the fitted BTFRs using different width measurements and sample selections. 

We note that, for the same sample, different width measurements result in slightly different BTFRs, and the slope decreases from 4.1 for \wfvty to 3.8 for $v_r$ for the full sample, though their posterior distributions largely overlap.
The fits also produce different intrinsic scatters, and the``intrinsic scatter'' can also be interpreted as the excess of uncertainty that is not accounted for the error bar for either the mass or width measurements. 
Thus a decreasing intrinsic scatter for different fits could mean either a tighter relation or a decreasing amount of unaccounted uncertainty in the mass or width measurements.  
We therefore caution against comparing the intrinsic scatter across different width measurements as the errors carry different systematics, and thus the intrinsic scatters have different statistical meanings.
However, comparison of BTFR fits using the same measurement sets but different samples is valid since it is not affected by the missing uncertainty problem.
Comparing the BTFR fit of the full sample (plus signs in the lower panels of Fig.~\ref{f:p16_btfr}) to the model-selected sample BTFR (cross signs in the figure) of the same width measurement, the intrinsic scatter also shrinks, suggesting a more constrained BTFR fit. 

Another point worth noting is that a lower scatter is always correlated with a higher slope as a result of selection effect. 
This is because, at the lower mass end, narrow profiles are preferentially flagged, and the opposite selection also holds weakly at the high mass end. 
At the low mass end, our selection criteria tend to flag the single-peaked profiles which are often narrow, comparable to 2$\sim$3 times $v_\sigma$; while at the high mass end, the asymmetric or confused profiles are preferentially wider, and the spectra often have higher S/N so that they can be identified in the model fit (as discussed in Sec.~\ref{sec:demo}). 
For comparison, we plot the BTFR fit of the low kurtosis sample in Fig.~\ref{f:p16_btfr}; it manifests an even stronger selection effect on the line width by excluding almost all profiles with $W50/2 < 50 \ \mathrm{km/s}$.
The selection effect is weaker and more physically uniform for the model selection method that essentially limits the sample to the strongly double-peaked spectra, which is more biased towards higher S/N profiles instead of the larger width ones by the kurtosis cut.

\subsection{Comparison with the flat rotational velocity}
\label{sec:sparc}

\begin{figure*}
	\centering
	\includegraphics[width = 0.9 \textwidth]{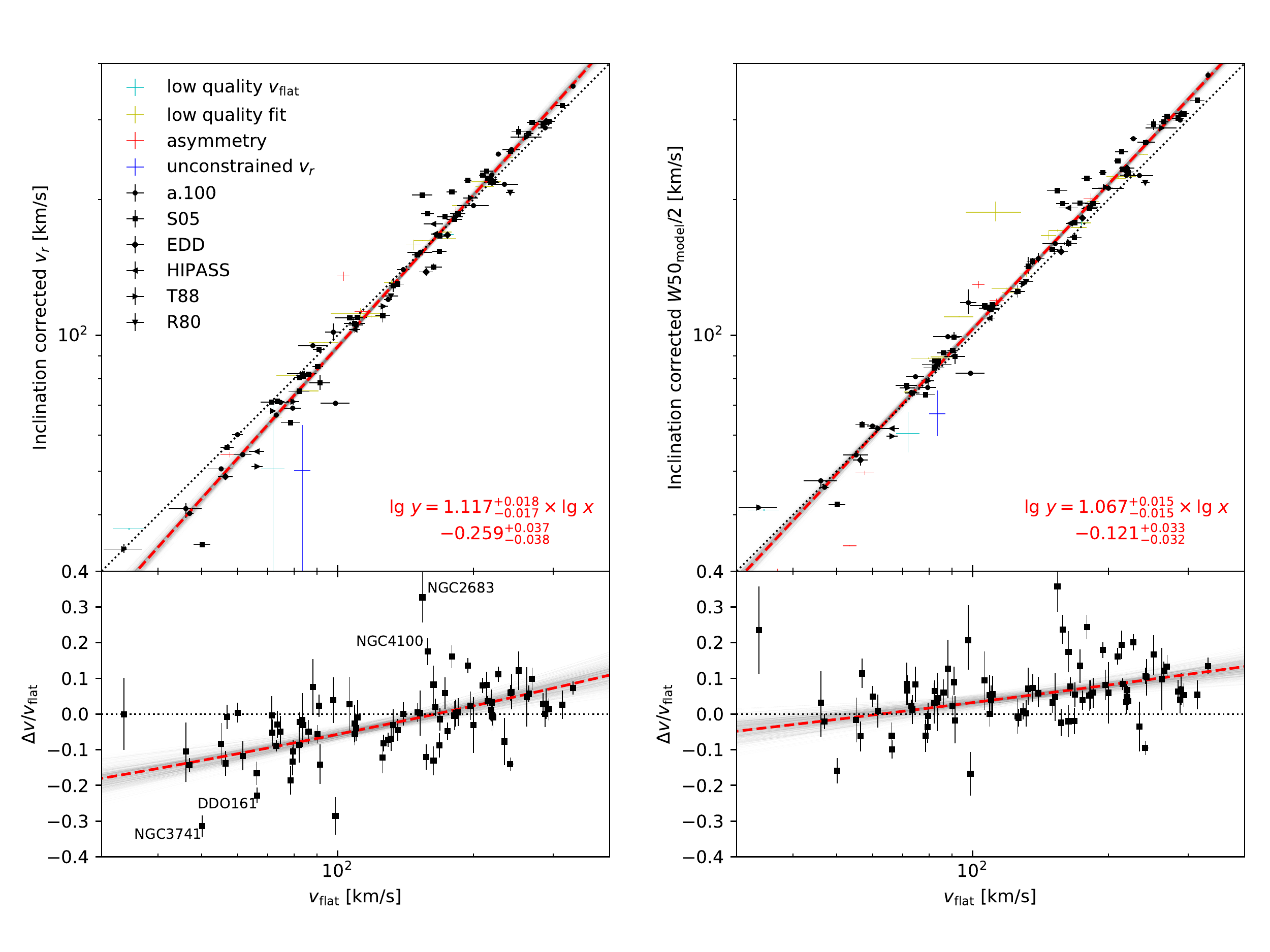}
	\caption{Comparison of the inclination-corrected $v_r$ (left) and \wfvtym (right) with the flat rotation velocity \vf in \citetalias{lelli2016}. In the upper panels, the data flagged based on \vf quality, model fit quality, asymmetry and disk fit quality are shown in cyan, yellow, red and blue colors without markers, while reliable fit results are displayed in black with marker shapes indicating their data sources: a.100 \citep{haynes18}, S05 \citep{springob2005}, EDD \citep{courtois2009}, HIPASS \citep{koribalski2004}, T88 \citep{tifft1988}, R80 \citep{rots1980}. In the lower panels, the relative width measurements differences normalized by \vf are plotted to show the trend and offset. In all panels, the one-to-one relation is plotted as the black dotted line, and a relation corresponding to the median MCMC fit is shown as the red dash line, embedded in random drawn MCMC samples as the faint gray lines. The fitted scaling relation is printed in red in the upper panel. In the lower left panel, four galaxies showing large deviations from the one-to-one relation are labelled, as examples for detailed study in Sec.~\ref{sec:sparc} and Fig.~\ref{f:sparc_outlier}.}
	\label{f:sparc}
\end{figure*}

Because the galaxies in the \citetalias{lelli2016} sample already have the flat rotational velocity \vf measured, and the majority have global \hi spectra available, it forms a good sample to test the physical meaning of $v_r$.
The model fitting is applied to the whole sample of 158 galaxies with integrated \hi observations available in the literature. 
For the \vf - $v_r$ comparison, we further restrict the sample in several ways. 
First, only the galaxies with \vf ($v_\mathrm{flat} > 0$ in \citetalias{lelli2016} Table 1) measurements and significant inclination angles ($i \geq 30^\circ$) are used.
We also drop five galaxies in the matched sample with absolute heliocentric velocity less than 100 km/s, due to confusion with galactic \hi.
This leaves 111 galaxies in the sample analyzed here.
We then flag three low quality \vf measurements, corresponding to $Q > 2$ in \citetalias{lelli2016}.
As in the previous section, we flag the spectra with low quality fits, asymmetric profiles and disk fit quality. 
This process leaves 84 galaxies for our analysis.

Both $v_r$ and \wfvtym measured by PANDISC are corrected for the inclination and then compared with \vf, shown in Fig.~\ref{f:sparc}. 
We also perform a MCMC fit for the scaling relation to aid a quantitative comparison. 
The fitting result can be found in Fig.~\ref{f:sparc}. 
$v_r$ shows better agreement with \vf, while \wfvtym values show a distribution systematically larger than \vf partly due to the widening effect of the velocity dispersion. 
However, when comparing the relative difference plotted in the lower panels in Fig.~\ref{f:sparc}, $v_r$ shows a trend such that it underestimates \vf at the low end, and overestimates at the high end. 
A similar trend shows up for \wfvtym but to a smaller degree. 
The fit of this trend gives a slope of 1.117 dex for $v_r$, and a slope of 1.067 for \wfvtym.
The different trends agree with the different  BTFR slopes, with the $v_r$ BTFR slope being $\sim$0.1 smaller than that of \wfvtym in Sec.~\ref{sec:p16}.

\begin{figure*}
	\centering
	\includegraphics[width = 1.0 \textwidth]{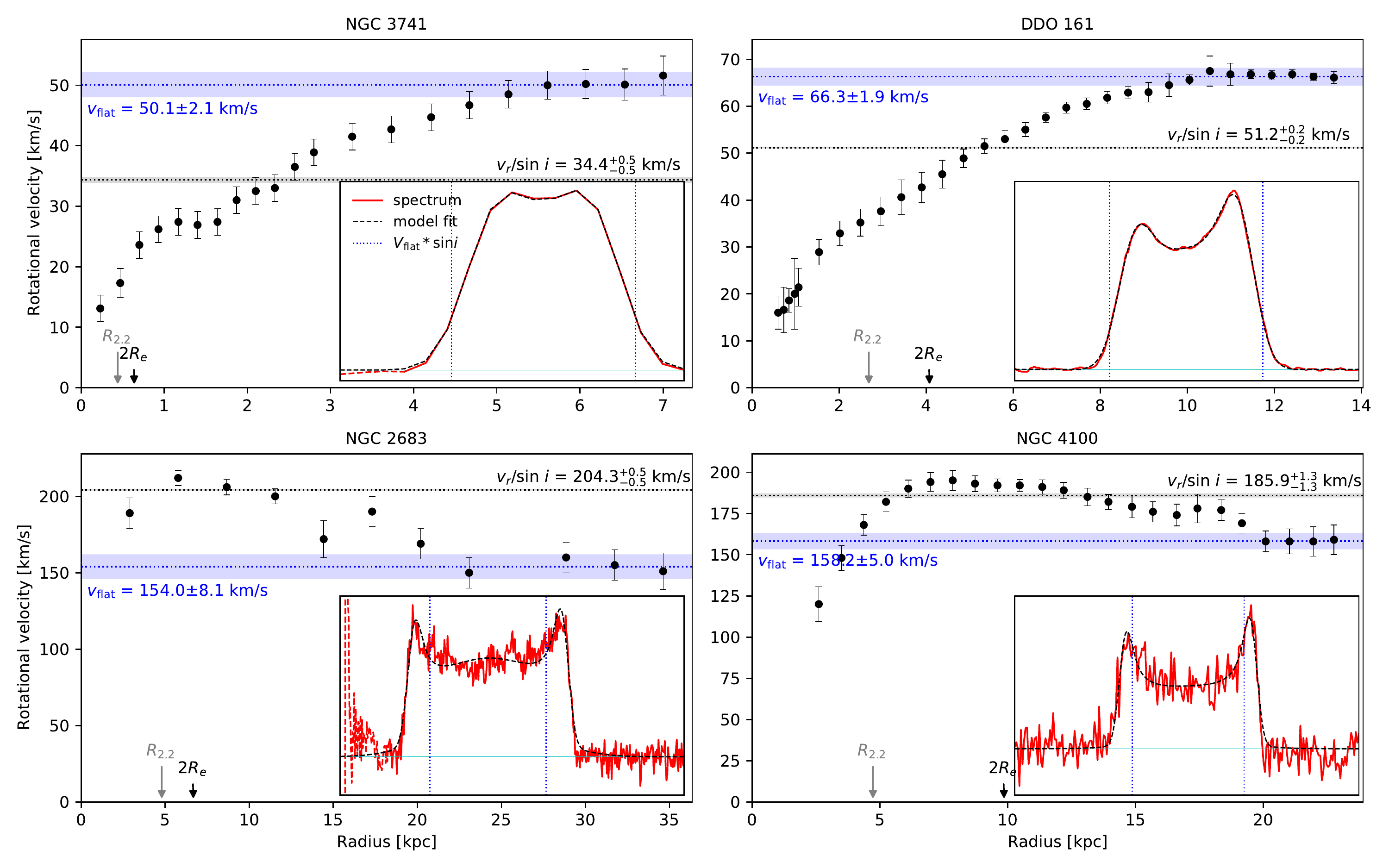}
	\caption{Rotation curves and global profiles of the galaxies labelled as outlier examples in Fig.~\ref{f:sparc}. The rotation curves are plotted as black dots on the main axes, with \vf and inclination corrected $v_r$ shown as grey and blue dotted horizontal lines with the shaded portion indicating the uncertainty. $2.2~\times$ disc scale length and $2~\times$ effective radius are labelled on the x-axis as $R_{2.2}$ (gray) and $2R_e$ (black), marking the characteristic sizes of the luminosity distribution, and where $V_{2.2}$ and $V_\mathrm{2R_e}$ are measured. The integrated \hi spectra are shown in the insets on the lower right in each panel, with the profile shown as the red solid line and the median fit as the black dashed line. \vf is also plotted as the blue dotted vertical lines to aid the comparison.}
	\label{f:sparc_outlier}
\end{figure*}

To better understand the cause of this trend, four galaxies showing large discrepancies in  Fig.~\ref{f:sparc} are selected for further inspection, namely NGC 3741, DDO 161, NGC 4100 and NGC 2683. 
Their rotation curves are plotted in Fig.~\ref{f:sparc_outlier}, along with the labels of characteristic sizes including $2.2~\times$ disc scale length $R_d$ and $2~\times$ effective radius $R_e$ taken from \citetalias{lelli2016}, as well as plots of their global \hi profiles used for model fitting. 
The rotation curve data were measured by \citet{gentile07} for NGC 3741, \citet{cote00} for DDO 161, \citet{sanders96} for NGC 2683, and \citet{verheijen01} for NGC 4100.

At the low \vf end, NGC 3741 and DDO 161 both have slowly rising rotation curves, while at the high \vf end, NGC 4100 and NGC 2683 have rotation curves that rise to a higher value before flattening at the outermost radii.
In all cases, the rotation curves only flatten at the very edge of the detected region, beyond $4 \times R_e$.
However, the model-fitted $v_r$ yields a value more consistent with the rotation velocity at smaller radius, typically at $2R_e$. 
These galaxies demonstrate circumstances where the co-rotation assumption of the model can break down. 
In practice, any line width measured on the integrated \hi spectral profile, either $v_r$ or \wfvty, are weighted averages of the maximal velocity of the \hi gas rings, with the weighting factors differ by the bias of the measuring method. 
$v_r$ is intensity-weighted, thus if structure exists in the rotation curve, what $v_r$ measures is the rotation velocity of the ring in which most of the \hi gas resides.
In addition to being weighted by intensity, \wfvty is also velocity-weighted, hence it is more susceptible to the gas moving at the highest L.o.S. velocity.

The trend seen in the comparison figure also suggests the dependence of the rotation curve shape on the rotational velocity, or equivalently, the mass. 
At the lower \vf and hence lower mass end, galaxies tend to show slowly rising rotation curves, most likely due to the  fact that the \hi disks do not extend far enough out to sample \vf. 
At the high mass end, some galaxies exhibit rotation curves that peak at relatively small radii, suggesting the dynamical mass is more concentrated in the inner galaxy. 
The trend at the high mass end also confirms the $v_\mathrm{max}$-to-\vf offset found in \citet{ponomareva17,lelli19}.
We argue that this trend of varying rotation curve shape is the primary cause of the differences between global profile line width measurements and consequently, varying slopes and intrinsic scatter in BTFR fits. 

\begin{figure*}
	\centering
	\includegraphics[width = 1.0 \textwidth]{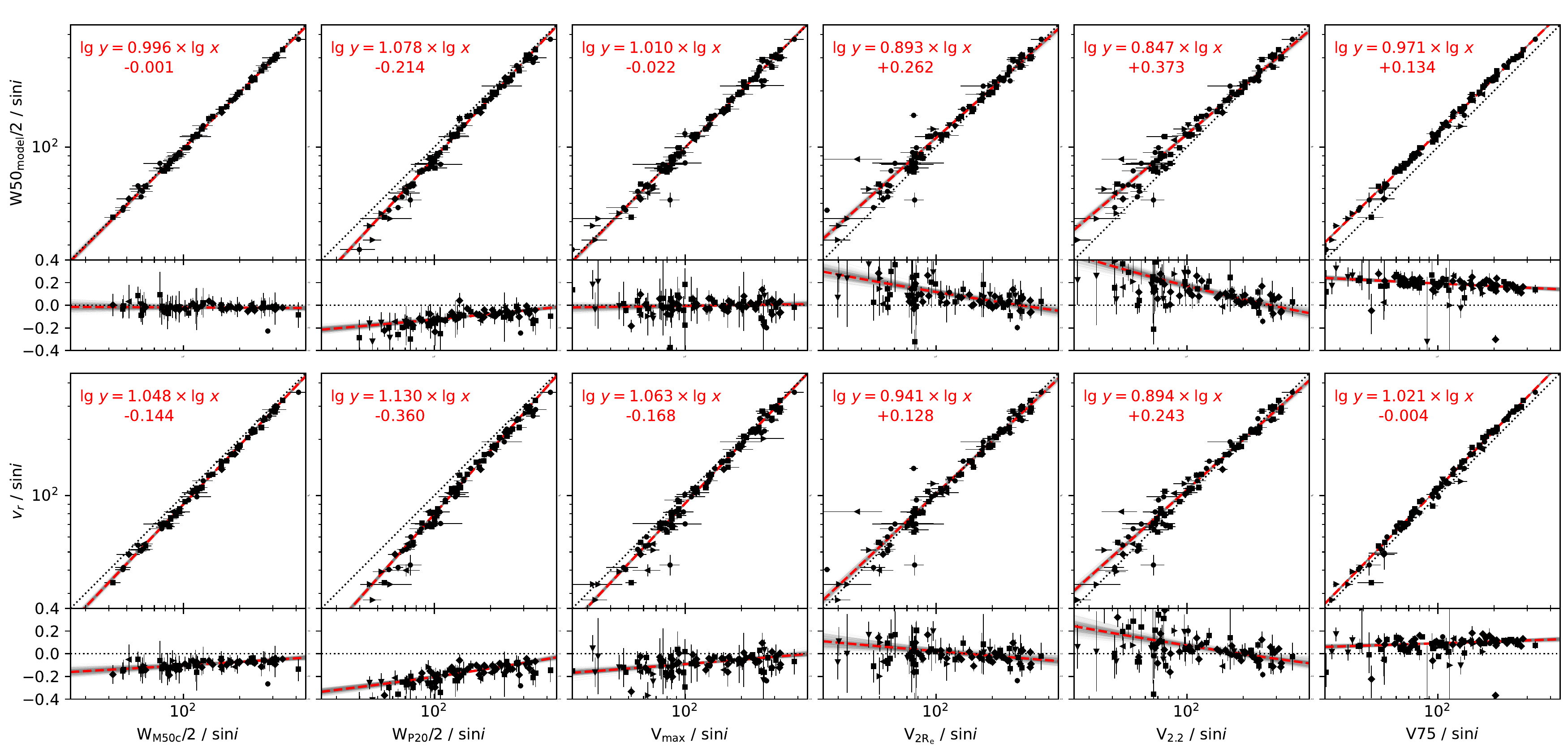}
	\caption{Comparison of the model fitted width in y-axis (upper to lower: \wfvtym and $v_r$) with other width measurements in x-axis. All the width values have been corrected for inclination, and filtered by the same sample control criteria as those used in the \vf comparison. The format of each panel is the same as that of Fig.~\ref{f:sparc}.}
	\label{f:sparc_compare}
\end{figure*}

For the completeness, we also compare the model fitted widths to several other width measurements in Fig.~\ref{f:sparc_compare}.
The widths compared here are (1) corrected 50\% mean flux width $W_\mathrm{M50c}$; (2) width at 20\% peak flux $W_\mathrm{p20}$; (3) maximal circular velocity $W_\mathrm{max}$; (4) circular velocity measured at $2~\times$ effective radius $R_e$ as $V_\mathrm{2R_e}$; (5) circular velocity measured at $2.2~\times$ disk scale length $R_d$ as $V_\mathrm{2.2}$; (6) 75\% curve of growth width $V75$.
$V75$ is measured by the method detailed in \citet{ball22}, all the other measurements are taken from \citet{lelli19}. 
A scaling relation is also fitted to each width comparison, also presented in the figure.

Some of the comparisons provide us with more insight into the meanings of \wfvtym and $v_r$.
For \wfvtym, its tight one-to-one relation with $W_\mathrm{M50c}$ bolsters the \wfvtym to \wfvty agreement demonstrated in Sec.~\ref{sec:demo}. 
\wfvtym also displays a good agreement with $V_\mathrm{max}$, consistent with our argument that \wfvty is weighted towards the fastest moving gas. 

For $v_r$, the width showing the best agreement, or the least offset, is $V_\mathrm{2R_e}$, followed by $V_\mathrm{2.2}$. 
The $v_r$ to $V_\mathrm{2R_e}$ relation further supports the observation in Fig.~\ref{f:sparc_outlier} and the aforementioned interpretation that $v_r$ is intensity-weighted, being more representative of the rotational velocity at a smaller radius such as $2R_e$. 
The better agreement with $V_\mathrm{2R_e}$ instead of \vf also implies that it is common for both low mass and high mass galaxies to have most of their \hi gas residing in a smaller radius than where the rotation curve flattens. 
So the width and profile of the integrated \hi line is more strongly affected by the inner structure of rotation curves than what was previously thought. 

We also notice that another fully automated method $V75$ shows tight relations to both \wfvtym and $v_r$, with some scaling offsets and trends. 
This is partly due to the fact that both the PANDISC model and $V75$ are applied on the same set of spectral data. 
But the small scatter of the relations, especially when comparing the errorbar to that of other width measurement comparisons, signifies the consistency and statistical robustness of these newly developed width measuring methods.

\section{Discussion}
\label{sec:discussion}

\subsection{Caveats of the model}
\label{sec:model_caveat}

The most important assumption of this model is that of the co-rotating disk, but such idealized \hi disks don't exist in reality.
Sec.~\ref{sec:sparc} shows how $v_r$ deviates from \vf due to the structural variations evident in the rotation curves of some galaxies. 
However, such deviations seem to follow a trend as a function of rotational velocity.
This trend suggesting the use of $v_r$ as a scaled approximation of \vf. 
It also hints a common dependence of the inner structure of rotation curves, or equivalently the distribution of dynamical mass, on the rotational velocity, in another word, the total baryonic mass of the galaxy.
Much more could be learned about the distribution of baryons, \hi gas and dark matter in a galaxy as well as the dynamics by understanding this trend, though it requires more detailed theoretical and observational studies that are beyond the scope of this work.

The model assumption of the velocity dispersion is also over-simplified for the purpose of parametrization.
The typical dispersion velocity of \hi is $\sim$10 km/s, but the value generally declines with radius \citep{ianjamasimanana15}. 
The physical origin of the velocity dispersion includes the random motions of the gas within \hi clouds, the random motions of \hi clouds in the disk, turbulence related to star formation or galactic shear, non-circular motions, etc. 
Moreover, the velocity dispersion is found to be better described by a two component model \citep{ianjamasimanana12}, further complicating the interpretation of this parameter. 
We speculate that $v_\sigma$ is likely a flux weighted estimate, or upper limit, of the velocity dispersion in the part of the \hi disk where the rotational velocity maximizes, and should be interpreted on an ensemble basis instead of for an individual galaxy.
The reason that the measured $v_\sigma$ is sometimes only an upper limit is the effect of the beam smearing effect. 
As in integrated spectrum, if a significant amount of gas exists moving at velocity higher than model fitted $v_r$, typically the rising part of the rotation curves, this part of gas would smoothen the line edge and increase the measured value of $v_\sigma$.
This is more significant for low mass galaxies with slowly rising rotation curves, as the only way to account for the disk component fluxes beyond $v_r$, which underestimates \vf in such cases, is $v_\sigma$.
Therefore, in the case where where $v_r$ underestimates \vf, $v_\sigma$ may be further inflated beyond the true velocity dispersion.

The asymmetry variable is assumed to be the gradient of the radial density from one side of the disk to the other end, similar to lopsidedness. 
However, there is no physical reason that the radial density increases in a linear way, as truly lopsided galaxies typically display more complicated radial variations. 
It is also a simplification to assume that the two extremes of the radial density variation coincide with the major axis projected onto the sky. 
Furthermore, the asymmetry of the \hi in galaxies is much more complex, as shown by numerous studies \citep[e.g.][]{richter94,haynes98}. 
Many other possible causes of asymmetry have been proposed, including beam confusion, non-circular motions, and distortions in the \hi distribution, but a universal picture of what dominates the observed asymmetry is still missing. 
Nevertheless, the outliers in our demonstration sample are preferentially highly asymmetric, as shown in Sec.~\ref{sec:p16} and Appendix~\ref{sec:unusual}. 

The physical interpretation of the gaussian component is even more uncertain, because of the diversity of its potential contributions and the lack of spatially-resolved interferometric data. 
The most likely origin of this component is the rising part of the rotation curve in the inner galaxy where the \hi profile is rotation-dominated \citep{deblok14}. 
However, for a few galaxies discussed in the next section, some \hi profiles show broad wings extending far beyond $v_r$ which are probably associated with unusual gas dynamics. 

In addition to the simplification of the variables, the model doesn't take into account any radiative transfer effects, such as absorption or intrinsic line broadening. 
A better solution would be to convolve with a Voigt profile instead of a gaussian, but to do so would add significantly to the degrees of freedom of the model. 

Another consideration for the application of the PANDISC model is the computational cost. 
Despite the facts that one integration in the model needs to be evaluated numerically for each channel, and that MCMC is intrinsically computationally-heavy, it takes about 150 seconds to fit one ALFALFA spectrum on a dual-core 3.1 GHz CPU. 
Therefore it is well prepared for applications on large databases such as ALFALFA and the on-going next generation surveys such as  MIGHTEE-HI \citep{maddox21}, WALLABY \citep{koribalski2004} and CRAFTS-HI \citep{zhang21}.

\subsection{Broad-wing features}
\label{sec:broad}

Several of the \hi profiles shown in Fig.~\ref{f:atlas} display very broad gaussian components extending well beyond the range of velocities associated with the disk, thus appearing as ``broad-wing'' features.
Although in low S/N spectra the majority of these wing-like features are fitting artifacts arising from either noise or potential residual baseline ripple, some high S/N spectra are also found to have very broad gaussian components. 
These could represent a distinct and potentially-interesting category of \hi profiles.
To survey the prevalence of broad-wing features, we focus on the high S/N sample with good fits, excluding low quality fits or those with unconstrained $v_r$.
We then select broad-wing candidates by two criteria: (1) the flux density of the gaussian component at channels beyond the disk part is significant, such that $F_\mathrm{\nu, gaus}(v=\mathrm{FWHM}_\mathrm{disk}/2 + v_\sigma) > 3\ \mathrm{rms_{blank}}$; (2) the gaussian component FWHM is wider than the disk projected rotational velocity $2.355 \times v_g > 2 \times v_r$. 

\begin{figure}
	\centering
	\includegraphics[width = 0.45 \textwidth]{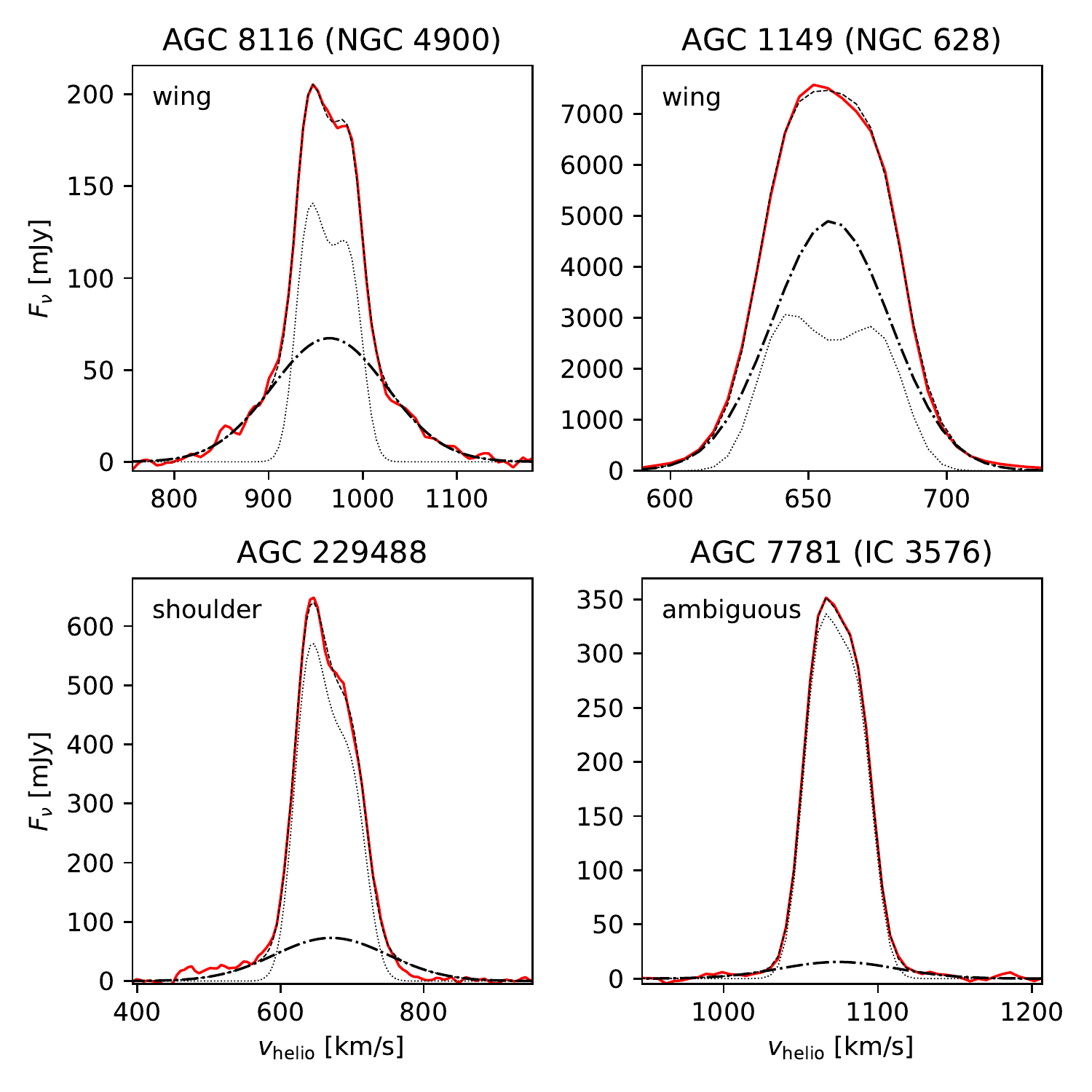}
	\caption{Example atlas of broad-wing candidates. The two panels in the upper row show two candidates with``wing-like'' features, the lower left panel is a candidate with a ``shoulder-like'' feature, and the lower right panel is an ``ambiguous'' candidate. In each panel, the global profile is plotted as the red solid line, with the model of the median fit as the thin black dashed line, the gaussian component fit as the thick black dot-dashed line and the disk fit as the thin black dotted line.}
	\label{f:bw_atlas}
\end{figure}

Among 301 galaxies with good model fits in the high S/N sample, 44 are selected as broad-wing candidates.
After visually checking the candidate spectra, we conclude that half of the selected spectra indeed show wing-like features, with significant flux excess beyond the central disk part on both sides, well fitted by a Gaussian peak (e.g. NGC 628 or NGC 4900 in Fig.~\ref{f:bw_atlas}).
Half of the remaining candidates show flux excess only on one side, resembling a ``shoulder'' like feature. 
The remaining quarter of the candidates are deemed as ``ambiguous'', as the disk fraction is so high that the fitted gaussian components are not clearly distinguishable from the extension of the line edge. 
Two examples are also shown in Fig.~\ref{f:bw_atlas} for the shoulder-like and ambiguous candidates.

For the wing-like features, we postulate that they are associated with gas components that are dynamically distinct from the rotating disks.
After checking the optical images, we identified one third of the galaxies showing wing-like feature to have a close companion or an irregular morphology, highlighting the potential effect of interactions.
Judging from the asymmetric shape of the ``shoulder'' features, we suggest that they could either be confused with companion galaxies, or reflect clumps of \hi gas that are dynamically-separate from the disk in one direction in velocity space through processes like tidal interaction or counter-rotation of the disk \citep{jore96}. 
An even higher fraction of these galaxies have likely companions in optical images, and additional examples of known confused spectra are discussed in Appendix.~\ref{sec:confusion}, supporting the confusion origin for the shoulder-like features. 
The ambiguous candidates are difficult to interpret, as their gaussian component is significant compared with the noise according to our $3 \mathrm{rms_{blank}}$ criterion, and the flux density is also above the typical baseline uncertainty \citep{haynes98}, making them very likely to have real flux excess beyond the rotation velocity. 
We thus speculate these weak features have a similar origin as the wing-like or shoulder-like features, but the fraction of gas contributing to the high velocity wings in these galaxies is very small.

\begin{figure}
	\centering
	\includegraphics[width = 0.45 \textwidth]{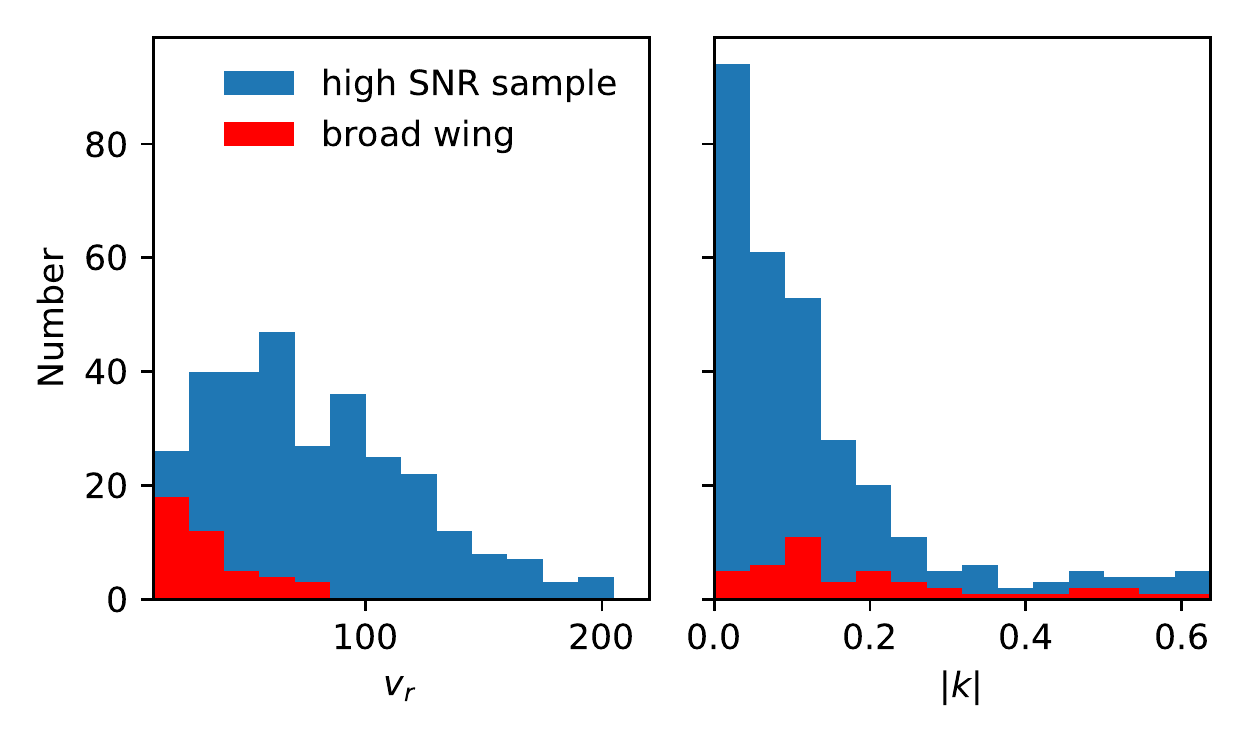}
	\includegraphics[width = 0.45 \textwidth]{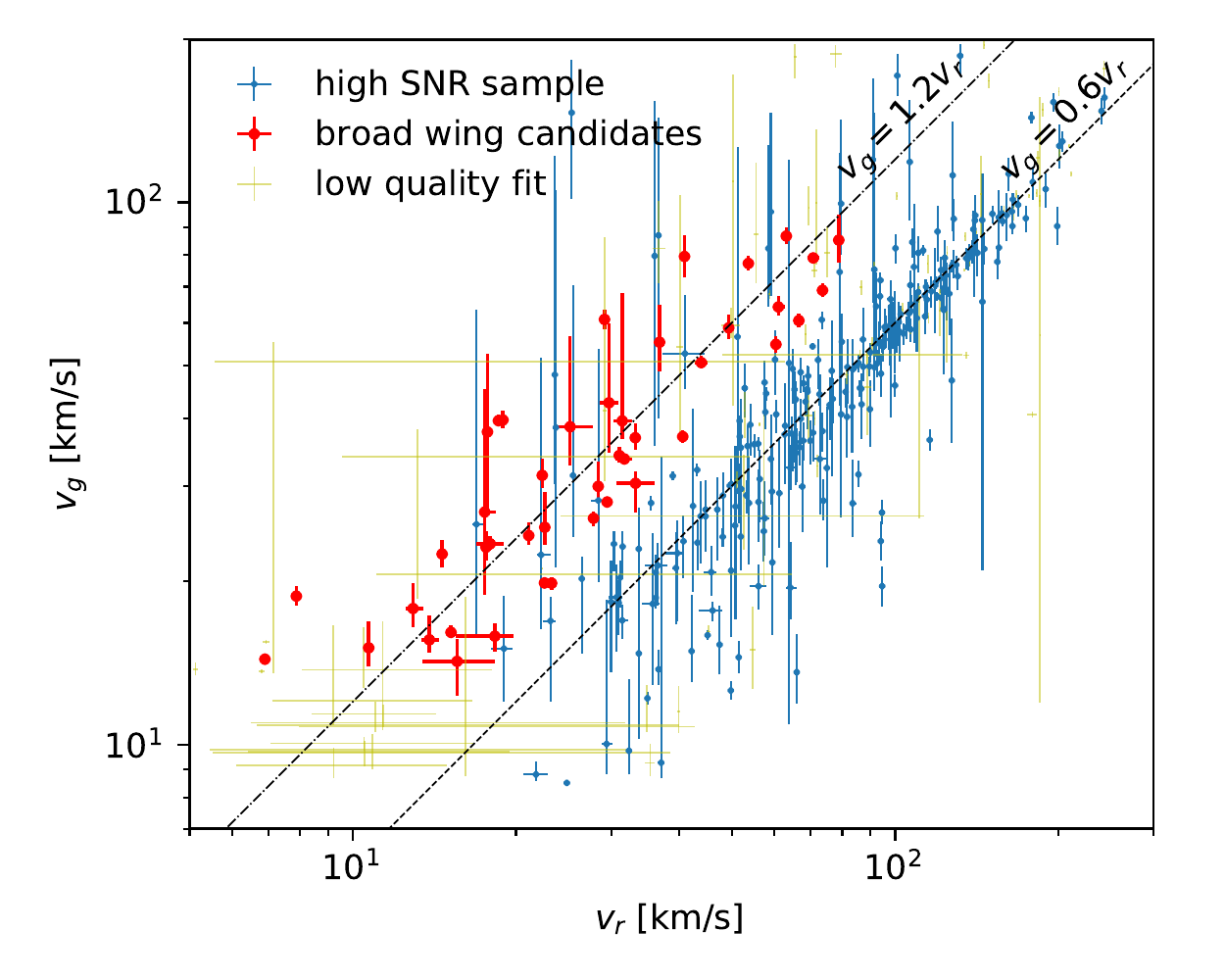}
	\caption{Comparison of the broad-wing candidates to the whole high S/N sample in parameter space. The upper left and right panels show the histogram of $v_r$ and $|k|$, for the broad-wing candidates (red) compared with all the reliable disk fits for the whole sample (blue). The lower figure displays the distribution of $v_g$ against $v_r$, with the high quality fits of the whole sample in blue, the broad-wing candidates in red, and the sources flagged by fit quality $q$ or disk fit quality $\lg v_{r, 84}/v_{r, 16}$ criterion in yellow. Two lines of $v_g=0.6 v_r$ and $1.2 v_r$ are shown in dashed and dot-dashed lines to aid the view.}
	\label{f:bw_dist}
\end{figure}

These broad-wing candidates are also distinctive in their distribution of other parameters, as shown in Fig.~\ref{f:bw_dist}. 
First, when compared with the total high S/N sample, the broad-wing candidates all exhibit relatively narrow profiles with systematically lower $v_r$ and \wfvtym, with a typical line width of $\sim$60 km/s. 
This could be due in part to the selection bias that the broad-wing features are easier to identify when the disk profile is narrow. 
The broad-wing candidates are also preferentially asymmetric, with the distribution of the absolute disk asymmetry $|k|$ being more extended while that of the entire sample is clustered close to zero. 
Together with the narrow width of the profile, we suspect that the broad-wing selection criteria identify a population of gas components that are dynamically different from the majority rotating disks.
Furthermore, they are preferentially identified in systems with low inclinations and non-uniform \hi distributions.

Moreover, the broad-wing candidates represent a distinct population of \hi profiles in the $v_r$-$v_g$ diagram in Fig.~\ref{f:bw_dist}.
The $v_r$ versus $v_g$ plot displays a clearly bimodal distribution: while most of the galaxies in the high S/N sample are distributed along the relation $v_g = 0.6 \times v_r$, consistent with our hypothesis that the gaussian component describes the rising part of the rotation curve, another population of galaxies cluster around the line $v_g = 1.2 \times v_r$ with larger scatter.
Moreover, the population of galaxies at higher $v_g$ is dominated by the broad-wing candidates.
Although the offset in the $v_r$-$v_g$ relation for the broad-wing candidates could be affected by selection bias and confusion, the bimodal distribution for $v_r$-$v_g$ and the dominance in the higher $v_g$ population suggest the existence of previously-unexplored but prevalent gas dynamics which becomes identifiable only when the gas disk appears face-on. 

Judging from the preferentially narrow line widths, typical gaussian component widths, the higher degree of asymmetry, and the prevalence of such features, possible origins of the wing-like feature include tidal tails, bulge gas, halo or circum-galactic gas, weak outflow by stellar feedback, and high velocity clouds. 
Robustly studying the nature of the excessive flux requires modelling and decomposing resolved interferometric observations. 
Although interferometric observations exist for several of the broad wing candidates, further analysis is beyond the scope of this study but represents a promising direction for future spatial-spectral disk modelling and galactic dynamics studies.

\subsection{Application of PANDISC to CO or [C II] profiles}

Spectroscopic studies of other ISM tracers, notably the CO vib-rotation lines and the \cii 158 $\mu$m fine-structure line, have also contributed greatly to our understanding of the dynamics of galaxies \citep[e.g.][]{rizzo20,lelli21}. 
At redshifts above 0.2, CO and \cii are the most promising gas tracers at mm and submm wavelengths. 
Given their distance at high redshift, galaxies emitting those tracers typically extend across fewer than three resolving elements, especially for lower mass systems.
Thus the integrated line profile is often the only way to extract the dynamical information of galaxies, suggesting the potential application of the PANDISC model to global profiles beyond \hi line. 

Significant effort has also gone toward establishing the Tully-Fisher relations for both CO and \cii lines \citep[e.g.][]{dickey92,ho07,davis16,fraternali21,wu22}.
We emphasize that the CO and \cii lines can also be fitted very well by the PANDISC model, but the interpretation of the derived parameters must be taken with extra consideration.
Because of the compact distribution of molecular gas and the ISM surrounding regions of star formation, the co-rotating assumption in particular may no longer hold for CO and \cii. 
\citet{blok2016} also found a difference between the line width of the \hi, \cii and CO lines.
Although the utilization of PANDISC beyond \hi is of great interest, it should be treated in the first place as a parametrized description of the line profile, instead of a conclusively physical interpretation.

\section{Summary}
\label{sec:summary}

In this paper, we present a physically-motivated parametric model for the integrated \hi spectrum. 
The model is comprised of a co-rotating disk and a distinct Gaussian component.
The shape of the model is controlled by 5 parameters: $v_r$, $k$, $v_\sigma$, $v_g$, $r$, plus two other parameters,  the line flux $F$ and line center $v_c$. 
The model is designed to extract information from the integrated \hi line profile, such as the width of different components, asymmetry, and the line edge steepness. 
We use MCMC to fit the line model on observed \hi spectra, taking account of the correlation between channels. 
This fitting method produces a statistically-robust description of the \hi spectral line. 

The model is applied on various samples to test and demonstrate its use. 
We found that:

\begin{itemize}
\item The model is a good description of \hi line profiles of various shapes and is able to fit structures including the trough, peaks, edges, and wings if present.
\item Model fitting provides an automated measurement of the velocity width \wfvty, making it a useful tool for checking published global \hi line widths, and for application to large \hi profile datasets. 
\item The \wfvtym and flux derived from PANDISC agree with the ALFALFA \wfvty and flux within the uncertainty for profiles of S/N down to $\lesssim 6$.
\item The model-based SRD agrees well with the ALFALFA S/N.
\item The model provides another line width measurement $v_r$ which can be a proxy of the flat rotational velocity \vf. The comparison with \vf for the \citetalias{lelli2016} sample shows good agreement despite a trend of deviation at the lowest and highest \vf. 
\item The $v_r$ to \vf scaling trend is caused by the rotation curve structures in both the low mass and high mass galaxies, which also explains the agreement between $v_r$ and $V_\mathrm{2R_e}$. This suggests that the majority of the \hi gas in galaxies may reside in radius smaller than where the rotation curve flattens, and the inner structure of rotation curves causes the differences between different line width measurements. But such structure is also a function of galaxy mass, so that a trend emerges. 
\item We fit BTFR using different line width measurements. The difference in the fitted slope is consistent with the $v_r$--\vf and \wfvty--\vf trends.
\item We use model-fitted parameters to control the sample used to derive the BTFR. Restriction to the model-selected rotation-dominated disk sample improves the BTFR fit and introduces less bias on the line width compared with the kurtosis-based selection suggested by \citetalias{papastergis16}. 
\item Inclusion of the Gaussian component reveals interesting structures in \hi profiles. We select spectra which display high S/N broad Gaussian wings that are probably affected by confusion or dynamically distinct \hi gas. Such broad-wing features are worthy of further spatially-resolved investigations.
\end{itemize}

At the same time, we point out limitations in the PANDISC model fitting and interpretation:

\begin{itemize}
\item The physical assumptions associated with many parameters are over-simplified. We already see that $v_r$ deviates from the assumed projected rotational velocity due to the inner structures in rotation curves.
\item Model fitting loses constraining power for $k$, $v_\sigma$, $r$ for individual spectra with S/N$<$8. 
\item The unusual profiles that are probably affected by confusion pose challenges in fitting. Some special terms are included in the prior function to handle these cases. 
\item The model is a linear mixed model with different dimensionalities, so special care needs to be taken in setting the prior function to normalize the parameter space volume.
\end{itemize}

Other than the applications demonstrated in the paper, the model can also be used to explore the potential to extract more dynamical information in the \hi spectra for large observational datasets. 
It also provides a framework to compare with and aid the disk modelling for interferometrc data, and to develope similar tools for other gas tracers like CO and \cii. 
Furthermore, the parametrized PANDISC model makes it possible to perform ensemble studies of the \hi line profile. 
The distribution of line width, asymmetry, line edge steepness, and their correlation with other physical quantities such as galaxy mass, morphological type and star formation rate could give us an enriched view of \hi dynamics and properties. 
We plan to apply the model to the full ALFALFA sample in order to study the aforementioned topics, and the results will be described in a future paper.


\begin{acknowledgments}

We gratefully thank the anonymous referee for the constructive comments and suggestions. 
We also thank Tom Loredo for the advice and clarifications on the statistics used in the paper. 
We acknowledge support from NSF/AST-1714828 and grants from the Brinson Foundation. 
B.P. acknowledges the support of NRAO SOS 1519126.

\end{acknowledgments}

\software{\texttt{AstroPy} \citep{astropy13,astropy18}, \texttt{NumPy} \citep{numpy}, \texttt{emcee} \citep{emcee}, \texttt{george} \citep{george}}



\appendix

\section{Derivation of the model line widths}
\label{sec:width}

As the most common application of the integrated \hi spectrum relies on the line width, we will provide here some recipes for estimating the commonly used width measurements based on the model parameters. 

\subsection{Peak-to-peak width}
The peak width of the disk profile is the easiest to estimate. 
In the limit that $v_r = 0$ so that the disk profile is just a Gaussian peak representing the velocity dispersion, the peak width is 0; while at the other end $v_r \gg v_\sigma$, the width converges to $W_\mathrm{peak} \rightarrow 2 (v_r - 0.75 v_\sigma)$. 
The peak width is hence derived by gluing the two limits together, taking into consideration that the two peaks only appear when $v_r \geq 1.7 v_\sigma$, as well as the edge-narrowing effect, and that the value approaches the higher end limit in an exponential manner. 
\begin{equation}\begin{split}
\label{equ:w_peak}
	W_\mathrm{peak} = 
	\begin{dcases}
	0, & v_r < 1.7 v_\sigma \\
	(2 v_r - 1.5 v_\sigma) \cdot \left\{1 - \exp\left[-\left(\frac{v_r}{v_\sigma}\right)^2 + 3\right]\right\}, & v_r \geq 1.7 v_\sigma
	\end{dcases}
\end{split}\end{equation}

\subsection{\texorpdfstring{$W50_\mathrm{disk}$}{W50\_disk}}
A similar procedure can be applied to approximate $W50_\mathrm{disk}$, defined as the full width half maximum of the disk component of the model. 
In the narrowest limit, $W50_\mathrm{disk}$ is largely affected by the Gaussian profile of the velocity dispersion, approximating $W50_\mathrm{disk} \sim 2 v_r + 2.355 v_\sigma$. 
At the other end when disk profile is wide, \wfvty converges as $W50_\mathrm{disk} \rightarrow 2 v_r + 1.4 v_\sigma$. 
To estimate $W50_\mathrm{disk}$ for varying $v_r$ and $v_\sigma$, the values at the two ends are combined as an exponential transition happening around $v_r \cong v_\sigma$, and the best fit shows $\leq 3\%$ deviation 
\begin{equation}\begin{split}
\label{equ:w50_disk}
	W50_\mathrm{disk} = &(2 v_r + 2.4 v_\sigma) \cdot \exp\left(- 1.8 \frac{v_r}{v_\sigma}\right) + \\
	&(2 v_r + 1.4 v_\sigma) \cdot \left[1 - \exp\left(- 1.2 \frac{v_r}{v_\sigma}\right) \right]
\end{split}\end{equation}

\subsection{Estimating the peak flux density}
In order to estimate \wfvty for the whole model, a weight is needed to co-add $W50_\mathrm{disk}$ with the FWHM of the Gaussian peak. 
The weight we use is the flux density contrast between the two components at the edge of the disk profile, so we first describe how to approximate the peak flux density of the disk $F_\mathrm{\nu, peak}$. 
Because the peak is also affected by the asymmetry, $F_\mathrm{\nu, peak}$ should be treated as the average flux density of the two peaks in the case of an asymmetric disk. 
Again we start by looking at the narrowest and widest ends of the disk profile. 
At the narrow end, the disk peak flux density is equivalent to the Gaussian peak maxima, namely $F_\mathrm{\nu, peak} \sim 1/\sqrt{2 \pi}v_\sigma$. 
When the disk is very broad, the peak contains the flux at the edge of a perfect rotating disk $F \approx \arccos(1 - \Delta v/v_r) \Delta v$, spread out by the velocity dispersion. 
Hence the value converges as $F_\mathrm{\nu, peak} \rightarrow \frac{1}{\pi v_\sigma}\arccos\left(1 - 0.27 \frac{v_\sigma}{v_r}\right)$. 
The two limits are again stitched via an exponential transition at around $v_r \cong v_\sigma$, and the best fit is
\begin{equation}\begin{split}
\label{equ:fpeak}
	F_\mathrm{\nu, peak} = & \left(\frac{1}{\sqrt{2\pi}v_\sigma} - 0.13 \right) \cdot \exp\left[-\frac{1}{2}\left(\frac{v_r}{v_\sigma}\right)^2\right] + \\
	&\frac{1}{\pi v_\sigma}\arccos\left(1 - 0.27 \frac{v_\sigma}{v_r}\right) \cdot \left[1 - \exp\left(-1.8 \frac{v_r}{v_\sigma}\right)\right]
\end{split}\end{equation}

\subsection{\texorpdfstring{\wfvtym}{W50\_model}}
Now with all the tools ready, we can estimate the \wfvty for the whole model. 
The model \wfvty is derived by combining the widths of both the disk and the Gaussian components by the weight $\mathit{w}$, such that $W50_\mathrm{model} = \mathit{w} \cdot W50_\mathrm{disk} + (1 - \mathit{w}) \cdot 2.355 \cdot v_g$. 
The variable $\mathit{w}$ denotes the contribution of the two components to the line width, different from the disk flux fraction $r$. 
For example, in a line profile for which the disk and Gaussian components both share half of the flux, the latter may not affect the width if it is very narrow as a spike at the center, or very wide as a negligible wing sitting beneath the line. 
The weight $\mathit{w}$ is found to be best representative as the flux density contrast between the disk and the Gaussian at both the disk peak and half maxima, as
\begin{equation}\begin{split}
\label{equ:w}
	\mathit{w} = \frac{2F_{v, peak}}{2F_{v, peak} + F_{v, \mathrm{gaus}}(W_\mathrm{peak}/2) + F_{v, \mathrm{gaus}}(W_\mathrm{50, disk}/2)}
\end{split}\end{equation}
where $F_{v, \mathrm{gaus}}(\Delta v)$ denotes the flux density of the Gaussian component evaluated at the $\Delta v$ relative to the line center, and $F_{v, peak}$ is the disk peak flux density in Equ.~\ref{equ:fpeak}. 

\subsection{Asymmetric flux contrast}
Another property that is of interest is the quantitative description of the asymmetry. 
In the model, the flux ratio in the two halves can be easily derived using $k$ and $r$. 
Denoting the integrated flux in the blue- and red-shifted halves of the disk component as $F_\mathrm{b, disk}$ and $F_\mathrm{r, disk}$, their values are $F_\mathrm{b, disk} = F \cdot r \cdot (\tfrac{1}{2} + \tfrac{\pi}{8}k)$ and $F_\mathrm{r, disk} = F \cdot r \cdot (\tfrac{1}{2} - \tfrac{\pi}{8}k)$.

\section{Justification of the adopted prior}
\label{sec:priori}

The first part of the prior function as well as the allowed range of each variable are

\begin{equation}\begin{split}
\label{equ:priori_flat}
	p_1(\theta) = & \frac{1}{2} \qquad\qquad\qquad\qquad\qquad (\log 5 < \log v_r < \log 500)\\
	& \times \frac{1}{0.568} \exp\left(-3|k|\right)   \qquad (-2/\pi < k < 2/\pi)\\
	& \times \frac{1}{22}  \qquad\qquad\qquad\qquad (3 < v_\sigma < 25)\\
	& \times \frac{1}{0.786} 0.44^{(1-r)^2}  \qquad\qquad (0 \le r \le 1)\\
	& \times \frac{1}{1.4} \qquad\qquad\qquad\qquad (\log 8.5 < \log v_g < \log 200)
\end{split}\end{equation}

\begin{figure}
	\centering
	\includegraphics[width = 0.5 \textwidth]{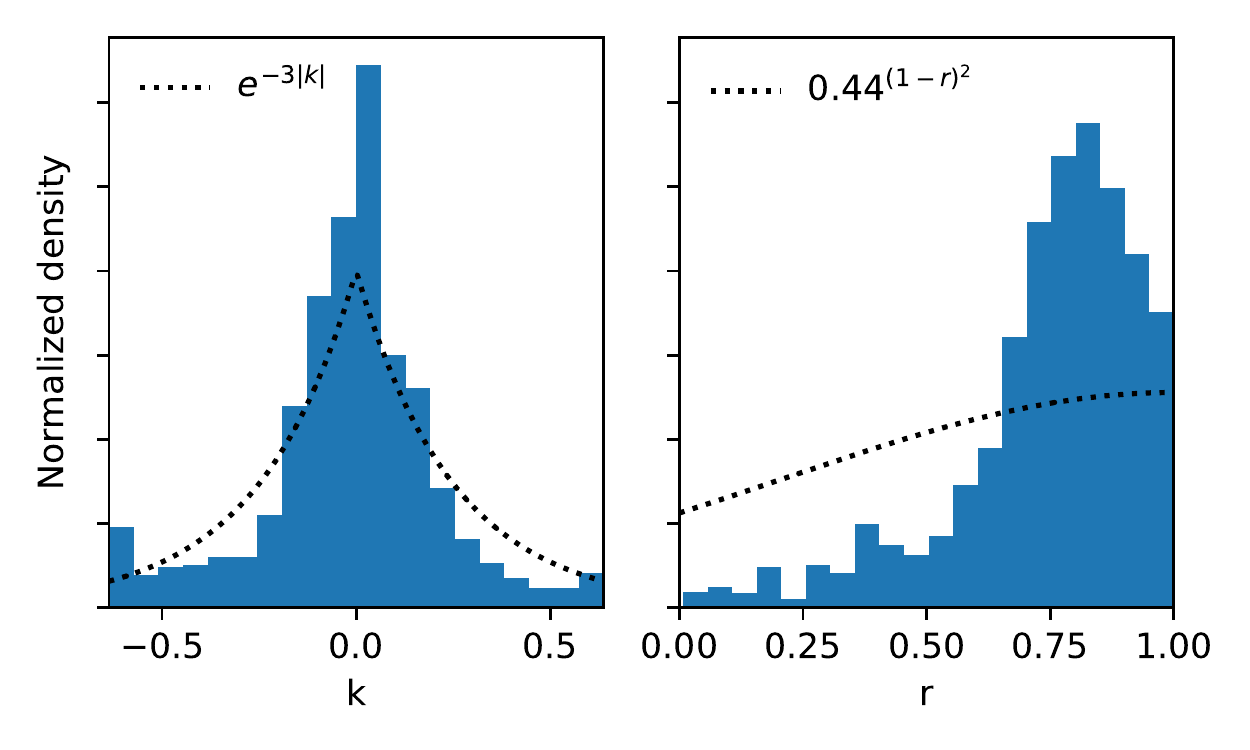}
	\caption{The histogram of $k$ and $r$ distribution for the whole ALFALFA high S/N sample, as the justification for the choice of prior function shown in dotted line.}
	\label{f:kr}
\end{figure}

The $\exp\left(-3|k|\right)$ term accounts for the fact that most of the \hi spectra are symmetric, and the value is chosen to balance between imposing a strong prior and the distribution of $k$ in the ALFALFA high S/N sample. 
The $0.44^{(1-r)^2}$ term is applied in order to account for different dimensions in the disk and Gaussian component's parameter space. 
The PANDISC model is a mixed model combined together by the variable $r$ as the weight.
Three variables, namely $v_r$, $k$, $v_\sigma$, control the disk model, but the Gaussian component is only described by the variable $v_g$. 
This difference in dimensionality makes $r$ much less constrained in the Gaussian-dominated region, and hence inflates the probability of $r$ in its marginalized distribution. 
Thus a factor $0.44^{(1-r)^2}$ is deduced as the parameter space normalization in the Gaussian-dominated regime. 
A comparison of the prior function to the fitted parameter distribution for the high S/N sample can be found in Fig.~\ref{f:kr}.

Because sometimes a shoulder appears at one edge of a double horn profile, probably due to confusion by a companion galaxy especially in the profiles of low S/N distant galaxies, the model tends to fit a highly asymmetric disk for one of the line peaks and the flat part in the trough, while using a narrow and high Gaussian component at the model center to fit the other line peak. 
This enables the extra flux in the fitted highly asymmetric disk to fit the shoulder, but the model fitting itself is unphysical.
To avoid such unphysical fitting, a special term is multiplied by the prior function:

\begin{equation}\begin{split}
\label{equ:priori_r}
	p_2(\theta) = \begin{cases}
	1-\frac{1-r}{r\sqrt{2\pi}}\frac{v_r}{v_g} & \left(\frac{1-r}{r\sqrt{2\pi}}\frac{v_r}{v_g} > \frac{v_r}{2v_g} + 0.3\frac{v_\sigma}{v_g} > 1\right)\\
	1-\frac{v_r}{2v_g} + 0.3\frac{v_\sigma}{v_g} & \left(\frac{v_r}{2v_g} + 0.3\frac{v_\sigma}{v_g} > \frac{1-r}{r\sqrt{2\pi}}\frac{v_r}{v_g} \ \& \ \frac{v_r}{2v_g} + 0.3\frac{v_\sigma}{v_g} >1 \ \& \ \frac{1-r}{r\sqrt{2\pi}}\frac{v_r}{v_g} > \frac{1}{2}\right)
	\end{cases}
\end{split}\end{equation}

The prior probability used is $p(\theta) = p_1(\theta) * p_2(\theta)$.

\section{Collection of unusual profiles}
\label{sec:unusual}

In our study, many spectra are flagged as having unusual profiles for various reasons. 
In this section, we show that some of the typical unusual profiles can be categorized by their shapes and potential causes. 
The prevalence of highly asymmetric profiles in these unusual profiles also helps to justify our criterion of flagging by asymmetry.

\subsection{Confusion}
\label{sec:confusion}

Due to the large beam size typical of radio single-dish observations, confusion plays an important role. 
For ALFALFA survey, $\sim4\%$ of the \hi profiles are estimated to be blends \citep{jones16}. 
Among the spectra flagged in our study, a large fraction of them can be attributed to likely confusion with neighbors.
The confusion-contaminated spectra appear in several different shapes and levels of confidence, though they all involve two components showing distinctively different properties, thus suggesting different origins. 

\begin{figure}
	\centering
	\includegraphics[width = 0.475 \textwidth]{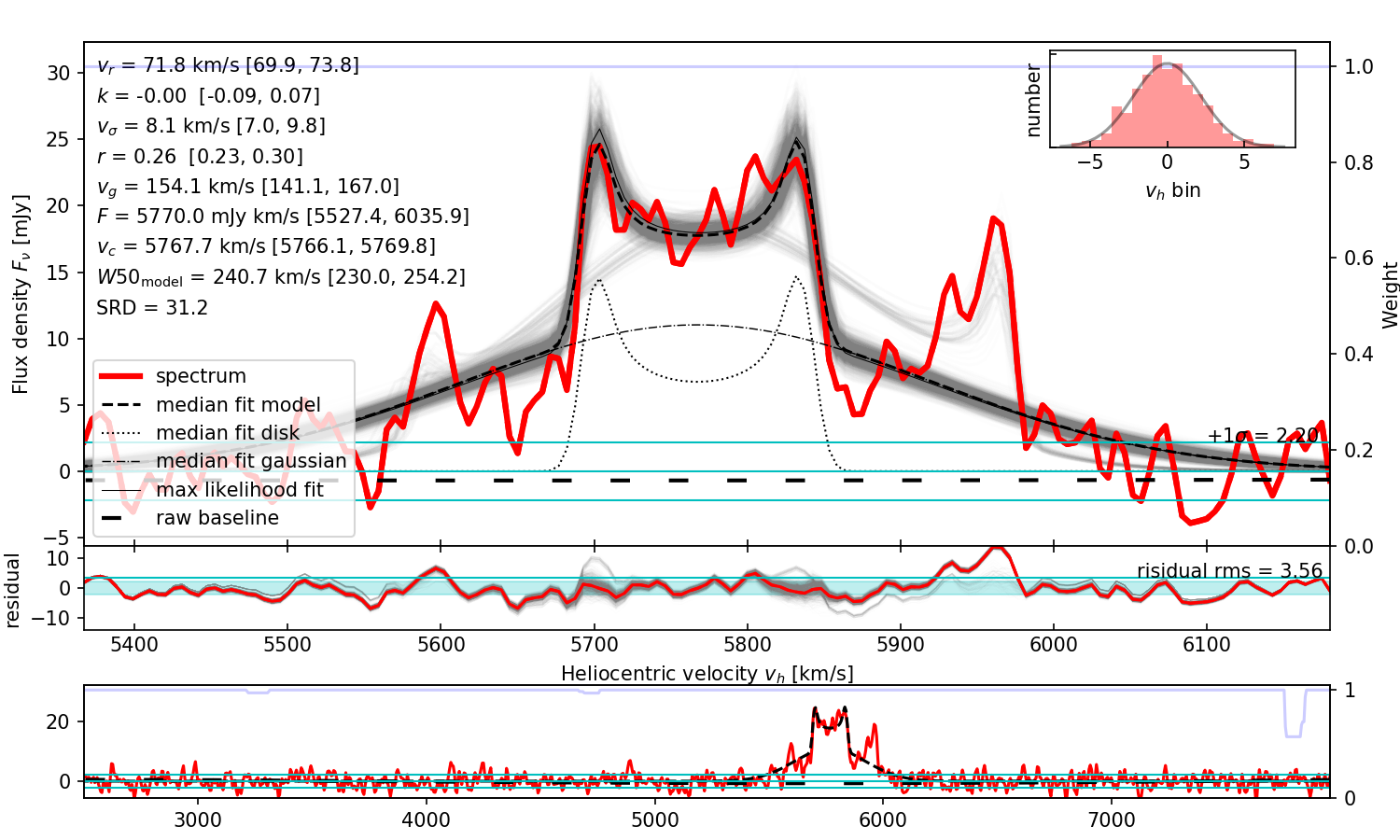}
	\includegraphics[width = 0.475 \textwidth]{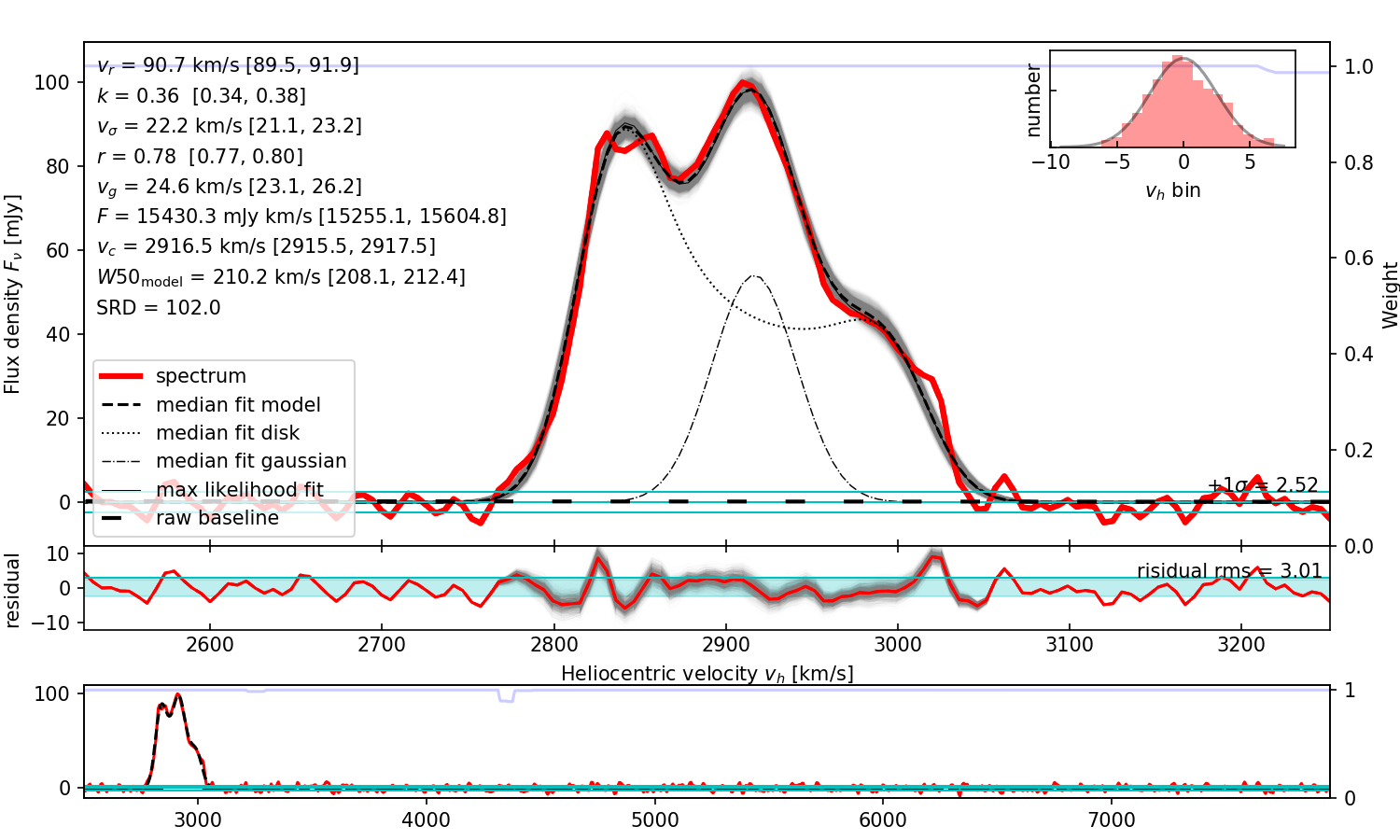}
	\includegraphics[width = 0.475 \textwidth]{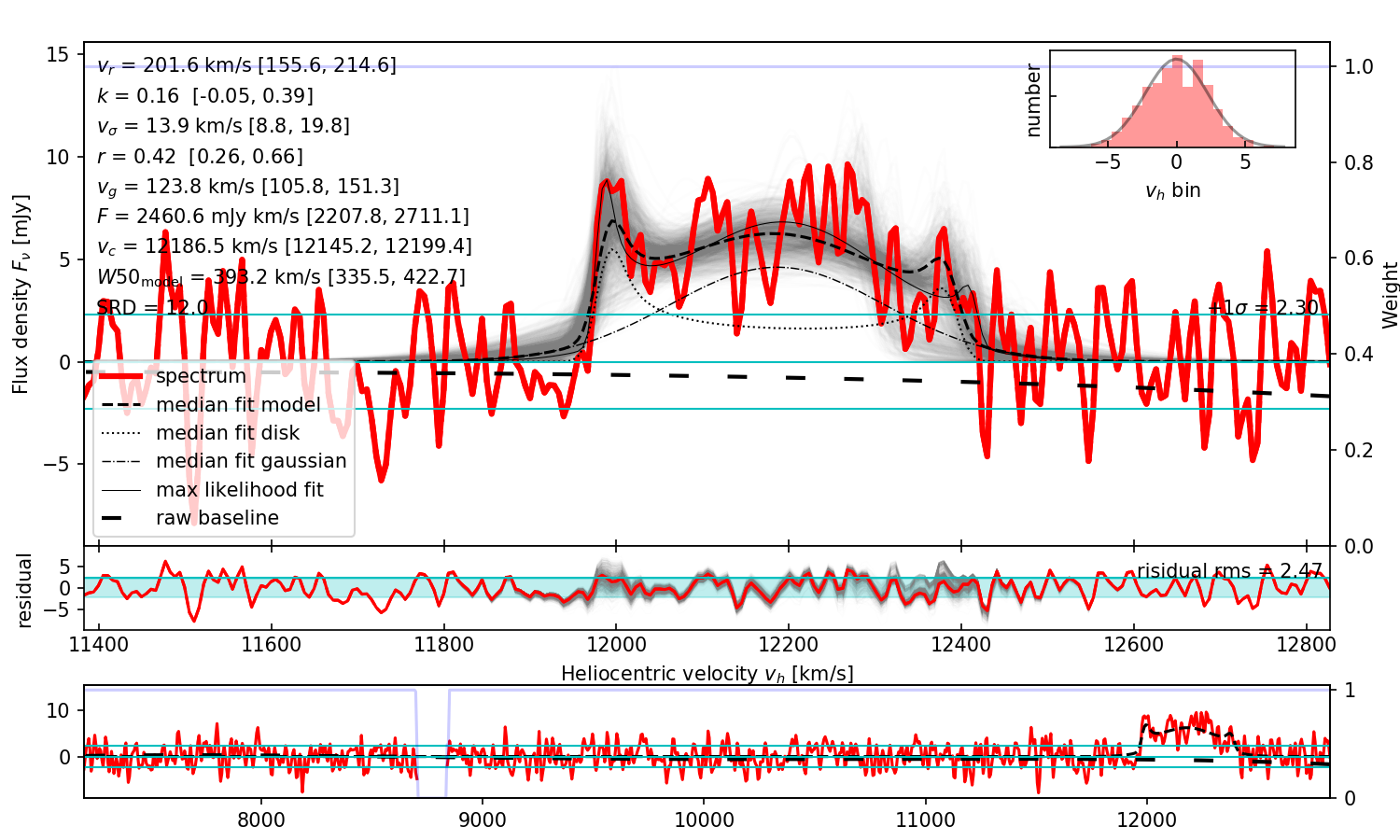}
	\includegraphics[width = 0.475 \textwidth]{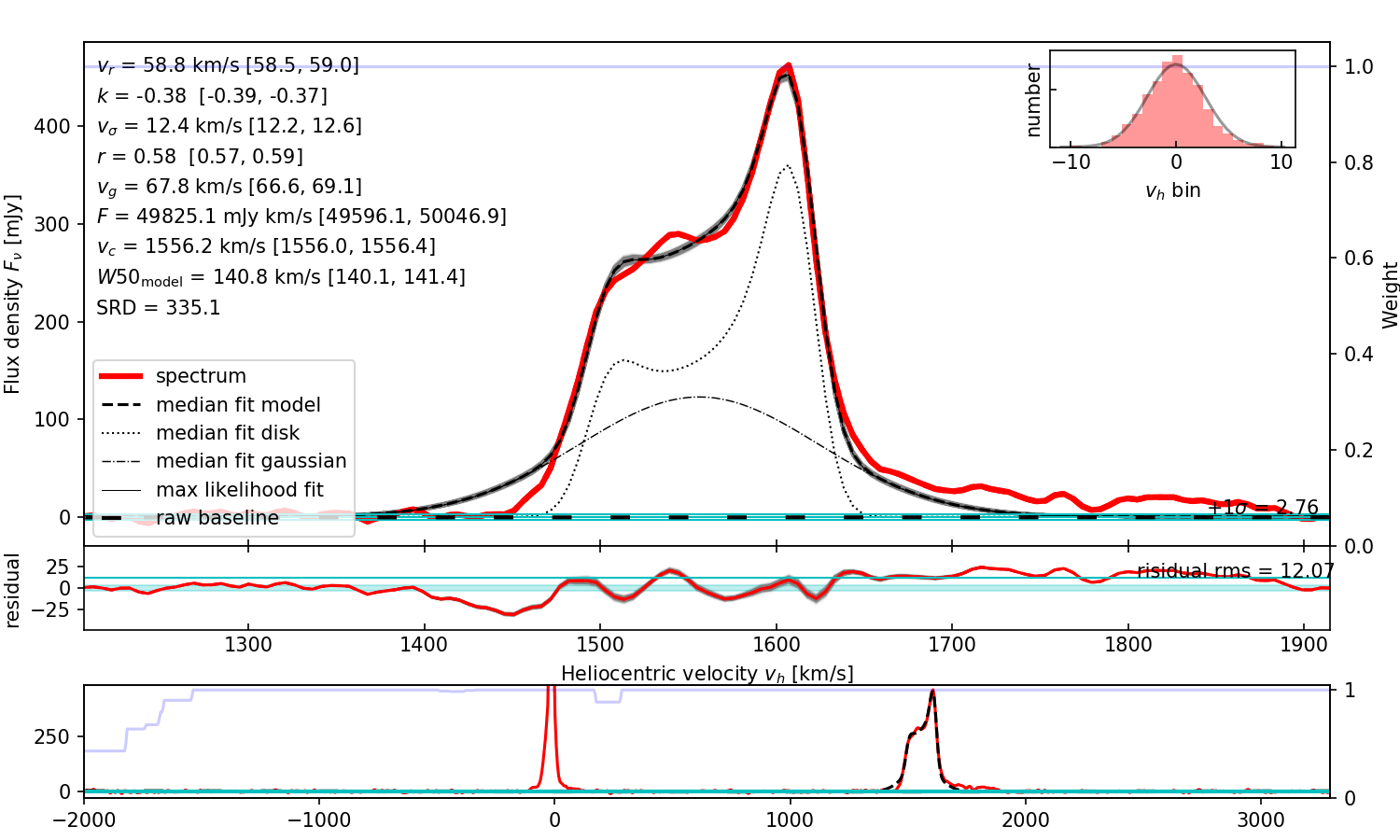}
	\caption{Examples of confusion spectra, left to right, upper to lower are AGC 201046 (F568-V01), AGC 12737 (NGC 7731), AGC 252877, and AGC 9576 (NGC 5774).}
	\label{f:confusion}
\end{figure}

The most obvious evidence of confusion involves one double-horned component lying on top of another, with an example shown in the first panel in Fig. ~\ref{f:confusion}. 
The example spectrum for AGC 201046 contains two galaxies NGC 3363 and VLSB F568-V01 at the same redshift and only 2 arcmin apart, the latter likely responsible for the confusion.
This spectrum is flagged by our criterion by the large discrepancy between the derived \wfvtym and the ALFALFA \wfvtym.
It is also unusual for having a multi-modal posterior distribution, and would be selected as a broad-wing candidate had the selection in Sec.~\ref{sec:broad} been applied on the \citetalias{lelli2016} sample.
However, such profiles are rare, requiring the host-to-companion mass ratio to be relatively low, exact redshift alignment, the existence of a disk component in both galaxies, and the the inclination to be just right so that the flux densities are comparable.
Only one such spectrum is found in the $\sim$900 spectra analyzed in this paper.

Another form of confusion likely appears as a narrow and high peak lying on top of a double horned profile. 
Such a profile is expected of a galaxy pair with a small mass contrast but very different inclinations. 
One example is shown for UGC 12737 in the upper right panel of Fig.~\ref{f:confusion}. 
In the spectrum of UGC 12737, two spiral galaxies separated by 1.5 arcmin, NGC 7731 and NGC 7732, are both present in the ALFA beam. 
Additionally, the spectrum of UGC 12737 is almost exactly the same as UGC 12738.
The fact that NGC 7731 is almost face-on but NGC 7732 has a high inclination results in the bright and narrow peak in the middle of the double-horned profile. 
The asymmetric shape of the double-horned profile could also be the result of the galaxy-galaxy interaction, but interferometric observations would be required for confirmation. 
This spectrum is flagged for being highly asymmetric as well as having a large discrepancy between \wfvty and \wfvtym because the central peak was considered as the line edge in the manual ALFALFA measurement.
This type of confusion is difficult to distinguish from normal spectra, especially in the case of low S/N with a small offset in the redshift, and such cases may simply be identified as highly asymmetric profiles. 
Nevertheless, such profiles are physically rarer than the overlapping double-horned profile because of the lower probability for the necessary very small inclination angle.
Furthermore, given the low mass contrast, their intrinsic line profiles have a higher chance to be intrinsically asymmetric due to interaction as in the example. 

A more common signature of confusion is a shoulder-like feature, arising in the case of a high host-to-companion (or host-to-confusion) mass ratio with a small redshift offset. 
The blending can be present to varying degrees, from a small extra plateau on one side of the spectral line (e.g. AGC 252877, lower left panel in Fig.~\ref{f:confusion}), to a small excessive flux on one side of the profile (UGC 9576, lower right panel in Fig.~\ref{f:confusion}).
Many of the latter features are also selected as broad-wing candidates in Sec.~\ref{sec:broad}.
Due to the large mass ratio, these potentially-confused sources are difficult to confirm in optical images, especially for low mass systems. For example, in the case of AGC 252877, no source is found at the potential confusion redshift by searching in the SDSS spectroscopic database \citep{einsenstein11}.
However, UGC 9576, or NGC 5774, is a galaxy in a pair with NGC 5775 (UGC 9579) over the range of heliocentric velocity from 1500 through 1900 km/s. 
Given the large size of the galaxies compared to the size of the ALFA beam, the shoulder-like feature may be caused by the flux of NGC 5775 in the side-lobes.
However, we cannot rule out the possibility of confusion by tidal interaction debris between the galaxies. 
Using UGC 9576 as an example, it is reasonable to conjecture that the ``shoulder'' and ``ambiguous'' sources in Sec.~\ref{sec:broad} are largely caused by such confusion.
These galaxies are flagged as low quality fits because of their high S/N but obvious mismatch in the shoulder feature, because of being highly asymmetric, and because of the \wfvty to \wfvtym discrepancy as the PANDISC model tries to treat the flux excess as part of the profile. 
The shoulder-like features can also be selected in many other ways like the broad-wing selection in Sec.~\ref{sec:broad}, or the integrated flux discrepancy. 
Moreover, on a physical bases, large host-to-companion or host-to-confusion mass ratios with small redshift offset should dominate the confused spectra for such integrated \hi observations.

Confirming the origin of these features as the result of confusion within the telescope beam requires detailed, spatially-resolved studies combining multi-wavelength data for individual galaxies, and is beyond the scope of this study. 
However, these confusion examples and the capability of identifying them in the integrated line profile prove the value of PANDISC model and its application for sample control as discussed in Sec.~\ref{sec:p16}.

\subsection{\texorpdfstring{\wfvty}{W50\,} discrepancy}
\label{sec:discrepancy}

\begin{figure}
	\centering
	\includegraphics[width = 0.475 \textwidth]{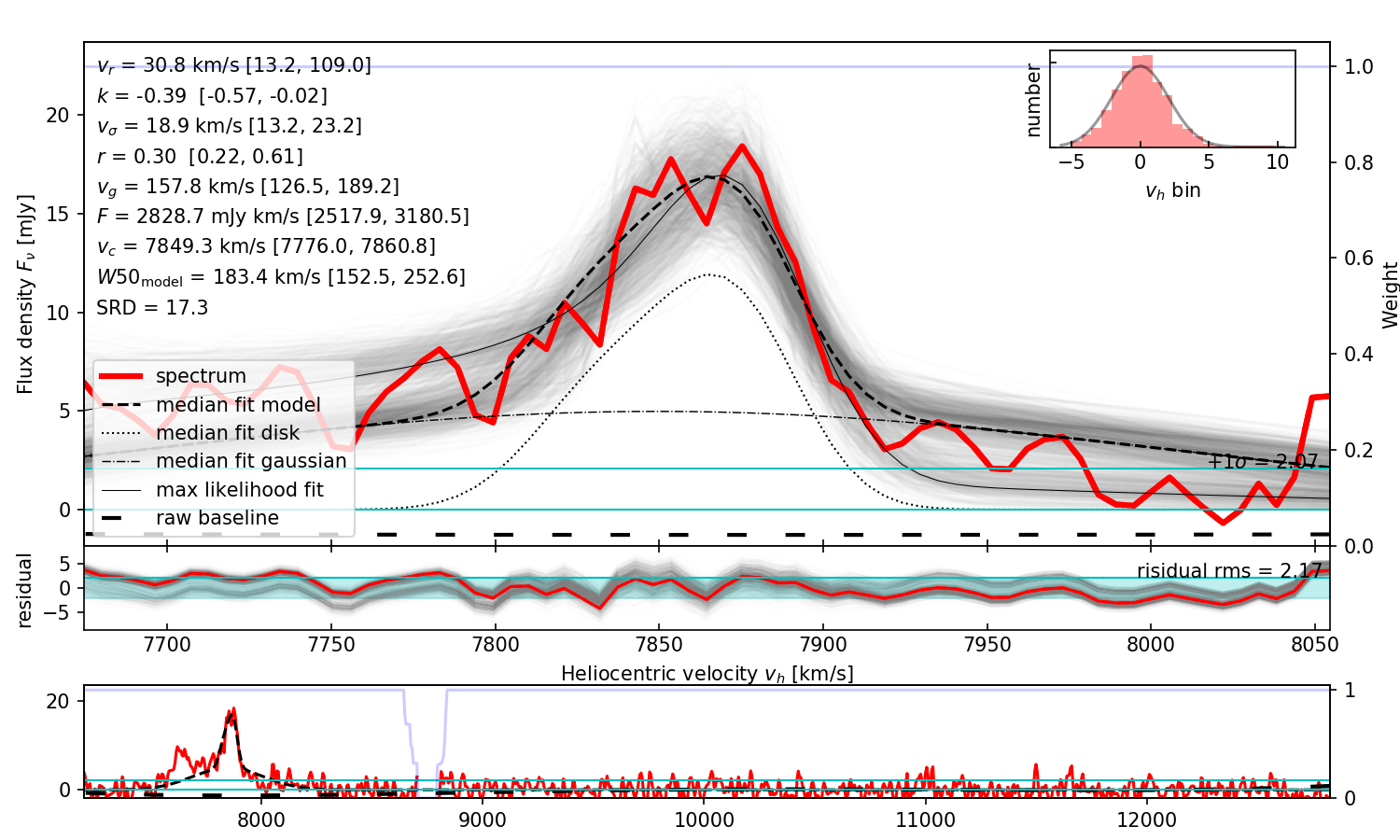}
	\includegraphics[width = 0.475 \textwidth]{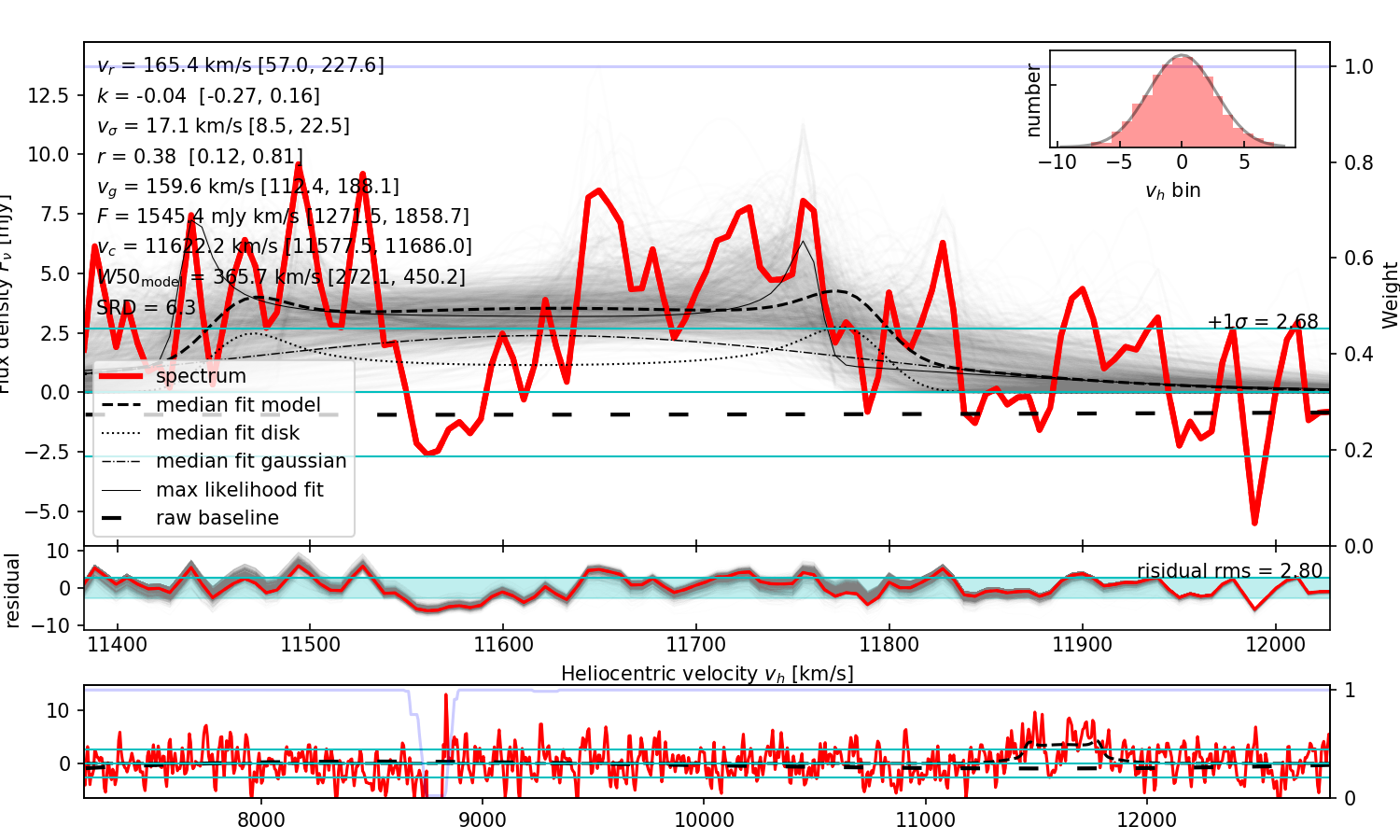}
	\caption{Example spectra of \wfvty--\wfvtym discrepancy, left: AGC 193902, right: AGC 728887.}
	\label{f:discrepancy}
\end{figure}

Many spectra used in the study are also flagged by their \wfvty to \wfvtym discrepancy, and some of them are likely not caused by the profiles themselves  but by the process of human-assisted data processing.
One example is AGC 193902 with $\wfvty = 79 \ \mathrm{km/s}$, shown in the left panel in Fig.~\ref{f:discrepancy}.
Because the model fitting only uses a $5 \times \wfvty$ bandwidth of the original ALFALFA spectrum to save  computation time (bottom panel in Fig.~\ref{f:discrepancy} left), it is clear that half of the spectral line is missing.
This is because the human inspection misidentified the higher peak of this asymmetric profile as a single-peak line, and missed the trough as well as the other smaller peak. 
However, in model fitting, the excessive flux on one side without the other peak data being input to the fitting routine forces the model to fit a very broad gaussian component, and results in the large discrepancy with \wfvty.
After checking the spectra flagged by such a \wfvty discrepancy, six spectra can be reliably categorized as having misidentified line peaks.
These spectra are not all asymmetric, but are preferentially low S/N, making it difficult to notice the rest of the flux except for the peak. 
The potential occurrence rate $\sim 2\%$ emphasizes the need for a fully automated and statistically robust method in reducing ALFALFA data, which we plan to undertake using the PANDISC model along with another line width measurement algorithm in \citet{ball22}.

But when the S/N is too low, if becomes even more difficult to distinguish misidentified lines from confusion.
Such is the case for AGC 728887 shown in the right panel in Fig.~\ref{f:discrepancy}.
There is a clear flux excess at $\sim 11500 km/s$ that is not identified in the ALFALFA measurement, but the small gap between the main emission and the extra bump at lower velocity makes it resemble a confused profile. 
Because the gap is very narrow and with low S/N, and the flux densities of the two components are almost the same, we cannot make any statistically convincing conclusion.
This case exemplifies the challenges to identifying unusual profiles in low S/N integrated spectra.

\subsection{Miscellaneous}
\label{sec:miscellaneous}

\begin{figure}
	\centering
	\includegraphics[width = 0.475 \textwidth]{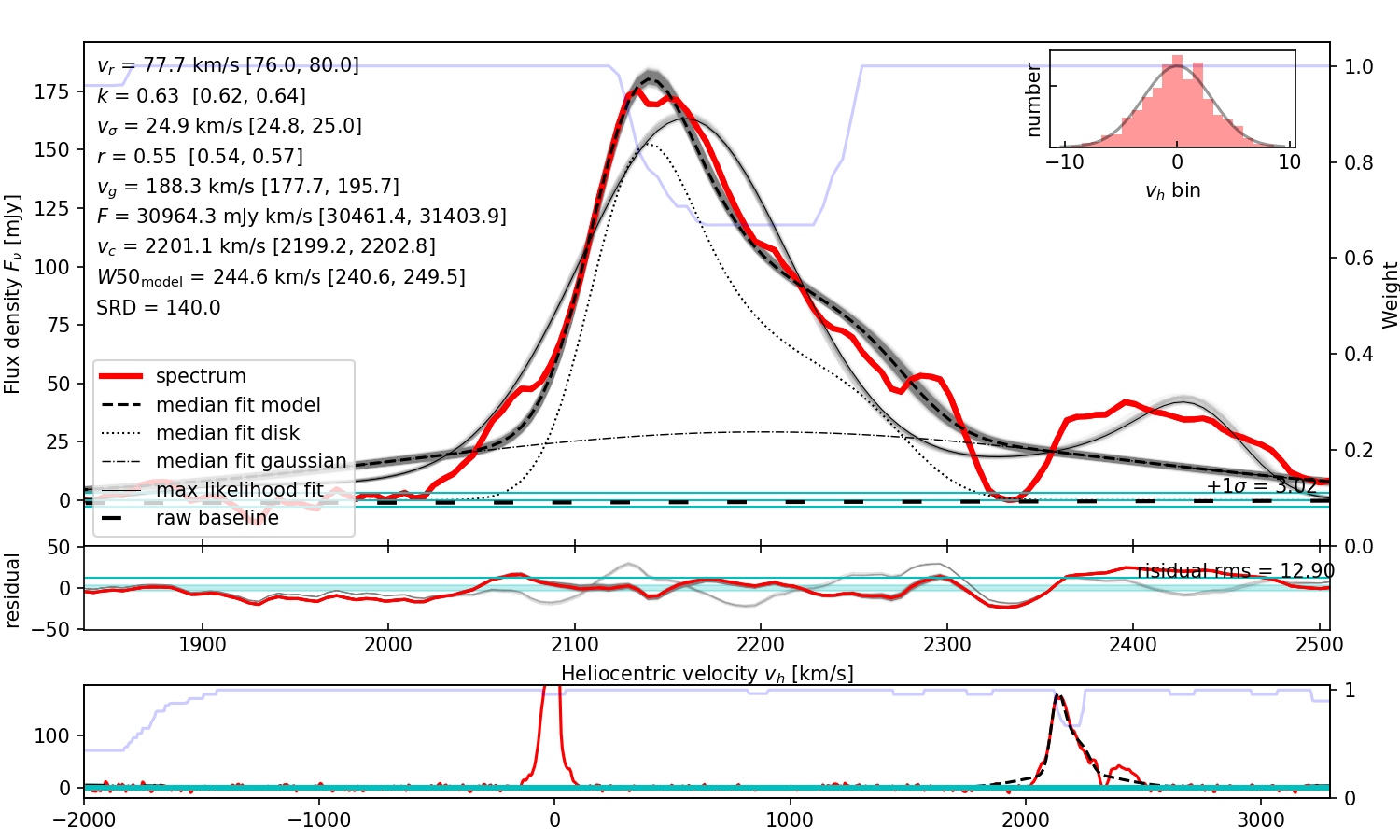}
	\includegraphics[width = 0.475 \textwidth]{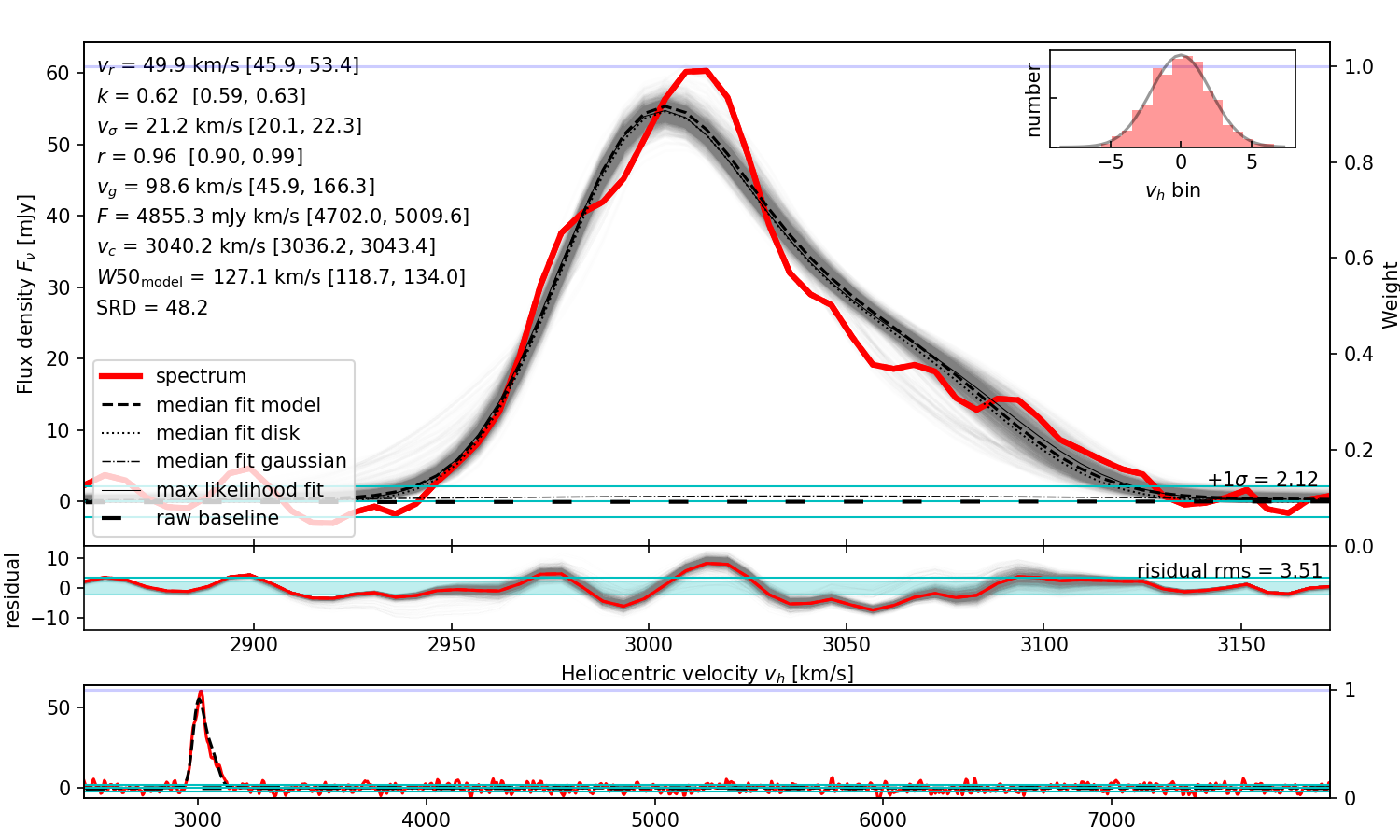}
	\caption{Example spectra of miscellaneous unusual profile, left: AGC 966 (NGC 520), right: AGC 8605 (UGC 8605).}
	\label{f:miscellaneous}
\end{figure}

There are also unusual spectra that can not be well fitted by the model, and are hard to be categorized. 
One spectrum that can be traced back to the effect of astrophysical process is AGC 966 (NGC 520) in the left panel in Fig.~\ref{f:miscellaneous}. 
The spectral resembles a confusion with one asymmetric line centered at 2200 km/s plus a smaller companion at 2400 km/s, both separated by a small gap at 2330 km/s, and this galaxy has long been suspected to be a merger. 
However, comparison with literature \citep{stanford90,beswick03} and archival data \citep{mirabel88,springob2005} suggests a much more complicated picture: the gas at 2330 km/s is most likely caused by the \hi absorption in the inner part of the galaxy; and a small companion UGC 957 does exist and may be connected by tidal tail, but it is at $\sim$ 2135 km/s. 
Neither ALFALFA nor model fitting give the correct width measurement. 
It is difficult to estimate the prevalence of such systems, but luminous infrared galaxies (LIRGs) like NGC 520 are known to have an elevated merger rate and display very complicated morphology and kinematics, and hence should be taken with extra caution. 

Another population of unusual spectra are the asymmetric profiles with unphysical fit. 
AGC 8605 (UGC 8605) is shown in Fig.~\ref{f:miscellaneous} as an example. 
Although the spectrum look like an ordinary asymmetric disk, it defies the model fitting by having too gentle the line edge on low velocity side, and too deep a trough, as well as the line peak offset from the expected position. 
This galaxy might be a complicated system as the optical image shows hints of companion and tidal tail. 
The spectrum is flagged by low quality fit flag, high asymmetry and \wfvty discrepancy.
Another example of unphysical fit is AGC 4115 in Fig.~\ref{f:atlas}. 
Although the model fitting agrees very well with the spectrum, the fitted broad and prominent Gaussian peak is difficult to interpret, and the SDSS image \citep{adelman07} shows a diffuse stellar component without obvious structure plus a spatially offset nucleus. 
AGC 4115 is flagged for high asymmetry and \wfvty discrepancy.
No conclusion can be made for these galaxy systems without optical spectroscopic or interferometric \hi data.
But they highlight the ability of integrated \hi spectroscopy in identifying potentially interesting sources, and again the high occurrence of asymmetry in these unusual spectra alerts the applicability of asymmetric \hi spectra in applications like \hi mass function or BTFR.

\section{Model fitting results}
\label{sec:dataset}

Table~\ref{t:combined} contains all the galaxies used in the study along with their model fitted parameters. 
The columns are (1) Galaxy Name, (2) alternative names, (3) galaxy sample used in the paper, (4) reference of the spectral data, (5)-(11) PANDISC model fitted parameters, (12) \wfvtym, (13) SRD, (14) blank (line-free) channel rms, (15) line channels residual rms. 
The alternative names are the names used in either the sample or the reference paper.
The reference codes use the same definition as in Sec.~\ref{sec:data}.
The superscripts in the ``Name'' column correspond to the following flags, $\ast$: low model fitting quality; $\star$: asymmetry; $\dagger$: low disk fit quality; $\ddagger$: \wfvty to \wfvtym $> 2 \sigma$ discrepancy; $\mathsection$: broad wing candidate.

Figure set 1 contains the fitted model and the MCMC posterior distribution of every galaxy used in the study.
Every galaxy is associated with two figures in the same name as listed in Table~\ref{t:combined}. 
The two figures are the model fitting result and MCMC ensemble corner plot.
Please refer to Fig.~\ref{f:fitting} for the example of figure set, and the format of the figures.
The complete figure set is available in the online version.

\figsetstart
\figsetnum{1}
\figsettitle{Model fitting result for all the \hi spectra}

\figsetend

\newpage
\clearpage

\movetabledown=2.9in
\begin{rotatetable}
\begin{deluxetable}{ccccccccccccccc}
\tabletypesize{\scriptsize}
\tablecolumns{15}
\tablecaption{Model fitted parameter of all the spectra used in the paper. Only a fraction of the table is shown here as an example, the complete table is available in the online version.\label{t:combined}}
\tablehead{
\colhead{(1)} & (2) & (3) & (4) & (5) & (6) & (7) & (8) & (9) & (10) & (11) & (12) & (13) & (14) & (15)\\ 
Name & Alternative Name & Sample & Reference & $v_r$ & $k$ & $v_\sigma$ & $r$ & $v_g$ & $F$ & $v_c$ & $W50_\mathrm{model}$ & SRD & rms$_\mathrm{blank}$ & rms$_\mathrm{residual}$ \\
 &  &  &  & $\mathrm{km\,s^{-1}}$ &  & $\mathrm{km\,s^{-1}}$ &  & $\mathrm{km\,s^{-1}}$ & $\mathrm{Jy\,km\,s^{-1}}$ & $\mathrm{km\,s^{-1}}$ & $\mathrm{km\,s^{-1}}$ &  &  &  
}
\startdata
AGC000027 & CGCG 408-020 & high S/N & a.100 & 99.93$^{+0.41} _{-0.41}$ & 0.024$^{+0.011} _{-0.011}$ & 13.15$^{+0.46} _{-0.44}$ & 0.919$^{+0.025} _{-0.025}$ & 58.56$^{+10.08} _{-10.09}$ & 13.69$^{+0.15} _{-0.14}$ & 3113.42$^{+0.36} _{-0.39}$ & 216.69$^{+1.12} _{-1.09}$ & 106.7 & 2.07 & 2.11 \\
AGC000075$^\star$$^\ddagger$ & NGC  14 & high S/N & a.100 & 45.84$^{+1.06} _{-1.38}$ & 0.557$^{+0.051} _{-0.072}$ & 17.53$^{+0.59} _{-0.56}$ & 0.741$^{+0.043} _{-0.062}$ & 20.82$^{+2.47} _{-2.0}$ & 18.06$^{+0.13} _{-0.13}$ & 868.95$^{+1.87} _{-2.03}$ & 104.82$^{+3.5} _{-4.02}$ & 164.5 & 2.55 & 2.8 \\
AGC000099 & M+201023 & high S/N & a.100 & 33.69$^{+0.42} _{-0.43}$ & 0.017$^{+0.014} _{-0.015}$ & 10.34$^{+0.48} _{-0.57}$ & 0.858$^{+0.093} _{-0.089}$ & 22.97$^{+4.37} _{-9.33}$ & 9.93$^{+0.12} _{-0.09}$ & 1740.0$^{+0.27} _{-0.27}$ & 78.8$^{+1.08} _{-0.98}$ & 117.3 & 2.18 & 2.63 \\
AGC000122$^\ddagger$ & CGCG 456-039 & high S/N & a.100 & 44.75$^{+0.31} _{-0.28}$ & 0.18$^{+0.023} _{-0.022}$ & 7.5$^{+0.36} _{-0.3}$ & 0.501$^{+0.046} _{-0.037}$ & 26.39$^{+1.23} _{-1.56}$ & 14.85$^{+0.13} _{-0.13}$ & 854.02$^{+0.3} _{-0.28}$ & 91.5$^{+1.48} _{-1.1}$ & 146.9 & 2.42 & 2.96 \\
AGC000156 & CGCG 433-041 & high S/N & a.100 & 57.31$^{+0.68} _{-0.78}$ & -0.17$^{+0.019} _{-0.02}$ & 12.57$^{+0.65} _{-0.63}$ & 0.74$^{+0.049} _{-0.057}$ & 24.77$^{+3.28} _{-2.99}$ & 13.19$^{+0.11} _{-0.12}$ & 1134.53$^{+0.47} _{-0.5}$ & 127.54$^{+1.2} _{-1.67}$ & 123.5 & 2.23 & 3.34 \\
AGC000191 & CGCG 434-001,UGC 191 & high S/N,SPARC & a.100 & 57.66$^{+0.28} _{-0.28}$ & -0.03$^{+0.011} _{-0.012}$ & 9.75$^{+0.35} _{-0.32}$ & 0.74$^{+0.032} _{-0.029}$ & 41.2$^{+2.3} _{-2.59}$ & 17.37$^{+0.16} _{-0.15}$ & 1144.03$^{+0.25} _{-0.25}$ & 125.25$^{+0.66} _{-0.65}$ & 138.7 & 2.58 & 3.0 \\
AGC000230 & NGC  99 & high S/N & a.100 & 58.44$^{+0.45} _{-0.45}$ & -0.063$^{+0.014} _{-0.014}$ & 16.13$^{+0.61} _{-0.69}$ & 0.929$^{+0.042} _{-0.051}$ & 82.34$^{+45.29} _{-17.94}$ & 13.48$^{+0.21} _{-0.2}$ & 5312.92$^{+0.48} _{-0.48}$ & 139.74$^{+1.77} _{-1.46}$ & 105.1 & 2.44 & 2.83 \\
AGC000231$^\ddagger$ & NGC 100 & high S/N,SPARC & a.100 & 94.77$^{+0.16} _{-0.17}$ & -0.087$^{+0.005} _{-0.005}$ & 10.97$^{+0.2} _{-0.2}$ & 0.712$^{+0.012} _{-0.012}$ & 54.47$^{+1.33} _{-1.34}$ & 44.85$^{+0.18} _{-0.18}$ & 842.35$^{+0.15} _{-0.16}$ & 198.43$^{+0.52} _{-0.49}$ & 280.1 & 2.66 & 5.97 \\
AGC000260$^\ddagger$ & CGCG 434-013 & high S/N & a.100 & 116.79$^{+0.28} _{-0.28}$ & -0.118$^{+0.006} _{-0.006}$ & 17.05$^{+0.35} _{-0.35}$ & 0.853$^{+0.017} _{-0.018}$ & 68.64$^{+4.63} _{-4.74}$ & 34.33$^{+0.18} _{-0.18}$ & 2133.87$^{+0.27} _{-0.28}$ & 253.11$^{+0.94} _{-0.87}$ & 226.8 & 2.25 & 3.47 \\
AGC000369 & NGC 173 & high S/N & a.100 & 140.48$^{+0.44} _{-0.46}$ & 0.001$^{+0.011} _{-0.011}$ & 13.9$^{+0.62} _{-0.57}$ & 0.859$^{+0.031} _{-0.029}$ & 92.74$^{+10.59} _{-13.04}$ & 21.7$^{+0.26} _{-0.25}$ & 4367.77$^{+0.42} _{-0.41}$ & 297.4$^{+1.39} _{-1.3}$ & 107.6 & 2.77 & 2.86 \\
AGC000477$^\ddagger$ & CGCG 458-004 & high S/N & a.100 & 110.4$^{+0.3} _{-0.3}$ & 0.042$^{+0.009} _{-0.009}$ & 12.36$^{+0.39} _{-0.39}$ & 0.714$^{+0.021} _{-0.019}$ & 68.39$^{+2.55} _{-2.91}$ & 27.4$^{+0.2} _{-0.18}$ & 2647.99$^{+0.29} _{-0.28}$ & 230.97$^{+0.89} _{-0.8}$ & 169.4 & 2.48 & 3.35 \\
AGC000499$^\mathsection$ & NGC 262 & high S/N & a.100 & 22.36$^{+0.31} _{-0.31}$ & 0.169$^{+0.026} _{-0.024}$ & 11.76$^{+0.57} _{-0.57}$ & 0.568$^{+0.076} _{-0.079}$ & 31.41$^{+2.31} _{-1.71}$ & 14.37$^{+0.11} _{-0.12}$ & 4541.76$^{+0.34} _{-0.33}$ & 62.48$^{+0.74} _{-0.71}$ & 176.2 & 2.13 & 3.03 \\
AGC000521$^\ddagger$ & CGCG 435-014 & high S/N & a.100 & 56.03$^{+1.82} _{-2.06}$ & 0.077$^{+0.031} _{-0.034}$ & 14.32$^{+1.54} _{-1.29}$ & 0.625$^{+0.06} _{-0.06}$ & 19.61$^{+1.9} _{-1.84}$ & 13.95$^{+0.13} _{-0.13}$ & 663.96$^{+0.75} _{-0.74}$ & 126.43$^{+1.59} _{-1.89}$ & 131.2 & 2.31 & 3.71 \\
AGC000634 & CGCG 410-023,UGC 634 & high S/N,SPARC & a.100 & 59.31$^{+0.51} _{-0.53}$ & 0.118$^{+0.012} _{-0.012}$ & 14.86$^{+0.45} _{-0.44}$ & 0.934$^{+0.029} _{-0.035}$ & 30.12$^{+6.37} _{-7.82}$ & 14.89$^{+0.14} _{-0.13}$ & 2213.76$^{+0.4} _{-0.42}$ & 137.13$^{+0.96} _{-1.01}$ & 124.9 & 2.37 & 2.88 \\
AGC000668$^\mathsection$ & IC 1613 & high S/N & a.100 & 6.88$^{+0.01} _{-0.01}$ & -0.158$^{+0.002} _{-0.002}$ & 6.52$^{+0.01} _{-0.01}$ & 0.578$^{+0.004} _{-0.004}$ & 14.38$^{+0.06} _{-0.06}$ & 337.46$^{+0.19} _{-0.19}$ & -232.15$^{+0.01} _{-0.01}$ & 23.92$^{+0.02} _{-0.02}$ & 2798.4 & 5.17 & 20.09 \\
AGC000685$^\ast$ & CGCG 458-020 & high S/N & a.100 & 29.16$^{+0.27} _{-0.24}$ & 0.122$^{+0.014} _{-0.014}$ & 12.33$^{+0.36} _{-0.48}$ & 0.937$^{+0.041} _{-0.073}$ & 41.36$^{+44.9} _{-10.69}$ & 13.2$^{+0.16} _{-0.14}$ & 156.75$^{+0.28} _{-0.29}$ & 73.22$^{+0.97} _{-0.74}$ & 146.0 & 2.36 & 4.85 \\
AGC000763$^\ast$ & NGC 428 & high S/N & a.100 & 68.43$^{+0.18} _{-0.18}$ & -0.006$^{+0.004} _{-0.004}$ & 15.79$^{+0.23} _{-0.22}$ & 0.876$^{+0.015} _{-0.014}$ & 57.27$^{+2.38} _{-2.61}$ & 67.69$^{+0.24} _{-0.22}$ & 1153.15$^{+0.14} _{-0.14}$ & 156.69$^{+0.34} _{-0.33}$ & 398.3 & 3.12 & 14.79 \\
AGC000891$^\ast$ & CGCG 436-033,UGC 891 & high S/N,SPARC & a.100 & 51.12$^{+0.34} _{-0.36}$ & -0.068$^{+0.019} _{-0.019}$ & 9.15$^{+0.44} _{-0.41}$ & 0.539$^{+0.037} _{-0.032}$ & 30.73$^{+1.21} _{-1.39}$ & 17.05$^{+0.12} _{-0.12}$ & 642.55$^{+0.36} _{-0.36}$ & 105.91$^{+1.13} _{-0.97}$ & 163.0 & 2.31 & 5.04 \\
AGC000914$^\ast$$^\star$ & NGC 493 & high S/N & a.100 & 89.06$^{+1.31} _{-1.22}$ & -0.321$^{+0.016} _{-0.017}$ & 21.91$^{+0.68} _{-0.66}$ & 0.897$^{+0.032} _{-0.035}$ & 45.68$^{+7.98} _{-8.4}$ & 27.05$^{+0.28} _{-0.27}$ & 2368.17$^{+1.25} _{-1.35}$ & 204.42$^{+2.76} _{-2.52}$ & 165.2 & 2.69 & 5.07 \\
AGC000942$^\mathsection$ &  & high S/N & a.100 & 13.85$^{+0.58} _{-0.44}$ & -0.075$^{+0.042} _{-0.055}$ & 7.54$^{+0.78} _{-0.71}$ & 0.513$^{+0.201} _{-0.158}$ & 15.62$^{+1.73} _{-0.8}$ & 5.89$^{+0.08} _{-0.08}$ & 2334.88$^{+0.31} _{-0.32}$ & 36.19$^{+0.61} _{-0.56}$ & 99.8 & 2.11 & 2.24 \\
AGC000947$^\ast$$^\ddagger$ & NGC 514 & high S/N & a.100 & 115.03$^{+0.25} _{-0.25}$ & 0.016$^{+0.008} _{-0.008}$ & 11.62$^{+0.34} _{-0.32}$ & 0.765$^{+0.017} _{-0.017}$ & 77.21$^{+3.01} _{-3.19}$ & 25.54$^{+0.18} _{-0.18}$ & 2471.61$^{+0.25} _{-0.26}$ & 241.43$^{+0.66} _{-0.66}$ & 169.2 & 2.29 & 4.4 \\
AGC000957$^\star$$^\mathsection$ & KDG 5 & high S/N & a.100 & 25.17$^{+2.57} _{-1.72}$ & 0.494$^{+0.096} _{-0.124}$ & 17.26$^{+3.03} _{-4.28}$ & 0.56$^{+0.264} _{-0.29}$ & 38.6$^{+18.06} _{-5.81}$ & 10.15$^{+0.18} _{-0.15}$ & 2149.57$^{+4.28} _{-2.82}$ & 74.75$^{+5.83} _{-3.41}$ & 111.0 & 2.19 & 2.4 \\
AGC000966$^\ast$$^\star$$^\ddagger$ & NGC 520 & high S/N & a.100 & 77.68$^{+2.31} _{-1.71}$ & 0.633$^{+0.003} _{-0.009}$ & 24.92$^{+0.06} _{-0.15}$ & 0.555$^{+0.015} _{-0.016}$ & 188.34$^{+7.32} _{-10.62}$ & 30.96$^{+0.44} _{-0.5}$ & 2201.14$^{+1.65} _{-1.97}$ & 244.64$^{+4.89} _{-4.09}$ & 140.0 & 3.02 & 12.9 \\
AGC001102$^\star$$^\ddagger$ & CGCG 412-002a & high S/N & a.100 & 40.99$^{+3.64} _{-3.67}$ & 0.595$^{+0.031} _{-0.054}$ & 23.77$^{+0.93} _{-1.6}$ & 0.803$^{+0.138} _{-0.187}$ & 52.68$^{+14.9} _{-7.4}$ & 9.56$^{+0.16} _{-0.16}$ & 1971.4$^{+3.65} _{-4.49}$ & 110.21$^{+6.8} _{-6.45}$ & 96.1 & 2.13 & 2.3 \\
AGC001133 & M+105005 & high S/N & a.100 & 50.73$^{+0.37} _{-0.41}$ & 0.135$^{+0.012} _{-0.012}$ & 12.61$^{+0.39} _{-0.38}$ & 0.841$^{+0.042} _{-0.045}$ & 25.37$^{+3.64} _{-3.92}$ & 15.2$^{+0.09} _{-0.09}$ & 1964.94$^{+0.28} _{-0.27}$ & 115.48$^{+0.92} _{-1.23}$ & 171.4 & 1.92 & 2.56 \\
AGC001149$^\ast$ & NGC 628 & high S/N & a.100 & 22.37$^{+0.04} _{-0.04}$ & 0.057$^{+0.002} _{-0.002}$ & 8.48$^{+0.05} _{-0.04}$ & 0.383$^{+0.004} _{-0.004}$ & 21.14$^{+0.03} _{-0.03}$ & 421.25$^{+0.21} _{-0.2}$ & 658.21$^{+0.02} _{-0.02}$ & 52.28$^{+0.03} _{-0.03}$ & 2687.5 & 4.62 & 52.52 \\
AGC001175 &  & high S/N & a.100 & 39.46$^{+0.38} _{-0.54}$ & -0.018$^{+0.017} _{-0.017}$ & 9.44$^{+0.62} _{-0.62}$ & 0.665$^{+0.112} _{-0.102}$ & 21.19$^{+3.53} _{-4.53}$ & 14.82$^{+0.14} _{-0.12}$ & 729.17$^{+0.32} _{-0.31}$ & 86.24$^{+2.28} _{-2.67}$ & 142.2 & 2.59 & 3.47 \\
AGC001176$^\star$$^\mathsection$ & DDO 13 & high S/N & a.100 & 15.18$^{+0.31} _{-0.33}$ & -0.277$^{+0.036} _{-0.035}$ & 8.3$^{+0.35} _{-0.29}$ & 0.493$^{+0.09} _{-0.061}$ & 16.12$^{+0.47} _{-0.31}$ & 31.3$^{+0.1} _{-0.1}$ & 629.57$^{+0.11} _{-0.12}$ & 38.67$^{+0.24} _{-0.21}$ & 418.3 & 2.59 & 5.19 \\
AGC001192$^\ddagger$ & NGC 658 & high S/N & a.100 & 140.22$^{+0.47} _{-0.48}$ & -0.027$^{+0.009} _{-0.009}$ & 19.44$^{+0.66} _{-0.64}$ & 0.886$^{+0.028} _{-0.032}$ & 80.74$^{+12.01} _{-13.68}$ & 26.12$^{+0.21} _{-0.22}$ & 2989.65$^{+0.47} _{-0.47}$ & 304.25$^{+1.99} _{-1.76}$ & 138.4 & 2.57 & 3.5 \\
AGC001195$^\star$$^\ddagger$ & CGCG 437-010 & high S/N & a.100 & 57.55$^{+1.17} _{-1.24}$ & 0.537$^{+0.046} _{-0.039}$ & 17.47$^{+0.33} _{-0.35}$ & 0.861$^{+0.034} _{-0.045}$ & 26.18$^{+4.41} _{-3.87}$ & 25.91$^{+0.14} _{-0.14}$ & 779.88$^{+1.12} _{-1.11}$ & 133.78$^{+1.85} _{-1.65}$ & 216.7 & 2.52 & 4.68 \\
\multicolumn{15}{c}{$\cdots$}\\
UGC 7608$^\star$ &  & SPARC & T88 & 23.29$^{+0.92} _{-0.45}$ & -0.249$^{+0.029} _{-0.038}$ & 11.06$^{+0.44} _{-0.51}$ & 0.941$^{+0.043} _{-0.108}$ & 24.79$^{+102.06} _{-8.86}$ & 19.04$^{+0.45} _{-0.26}$ & 533.37$^{+0.5} _{-0.5}$ & 58.99$^{+1.36} _{-0.98}$ & 104.9 & 9.53 & 8.34 \\
IC 3687 & UGC 7866 & SPARC & T88 & 19.26$^{+0.31} _{-0.35}$ & -0.047$^{+0.024} _{-0.025}$ & 5.97$^{+0.56} _{-0.45}$ & 0.393$^{+0.062} _{-0.041}$ & 18.05$^{+0.33} _{-0.31}$ & 14.19$^{+0.08} _{-0.08}$ & 353.27$^{+0.18} _{-0.17}$ & 44.4$^{+0.38} _{-0.32}$ & 222.6 & 4.09 & 5.49 \\
NGC 5289 & UGC 8699 & SPARC & T88 & 174.85$^{+0.68} _{-0.68}$ & 0.013$^{+0.019} _{-0.019}$ & 12.83$^{+0.83} _{-0.72}$ & 0.961$^{+0.03} _{-0.051}$ & 125.48$^{+49.28} _{-82.21}$ & 6.15$^{+0.15} _{-0.15}$ & 2523.84$^{+0.72} _{-0.72}$ & 367.78$^{+2.07} _{-1.97}$ & 51.9 & 1.45 & 1.33 \\
UGC 8837$^\ast$$^\star$ &  & SPARC & T88 & 34.72$^{+0.16} _{-0.17}$ & 0.402$^{+0.007} _{-0.007}$ & 16.43$^{+0.14} _{-0.14}$ & 0.968$^{+0.013} _{-0.014}$ & 145.45$^{+38.02} _{-49.89}$ & 15.46$^{+0.19} _{-0.16}$ & 146.89$^{+0.2} _{-0.2}$ & 89.44$^{+1.07} _{-0.9}$ & 417.4 & 1.55 & 7.66 \\
UGC 9992 &  & SPARC & T88 & 16.79$^{+0.45} _{-0.32}$ & 0.166$^{+0.036} _{-0.027}$ & 6.9$^{+0.53} _{-0.54}$ & 0.665$^{+0.134} _{-0.11}$ & 17.58$^{+1.92} _{-1.08}$ & 6.51$^{+0.06} _{-0.06}$ & 427.95$^{+0.22} _{-0.23}$ & 41.51$^{+0.51} _{-0.49}$ & 157.1 & 3.11 & 3.03 \\
UGC 12632 &  & SPARC & T88 & 48.88$^{+0.07} _{-0.08}$ & -0.077$^{+0.003} _{-0.003}$ & 10.53$^{+0.1} _{-0.11}$ & 0.871$^{+0.014} _{-0.013}$ & 29.35$^{+1.48} _{-1.7}$ & 42.43$^{+0.08} _{-0.09}$ & 421.33$^{+0.07} _{-0.06}$ & 109.99$^{+0.34} _{-0.31}$ & 581.6 & 2.39 & 6.3 \\
Mrk 209$^\dagger$ & UGCA281 & SPARC & T88 & 18.39$^{+4.62} _{-4.09}$ & -0.086$^{+0.142} _{-0.167}$ & 14.15$^{+2.66} _{-2.16}$ & 0.722$^{+0.234} _{-0.388}$ & 18.12$^{+3.76} _{-4.48}$ & 6.23$^{+0.1} _{-0.09}$ & 280.75$^{+1.08} _{-1.61}$ & 49.13$^{+1.94} _{-2.44}$ & 81.0 & 4.4 & 4.4 \\
IC 2574 &  & SPARC & R80 & 44.6$^{+0.11} _{-0.09}$ & 0.102$^{+0.001} _{-0.001}$ & 17.11$^{+0.07} _{-0.07}$ & 0.955$^{+0.003} _{-0.003}$ & 13.86$^{+0.46} _{-0.4}$ & 406.13$^{+0.26} _{-0.25}$ & 57.77$^{+0.04} _{-0.04}$ & 108.91$^{+0.11} _{-0.11}$ & 1875.9 & 14.59 & 27.83 \\
NGC 2403 &  & SPARC & R80 & 108.83$^{+0.02} _{-0.02}$ & -0.08$^{+0.0} _{-0.0}$ & 16.93$^{+0.03} _{-0.03}$ & 0.769$^{+0.001} _{-0.001}$ & 60.64$^{+0.2} _{-0.2}$ & 1435.68$^{+0.43} _{-0.44}$ & 138.76$^{+0.02} _{-0.02}$ & 234.26$^{+0.08} _{-0.08}$ & 3649.1 & 18.56 & 151.63 \\
M 109 & NGC 3992 & SPARC & R80 & 171.78$^{+0.41} _{-0.39}$ & -0.139$^{+0.011} _{-0.01}$ & 12.85$^{+0.41} _{-0.41}$ & 0.999$^{+0.001} _{-0.004}$ & 119.51$^{+55.16} _{-78.19}$ & 38.5$^{+0.4} _{-0.41}$ & 1152.79$^{+0.41} _{-0.4}$ & 361.56$^{+0.96} _{-0.97}$ & 105.3 & 17.08 & 17.91 \\
\enddata
\end{deluxetable}
\end{rotatetable}


\bibliography{main}
\bibliographystyle{aasjournal}


\end{document}